\documentclass[trackchanges]{aastex701}
\usepackage{booktabs, longtable}

\begin{document}

\title{Learning Gaia: A Generative Model Trained on the Gaia Data Release 3 Source Catalog}

\author[orcid=0000-0002-0104-3593]{John Soltis}
\affiliation{Space Telescope Science Institute, Baltimore, MD 21218, USA}
\email[show]{jsoltisd@gmail.com}

\begin{abstract}

I train a conditional flow matching model on the positions, photometry, and astrometry of $\sim$ 579 million sources from the Gaia Data Release 3 Source Catalog. Using a probabilistic masking scheme during training, I develop a model capable of generating mock Gaia catalog sources with and without partial measurements of real Gaia sources. The model can produce coarse-grained mock Gaia Source Catalogs that reproduce the Milky Way disk and Magellanic Clouds, and can impute photometric and kinematic information. The model fails to reproduce small, distinct subpopulations like $\omega$ Centauri. This model is a necessary step towards a foundation model trained on the entirety of Gaia data products, and provides a blueprint for future work with the upcoming Gaia Data Release 4.

\end{abstract}

\section{Introduction}\label{intro}
Astronomy, through its data and its questions, provides an ideal proving ground for machine learning methods. Astronomers and machine learning scientists alike have recognized this, resulting in a substantial growth in machine learning applications to astronomy \citep[for a review of AI/ML applications in astronomy see][]{Historical_Review_AI_Astro}. The most ambitious among these applications are so-called ``Foundation Models" \citep[e.g.,][]{AstroCLIP,RadioZooFM,GaiaFM_1,AstroPT,SpectraFM,AION}, often understood to be large models trained on very large multi-modal datasets that are capable of performing a variety of tasks not defined in training. These models are thought to learn meaningful representations from the data that allow them to perform downstream tasks. Famous examples include commercial large language models which are trained on very large datasets of text. Through training on very large datasets, these large language models are apparently able to learn a variety of information (e.g., the appropriate syntax for a Python package). Astronomical foundation models seek to apply this formula to astronomical information (e.g., to learn the relationship between a source's spectrum and the astrophysical object that produces it). Inspired by this work, I seek to develop a foundation-model-adjacent model using Gaia Data Release 3's Source Catalog.

The Gaia mission has obtained angular positions and photometry for more than 1.8 billion sources and has measured parallaxes, proper motions, and colors for nearly 1.5 billion of those sources \citep{Gaia_Mission,eDR3,eDR3_phot,DR3}. This extensive survey of Milky Way and other nearby stars has led to advances in our understanding of the structure and evolution of the Galaxy \citep[e.g.,][]{Poggio_2025}, stellar astrophysics \citep[e.g.,][]{Dubus_2024}, exoplanets \citep[e.g.,][]{Dantas_2026}, and cosmology \citep[e.g.,][]{Reyes_2023}.\footnote{See \citet{Brown_2021} for a more thorough, albeit pre-Data Release 3, review of Gaia's contributions.} Data Release 3 (DR3) of the mission contains data from the first 34 months of observation, and includes new or substantially expanded data products like spectra, radial velocities, and time series data \citep{eDR3,DR3,DR3_RV}. Given the ease of access to its data, the size of its dataset, the multi-modality of its data, and the many well-defined fields of inquiry potentially probed by it, the Gaia mission is especially well-suited for machine learning applications.

Prior foundation model work using Gaia data has focused on training models on cross-matched pairings between Gaia data and other datasets \citep{GaiaFM_1, Zhao_2026, Zhang_2026}. This, along with a well-justified interest in Gaia spectra \citep{Gaia_XP_1,Gaia_XP_2,Gaia_XP_3}, has meant that the largest piece\footnote{Largest in terms of unique objects.} of Gaia data has been mostly untapped by foundation model research. In this work I seek to utilize the largest single dataset within Gaia DR3, the Gaia Source Catalog, to train a machine learning model for a variety of downstream tasks. Such a model, if precise and accurate, would learn the relationship between source position on the sky, parallax, proper motion, photometry, and radial velocity as found in Gaia. It would be able to simulate realistic Gaia surveys without conditioning inputs, would be able to reproduce specific subpopulations of sources as observed by Gaia (e.g., member stars in $\omega$ Centauri, or Milky Way field Red Giant Branch stars) when given conditioning information, and would be able to impute missing parameters for sources in the Gaia survey using information from the rest of the survey and the existing observables for those sources. Such a model would therefore be capable of producing data-driven simulations of Gaia DR3 (complementing existing forward model simulations like \cite{Rybizki_2020}), aid in follow-up selection by inferring likely source properties, and provide a trained representation of Gaia data that could be repurposed for downstream tasks. In this paper I describe my efforts to produce such a model, its current capabilities, and its limitations.

The paper is organized as follows. In Section \ref{methods}, I discuss the conceptual design of the model (Section \ref{concept}), the architecture of the model (Section \ref{architecture}), model training (Section \ref{training}), and model sampling (Section \ref{sampling}). In Section \ref{data}, I describe the source catalog parameters selected (Section \ref{params}), the normalization procedure (Section \ref{norms}), and the dataset partitioning (Section \ref{partition}). In Section \ref{results}, I test the model in three regimes: simulating the data distribution of Gaia (Section \ref{mw_field}), conditionally simulating $\omega$ Cen (Section \ref{om_cen}), and imputing radial velocity (Section \ref{im_RV}). In Section \ref{conclusion}, I discuss the results and their implication for future work. In Appendix \ref{appendix_stats}, I include more extensive analysis of the models performance as a posterior and point estimator.

\section{Methodology}\label{methods}

\subsection{Concept}\label{concept}
The core task addressed by machine learning for this work is to transform an easily sampled base probability distribution $p_0$, typically a Gaussian distribution, into some difficult to sample target, or data, probability distribution, $p_{1}$. Doing so allows one to generate realistic Gaia-like sources following the data distribution in Gaia. One method for learning this mapping between a base and target distribution is called conditional flow matching. In conditional flow matching \citep{CFM,CFM_OT}, one trains a model to learn a \textit{velocity field}, $v_{\theta}$, that is used to calculate a continuous mapping between the base and target probability distribution. This velocity field defines intermediate probability distributions, $p_t$, which transform continuously from $p_{t=0}=p_0$ to $p_{t=1}=p_{1}$. One trains a neural network to learn this flow using discrete pairs from the base and target distributions, i.e., $x_0$ and $x_1$. To do this, I use linear interpolation with independent couplings \citep{CFM_OT}. The model learns the velocity flow at time $t$ between randomly drawn pairs of $x_0$ and $x_1$ in straight line segments, defined by the displacement $\Delta x \equiv x_1 - x_0$. The model takes $t$ and $x_t$ as inputs, where $x_t$ is defined by Equation \ref{eqn:x_t}.\footnote{During training $x_t$ is defined as $x_t \sim \mathcal{N}\left(x_0 + t \times \Delta x, \sigma \right)$, where $\sigma$ is a hyperparameter selected by the user. Early empirical exploration suggested that high $\sigma$ negatively impacted model precision, leading me to set $\sigma=10^{-8}$, effectively reducing $x_t$ to Equation \ref{eqn:x_t}.}
\begin{equation}\label{eqn:x_t}
    x_t = x_0 + t\times\Delta x ; \ t\sim\mathcal{U}(0,1)
\end{equation}
Given inputs $x_t$ and $t$,\footnote{During training $t$'s are randomly generated from the user's choice of probability distribution. I found a uniform distribution between 0 and 1 was most effective.} a model with weights $\theta$ is trained by minimizing the mean squared error loss function below:
\begin{equation}\label{eqn:simple_model}
    \mathcal{L}(\theta) = \mathbb{E}_{x_0\sim p_0, x_1 \sim p_1, t \sim \mathcal{U}(0,1)}||v_\theta(x_t,t) - \Delta x||^2
\end{equation}
where $v_\theta \left(x_t, t\right)$ is the output of the model.
A trained model can then be used to sample from $p_1$ using samples from $p_0$ by solving the following ordinary differential equation:
\begin{equation}\label{eqn:sampling_integral}
    \frac{dx}{dt} = v_\theta(x_t,t); \ x(t=0) = x_0
\end{equation}

Note that in Equation \ref{eqn:simple_model} this hypothetical model only depends on $x_t$ and $t$. In this work, however, I would like to be able to simulate the model given some existing partial information about the source, $x_c$. This conditioning vector, which is a masked version of the possible observations for a given source, is an additional input to the model. Hereafter, when referring to conditioning information, conditioning parameters, or conditioning patterns, I am referring to $x_c$. To create a model capable of handling observations of varying completeness, I use a data masking scheme based on \citet{CFMI}.\footnote{The major modification between \citet{CFMI} and this work is that in this work the input and output masks are independent.} I introduce three masks: an initial mask, $m_0$, an input mask, $m_{in}$, and an output mask, $m_{out}$. Each mask is a boolean vector, equal in length to $x$, which is used to determine which parameters are visible to the model. The initial mask, $m_0$, is determined directly by the Gaia Source Catalog. For example, the initial mask value corresponding to the Galactic longitude for every source is 1 (unmasked), but the mask value corresponding to radial velocity is 0 (masked) for most sources (see Section \ref{params}). In training, both the input mask, $m_{in}$, and the output mask, $m_{out}$, are random subsets of the initial mask. For every source, a random real number, $p_{in}$, is drawn from a uniform random distribution between 0 and 1 to the $1/4$ power, i.e., $\mathcal{U}(0,1)^{1/4}$.\footnote{Note that masks are constructed in the normalized parameter space, thus $\ell$ is split into two parameters that are masked independently. See Section \ref{norms} for more information on the normalizations.}$^{\rm{,}}$\footnote{I choose a $\mathcal{U}(0,1)^{1/4}$ for the input mask to over-sample instances of higher masking. It is more difficult for the model to learn the unconditional case, or cases where a single parameter is provided as conditioning information, than it is for the model to learn the completely unmasked or mostly unmasked case.} Similarly, a second random real number, $p_{out}$, is drawn from a uniform random distribution between 0 and 1, i.e., $\mathcal{U}(0,1)$. These numbers are then compared to two random number vectors, $\hat{r}_{in}$ and $\hat{r}_{out}$, both drawn from $\mathcal{U}(0,1)$ with the same length as $x_t$. Everywhere that $\hat{r}_{in} \leq p_{in}$, $\tilde{m}_{in} = 0$, meaning that the corresponding parameter is masked. Likewise, everywhere that $\hat{r}_{out} \leq p_{out}$, $\tilde{m}_{out} = 0$. If $\hat{r}_{in} > p_{in}$ then $\tilde{m}_{in} = 1$, and similarly, if $\hat{r}_{out} > p_{out}$ then $\tilde{m}_{out} = 1$. These randomly generated masks are then multiplied with the initial mask, $m_0$, to produce the input ($m_{in}$) and output ($m_{out}$) masks (see Equations \ref{eqn:m_in} and \ref{eqn:m_out}).\footnote{In training, when multiplying the initial mask by $\tilde{m}_{out}$ results in an entirely masked source (i.e., $m_{out} = 0$ for all parameters), a random measured parameter from that source is chosen to be unmasked.} 
\begin{equation}\label{eqn:m_in}
    m_{in} = \tilde{m}_{in} \times m_0
\end{equation}
\begin{equation}\label{eqn:m_out}
    m_{out} = \tilde{m}_{out} \times m_0
\end{equation}
The input mask controls which existing observations the model sees during training. The conditioning variable for the model is defined as:
\begin{equation}\label{conditioning_variable}
    x_c \equiv x_1 \times m_{in}
\end{equation}
By setting the input mask randomly and training for many epochs, the model sees many variations in conditioning variables, and learns the relationships between many different parameter sets. Note that when $m_{in}=0$ for all observables, the model receives no conditioning information from $x_c$. I define this as as the unconditional case. 

The output mask is used to mask $x_t$ and to control which parameters define the loss function. In doing so, I ensure that the model learns different combinations of input and output masks, training the model for a variety of tasks. Combining the above information, I arrive at the final loss function below:
\begin{equation}\label{eqn:full_model}
    \mathcal{L}(\theta) = \mathbb{E}||v_\theta(x_t\times m_{out},t,x_c,m_{in},m_{out}) - \Delta x||^2_{m_{out}}
\end{equation}
where $\mathbb{E}||...||^2_{m_{out}}$ denotes that the mean squared error is averaged over values that are not masked by $m_{out}$.

\begin{figure*}[ht!]
\plotone{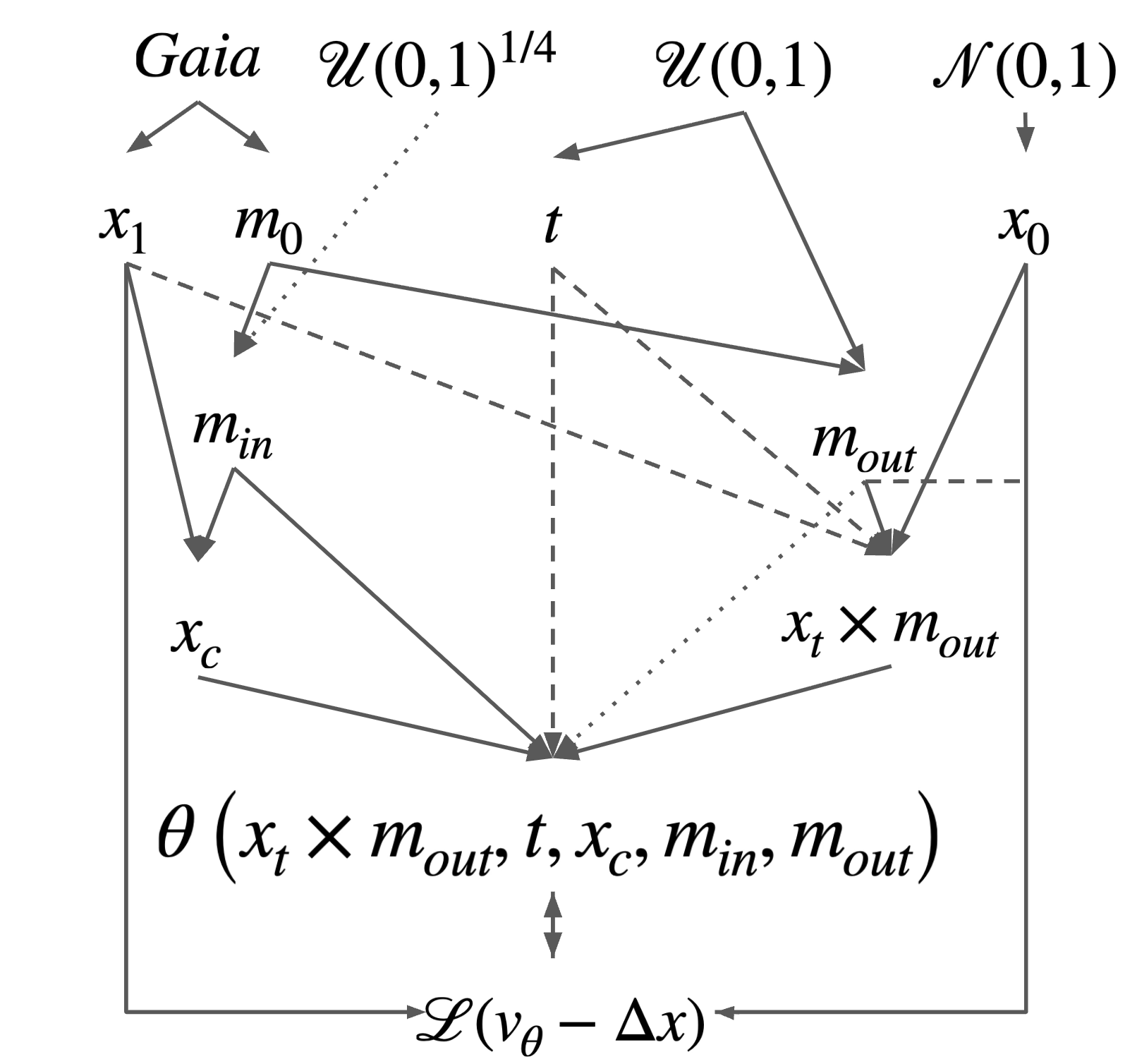}
\caption{A flow chart representing the complex flow of information during model training. Arrows represent the direction of travel for information, with dotted and dashed lines used for clarity. $x_1$ represents measurements taken from the Gaia Source Catalog. $m_0$ represents the ``initial mask", i.e., which measurements are available in the Gaia Source Catalog for a given source. $t$ represents the flow matching time variable, drawn from a uniform distribution. $x_0$ is drawn from a normal distribution. $m_{in}$ is the input mask, which is the product of a random mask constructed using a uniform distribution to the one fourth power, $\tilde{m}_{in}$, and $m_0$ (see Section \ref{concept}). $m_{out}$ is the output mask, the product of a random mask constructed using a uniform distribution, $\tilde{m}_{out}$, and $m_0$. The conditioning information for the model, $x_c$, is the product of the Gaia measurements, $x_1$, and the input mask, $m_{in}$. $x_1$, $x_0$, and $t$ combine to create $x_t$ (see Equation \ref{eqn:x_t}) which represents an intermediate point on the path between $x_0$ and $x_1$. $x_t$ is then multiplied by the output mask, $m_{out}$. The model, $\theta$, is then fed $x_t \times m_{out}$, $t$, $x_c$, $m_{in}$, and $m_{out}$. The model output, $v_{\theta}$, is then used to calculate the loss, $\mathcal{L}$, wherein the mean squared error of the model output compared to $\Delta x$ is calculated. The loss is written as $\mathcal{L}(v_\theta-\Delta x)$, unlike in Equations \ref{eqn:simple_model} and \ref{eqn:full_model}, to emphasize the role of $\Delta x$, the difference between $x_1$ and $x_0$. The loss is then masked by the output mask, $m_{out}$. This masked loss is then used to iteratively update the model weights.} 
\label{fig:model}
\end{figure*}

The model accepts an input mask, output mask, conditioning vector, time variable, and $x_t\times m_{out}$ vector. In turn, the model outputs a velocity vector, $v_\theta \left(x_t, t\right)$, which then allows the user to sample possible $x_1$ from $P(x_1|x_c,m_{in},m_{out})$. Figure \ref{fig:model} shows a flow chart representation of the model during training.
\subsection{Architecture}\label{architecture}

The model is constructed using eight fully connected hidden layers with 512 nodes each. The output of the input layer is added to the output of each subsequent layer, and the sum of these outputs is then used as an input to the next layer in the sequence (i.e., there is an additive skip connection between all layers and the input layer).\footnote{Due to a bug detected after training, the output of the input layer is added to the final layer twice.} For each hidden layer, a layer normalization \citep{LayerNorm} is applied to its input. A GELU activation function \citep{GELU} is used to supply the necessary nonlinearity for the input and hidden layers, but no nonlinearity is employed on the final output of the model. The time variable $t$ (defined in Equation \ref{eqn:x_t}) is decomposed into a 64-dimensional Gaussian Fourier time embedding \citep{Tancik_2020}. As noted in Section \ref{concept}, the model accepts the time variable $t$, the masked interpolated vector $x_t\times m_{out}$, the conditioning vector $x_c$, the input mask $m_{in}$, and the output mask $m_{out}$. Thus the initial input layer of the model accepts a 104-dimensional vector; 10 each for $x_t\times m_{out}$, $x_c$, $m_{out}$, $m_{in}$ (see Section \ref{data}) and 64 for $t$. The output layer produces a 10-dimensional vector. The model was constructed, trained, and sampled using the PyTorch Python package \citep{Pytorch}.

\subsection{Training}\label{training}
After experimentation with several different learning rate schedules, a flat learning rate of $10^{-4}$ was chosen. I used an Adam optimizer \citep{kingma2017adammethodstochasticoptimization}.\footnote{Technically an AdamW optimizer was used \citep{AdamW}, but weight decay for the final run was set to 0, making it equivalent to Adam.} The fiducial model was chosen after 3200 epochs of training, a number determined by computational and time constraints. I used a batch size of 50,000 sources. Batches were read directly from catalog hdf5 files, meaning that data was correlated by position. To mitigate the impact of this on training, the gradient updates of 10 batches were accumulated before weights were updated.\footnote{Due to a bug detected after training, if less than 10 batches are accumulated at the end of an epoch, the gradients of those batches are carried over to the next epoch.} Training was parallelized over 4 GPUs reading in independent files, meaning that updates contained sources from a variety of positions. An exponential moving average of the model weights was used during sampling to reduce noise in trained weight values. This has been found to improve model performance \citep[e.g.,][]{Song_2020}.

\subsection{Sampling}\label{sampling}
As discussed in Section \ref{concept}, one can obtain $x_1\sim P(x_1|x_c, m_{in}, m_{out})$ by solving Equation \ref{eqn:sampling_integral} (now including conditional information and masks). This can be done numerically. I implement a fourth order Runge-Kutta solution using 50 steps.\footnote{Increasing the step count did not noticeably change results.} During sampling the input mask is determined solely by the user, and the output mask is typically defined as the true initial mask of the source being conditioned on, unless otherwise stated.

\section{Data}\label{data}
\subsection{Source Catalog Parameters}\label{params}
This work uses data from Gaia DR3's Source Catalog \citep{eDR3,DR3}. I use 9 parameters from the catalog. The parameters are the Galactic longitude $\ell$,\footnote{$\ell$ in the Gaia Source Catalog ranges from 0 to 360 degrees, however for plotting purposes I convert it to 180 to -180 degrees, with 0 in the plot range corresponding to 0 in the original range. This re-framing centers the Galactic bulge.} the Galactic latitude, $b$, the parallax, $\varpi$, the proper motion in Galactic longitude, $\mu_{\ell*}\equiv\mu_{\ell} \cos\left(b\right)$, the proper motion in Galactic latitude, $\mu_b$, the Gaia $G$ magnitude, the Gaia $G_{BP}$ magnitude, the Gaia $G_{RP}$ magnitude, and the radial velocity $v_r$. In using Galactic coordinates for the angular position and proper motion of Gaia sources, I seek to make the relationship between observations and Galactic spatial and kinematic structure more obvious to the model. I opt to leave the parallax zero point \citep{Gaia_ZP} uncorrected, so that model estimates more directly reflect Gaia data and so that user-selected corrections can be applied to model samples.

Within the Gaia Source Catalog 100\% of sources have $\ell$ and $b$ positions, 99.7\% have $G$ magnitudes, 81\% have parallaxes and proper motions, 85.1\% have $G_{BP}$ magnitudes, 85.8\% have $G_{RP}$, and 2\% have radial velocities \citep{DR3}. For a full description of the limitations and inter-dependencies of the Gaia Source Catalog see \cite{eDR3} and \cite{DR3}.

\subsection{Data Normalization}\label{norms}
Data normalization is an essential aspect of model training. The distributions of input parameters must be transformed such that they cover roughly equivalent ranges and have roughly equivalent variances. For this work, I transform all parameters so that the bulk of their distributions fall between $-1$ and $1$. To transform Galactic longitude, $\ell$, while preserving information, I decompose the parameter into cosine and sine functions.
\begin{equation}\label{eqn:l_norm}
    \hat{\ell}_{norm} \equiv \left(\cos(\ell), \sin(\ell)\right)
\end{equation}

Galactic latitude, $b$, admits the simplest normalization, requiring a simple rescaling by $90^\circ$.
\begin{equation}\label{eqn:b_norm}
    b_{norm} \equiv \frac{b}{90^\circ}
\end{equation}
For the remaining parameters, I balance concerns about the total range of the normalized parameter distribution with concerns about the variance of that distribution. I use an empirical z-score-like transformation, subtracting the median value of the parameter from the training data and then dividing by the width of the central 95\% of the data.\footnote{Percentiles were calculated using 2\% of the total training set data.}
\begin{equation}\label{eqn:other_norm}
    a_{norm} \equiv \left(\frac{a-a_{50}}{a_{97.5} - a_{2.5}}\right)
\end{equation}
The result of these transformations is that the 9 Gaia Source Catalog parameters become 10 model input parameters (two for $\ell$, see Equation \ref{eqn:l_norm}). All training and sampling is done in normalized space, but results are given in physical units. NaN values in the source catalog are converted to 0 in normalized units. $\hat{\ell}_{norm}$ is converted into a unit vector before being transformed back to physical units.

\subsection{Dataset Partitioning}\label{partition}
For computational reasons, I limit the dataset to 40\% of the available data ($\sim725$ million sources), chosen at random using the source catalog's random index, creating an unbiased subset of the source catalog. Within that 40\%, I partition the data into training (80\%, $\sim579$ million sources), validation (10\%, $\sim$ 72 million sources) and testing (10\%, $\sim$ 72 million sources). The training partition was used for training the model, the validation set was used for training evaluation and model hyperparameter selection, and the test set was used for the results shown in Section \ref{results} (unless otherwise noted). Note that no data quality cuts were performed. The aim of this work is to produce a model capable of reproducing the native Gaia distribution of sources, matching its noisiness and selection function.

\section{Results}\label{results}
In the following section I present three tests of model performance: simulating the full catalog in Section \ref{mw_field}, simulating $\omega$ Cen in Section \ref{om_cen}, and imputing radial velocities in Section \ref{im_RV}. These three tests focus on three different modes of usage: simulating the full catalog, simulating subpopulations, imputing missing measurements. In Appendix \ref{appendix_stats}, I include TARP \citep{TARP} plots for a subset of the possible patterns and parameters. I also include tables with more comprehensive TARP and root mean squared error (RMSE) information (Tables \ref{tab:tarp:test:a}, \ref{tab:tarp:test:b}, \ref{tab:tarp:test:c}, \ref{tab:tarp:test:d}, \ref{tab:skill:test:a}, \ref{tab:skill:test:b}, \ref{tab:skill:test:c}, and \ref{tab:skill:test:d}).

\subsection{Simulating the Full Catalog}\label{mw_field}
The first test of the model is how well it is able to reproduce the Gaia survey on large scales. I do this by first creating a small sample of the test set, randomly drawing $\sim 1$\% of the total test set (723,048 sources). I use this sample as conditioning information for the model. For example, if I want to probe the model's performance when given only the positions of sources, I supply the model with all 723,048 ($\ell$, $b$) pairs as conditioning information. From each ($\ell$, $b$) pair, following the sampling procedure outlined in Section \ref{sampling}, I generate model samples for the full parameter set. In addition to the ($\ell$, $b$) pair, I also supply the model with an input mask, which in this case is masked for everything but $\ell$ and $b$, and an output mask, which matches the true output mask for that specific source. Given the size of the test sample, I generate only one model sample per real source. I repeated this procedure for many different possible conditioning patterns. In cases where a parameter is not available for a real source but is included in the conditioning pattern, e.g., radial velocity is part of the drop $\varpi$ pattern but most sources do not have a radial velocity, I choose to include these incomplete sources. 

\begin{figure*}[ht!]
\plotone{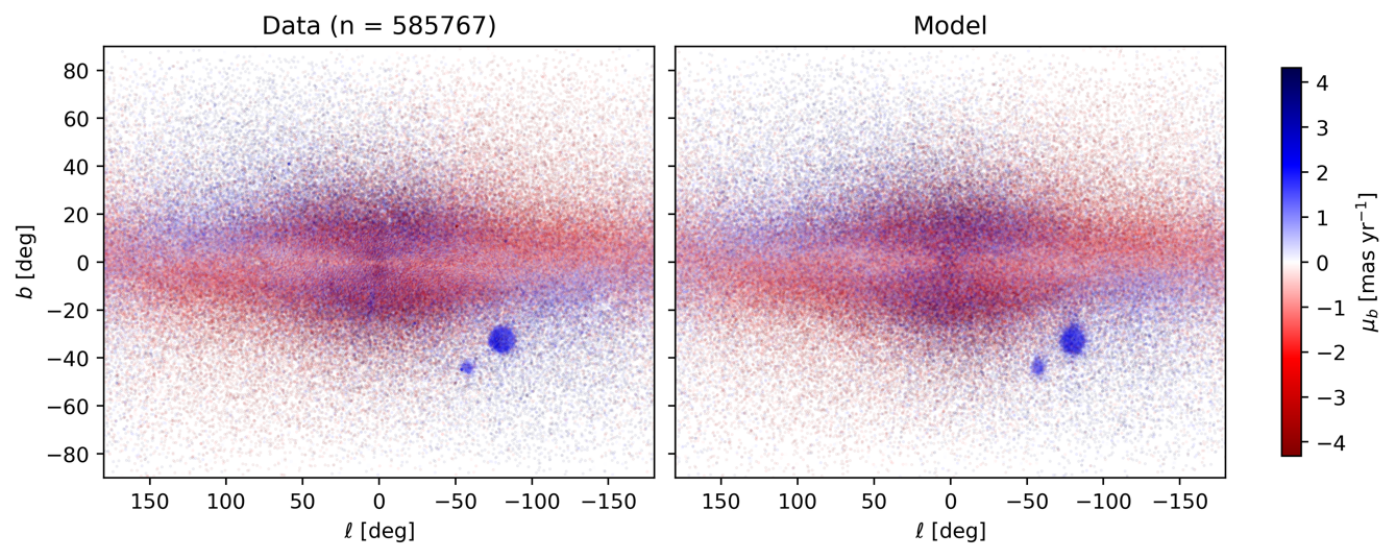}
\caption{Shown in the figure are the Galactic longitude ($\ell$), latitude ($b$), and proper motion in the latitude direction ($\mu_b$, in color) for a sample of sources from the Gaia Source Catalog (left), and an unconditionally generated sample of sources from the trained model (right). Only the fraction of sources in the subsample that have measured $\mu_b$ are used, leaving 585,767 out of 723,048. The model generated sample clearly shows features from the data like the Milky Way disk (center band) and the Magellanic Clouds (blue dots in the bottom right). $\ell$ is plotted from 180 to -180 degrees so that the center of the plot corresponds to the Galactic bulge. }
\label{fig:MW_field_uncond_pmb}
\end{figure*}

\begin{figure*}[ht!]
\includegraphics[width=\textwidth]{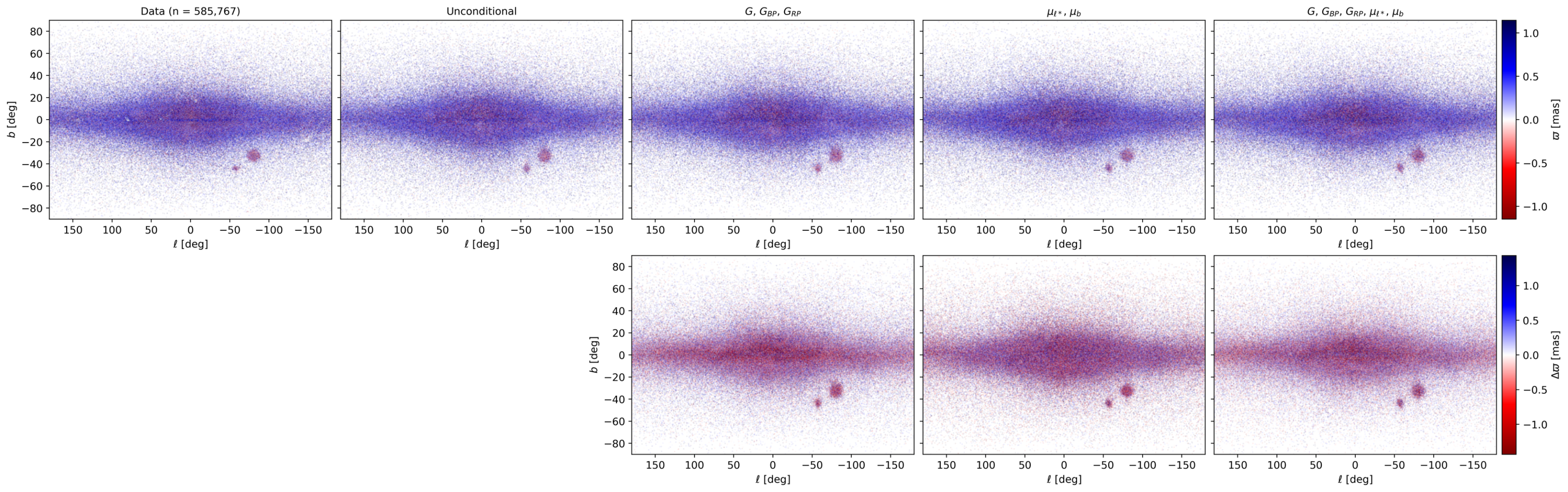}
\caption{Shown in the figure are the Galactic longitude ($\ell$), latitude ($b$), and parallax ($\varpi$, in color) for a sample of sources from the Gaia Source Catalog (leftmost column), and conditionally generated samples of sources from the trained model (right columns). The second leftmost column shows when the trained model is provided no conditioning information, as in Figure \ref{fig:MW_field_uncond_pmb}. The center column shows when photometry ($G$, $G_{BP}$, $G_{RP}$) is provided. The subsequent column shows model samples when proper motion ($\mu_{\ell*}$, $\mu_b$) is provided. The rightmost column shows when both photometry and proper motion information are provided as conditioning information. In instances where a parameter is unavailable in the real data (e.g., a source does not have a $G_{BP}$ in the real Gaia Source Catalog) that conditioning input is left masked, and the remaining conditioning information is provided. The second row shows the difference between the true data sources and the model predictions conditioned on those sources. No errors are shown for the unconditionally generated model samples, as there are no data sources to compare them to. Note the improvement in the fidelity of model generated samples between the unconditional and conditional cases. Including proper motion noticeably improves the sharpness of the Magellanic Clouds, while including both photometry and proper motion as conditioning information improves the overall sharpness of fine Milky Way disk features.}
\label{fig:MW_field_uncond_plx}
\end{figure*}

\begin{figure*}[ht!]
\includegraphics[width=\textwidth]{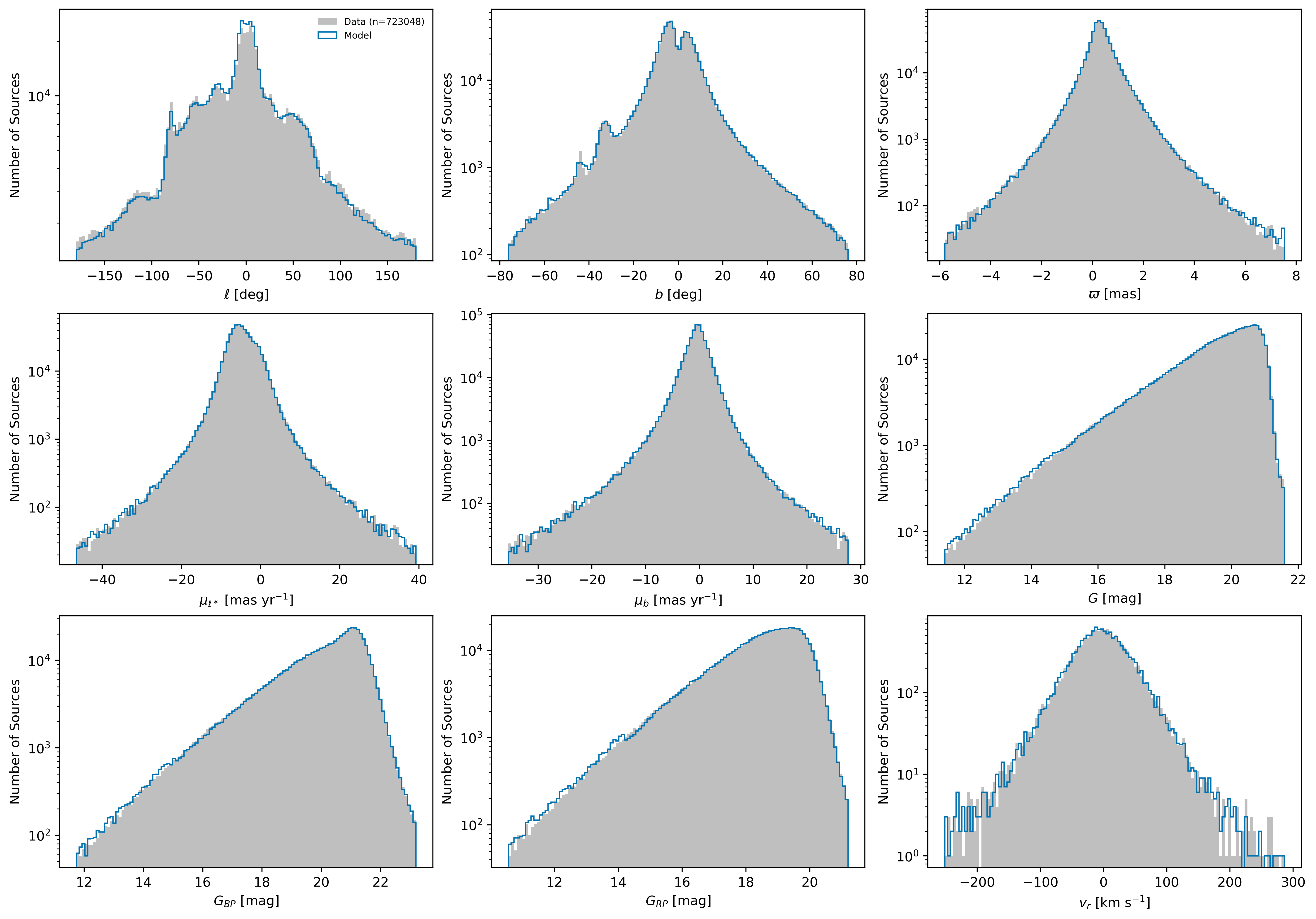}
\caption{The marginal distributions of Gaia parameters for a sample of the data (gray) and the unconditioned model (blue). Distributions are shown from their $0.1^{\rm{st}}$ to $99.9^{\rm{th}}$ percentiles. The model captures the shape and scale of the distributions for all parameters, albeit declining in quality for bright sources ($G$, $G_{BP}$, $G_{RP}$ $\lesssim 15$). Despite the limited number of radial velocities $v_r$ in the dataset, the model reproduces the distribution of the observed data.}
\label{fig:margins_uncond}
\end{figure*}

Figure \ref{fig:MW_field_uncond_pmb} shows position and Galactic latitude proper motions for a model generated sample (right) with no conditioning information provided,\footnote{Even the unconditional case is arguably conditional. As noted in Section \ref{concept}, the model takes the output mask as an input. This mask contains information on which parameters are available in the Gaia Source Catalog for a given source. The selection function for each parameter provides the model with useful information, and choosing an unrepresentative $m_{out}$ results in model generated samples that do not match the global distribution of sources in the Gaia catalog. In this sense, the model is conditioned on the Gaia selection function. Explicitly reproducing the selection function as a model prediction so that it could be compared to explicit estimates of the selection function, such as those in \citet{Everall_2022} or \citet{Castro_Ginard_2023}, is left to future work.} compared to a subset of the test data (left). The figure clearly illustrates the model's primary success; it has learned large scale features of the Gaia Source Catalog, like the Milky Way disk and the Magellanic Clouds. Figure \ref{fig:MW_field_uncond_plx} shows the model predictions in angular position and parallax space. It demonstrates the importance of different feature subsets for improving model performance, e.g., providing proper motions improves the model resolution of the Magellanic Clouds. Figure \ref{fig:margins_uncond} shows the marginal distributions for all the parameters included in model training, for both a sample of the test set and an unconditioned model generated sample. The model reproduces the parameter distributions well, even for radial velocity, of which it sees many fewer examples.

\subsection{Subpopulations: $\omega$ Cen}\label{om_cen}
\begin{figure*}[ht!]
\plotone{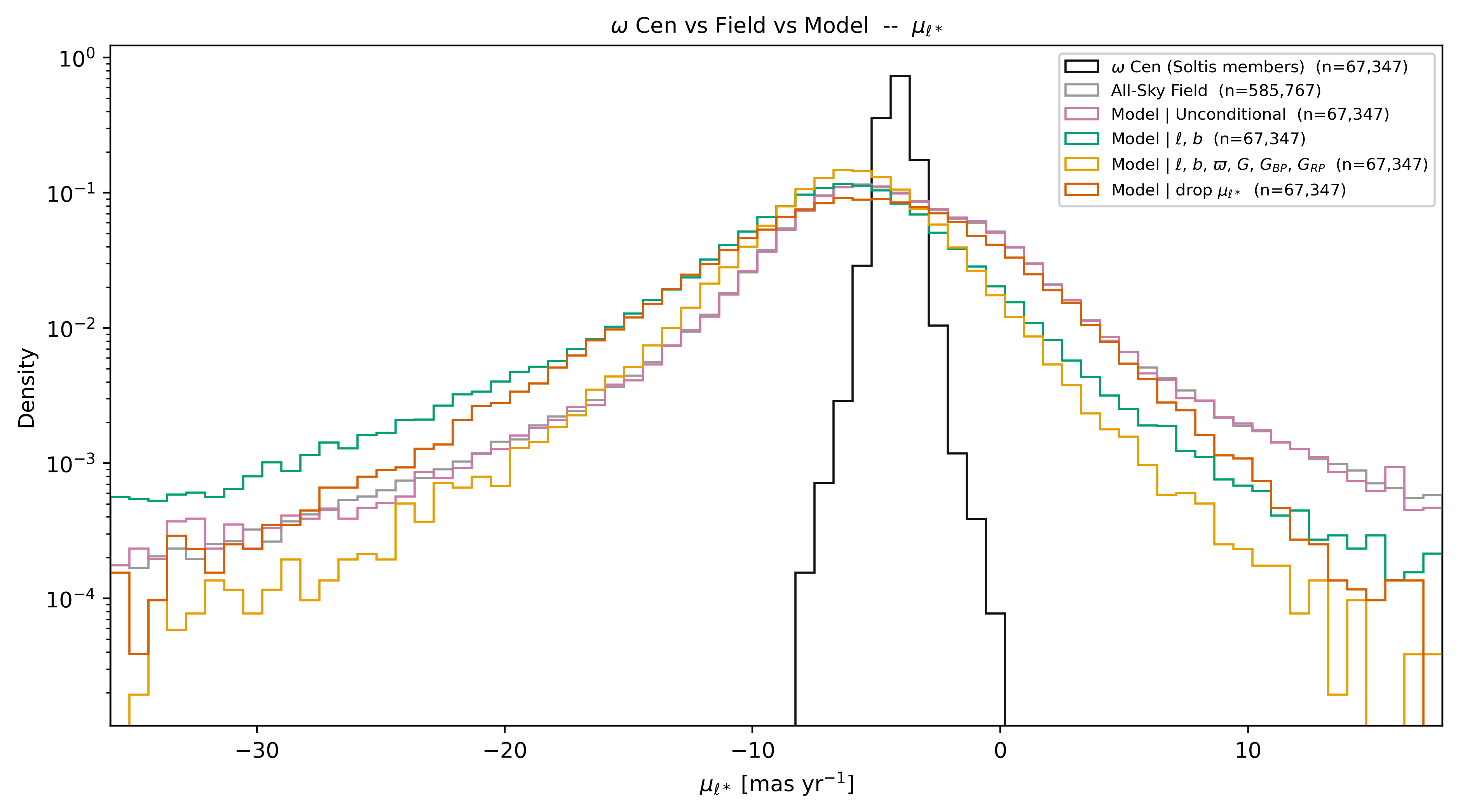}
\caption{Shown are marginal distributions of the proper motion in Galactic longitude for members of $\omega$ Cen (black), a sample of the full Gaia sky (gray), unconditionally generated sources (pink), and generated sources from models conditioned on position (green), position, parallax, and photometry (yellow) and all available parameters but $\mu_{\ell*}$ (orange). $\omega$ Cen forms a sharp distinct peak in $\mu_{\ell*}$, in part due to the selection criteria chosen in \citet{Soltis_21}, that the model fails to replicate in all cases. Instead, the model tends towards the all-sky field.}
\label{fig:Om_Cen_pmb_hist}
\end{figure*}

The second test of the model is its ability to faithfully reproduce subpopulations of sources found within the Gaia Source Catalog. I perform a simple test of the model's knowledge of subpopulations, identical to the procedure described in Section \ref{mw_field}, albeit using sources selected as members of $\omega$ Cen by \citet{Soltis_21}. Using this member selection,\footnote{I omit the final Color Magnitude Diagram (CMD) cut, thus leaving me with 67,347 sources. The CMD cut in \cite{Soltis_21} only removed 880 sources, which has negligible impact on the analysis done here.} I compare model generated sources to the cluster population. In Figure \ref{fig:Om_Cen_pmb_hist}, I show histograms of $\mu_{\ell*}$ for $\omega$ Cen members (black), a random sample of sources from the test set (gray), unconditioned model generated sources (pink), sources generated by a model conditioned on the angular positions of $\omega$ Cen members (green), sources generated by a model conditioned on the positions, parallaxes, and photometry of $\omega$ Cen members (yellow), and sources generated by a model conditioned on all valid\footnote{As in Section \ref{mw_field}, parameters that are missing for a source in the catalog are left masked.} parameters but $\mu_{\ell*}$ of $\omega$  Cen members.\footnote{Note that the set of sources associated with $\omega$ Cen includes sources in the model's training set. Despite this leakage, the model still performs poorly.} The model clearly fails to accurately reproduce the sharp true distribution of proper motion of $\omega$ Cen, even when it is provided with exact positions. Providing the model with additional conditional information fails to meaningfully improve the fidelity of generated sources. Given the size and brightness of $\omega$ Cen, and its distinctness in proper motion space, this suggests that smaller, less visible, and more subtly distinct populations will be more poorly resolved.

\subsection{Imputing Radial Velocity}\label{im_RV}
As discussed in Section \ref{data}, the Gaia Source Catalog does not provide measurements for all available parameters. Moreover, measurements of particular interest, like the radial velocity, are not available for the vast majority of sources in the catalog. A model that is able to impute said parameters could provide a useful prediction for the true distribution of those parameters for the whole catalog. There are theoretical limitations to the model's imputation powers, however. Missing parameters from the Gaia catalog are not missing completely at random \citep{Rubin_76}, but are missing because they come from sources with different values for other observed parameters. For example, very faint sources, $G>21$, tend to have only two parameter astrometry solutions (i.e., only positions) \citep{eDR3}. A model trained on the real Gaia survey will therefore impute the distribution of velocities for very dim stars using a prior derived from the distribution of a different population of stars, i.e., brighter stars. Even so, the model is capable of using the full 9 parameter input, therefore the model can learn the Milky Way kinematic structure as a function of position on the sky (see, e.g., Figure \ref{fig:MW_field_uncond_pmb}) and can use that information to produce better predictions for missing kinematic information. Moreover, the error on model estimates of parameters can only be measured when the true parameter value is known. Thus a model's imputation power can only be evaluated using the model's performance on sources with real observations as a proxy, sources which are known to come from a different population than sources without real observations. Previous work has demonstrated the importance of imputing radial velocity, and the utility of neural networks trained on the Gaia Source Catalog for accomplishing that task \citep{Dropulic_2023,Naik_2024}. A thorough comparison of this model's imputation power as compared to other imputations of $v_r$, such as \citet{Dropulic_2023} and \citet{Naik_2024}, is left to future work.

\begin{figure*}[ht!]
\includegraphics[width=\textwidth]{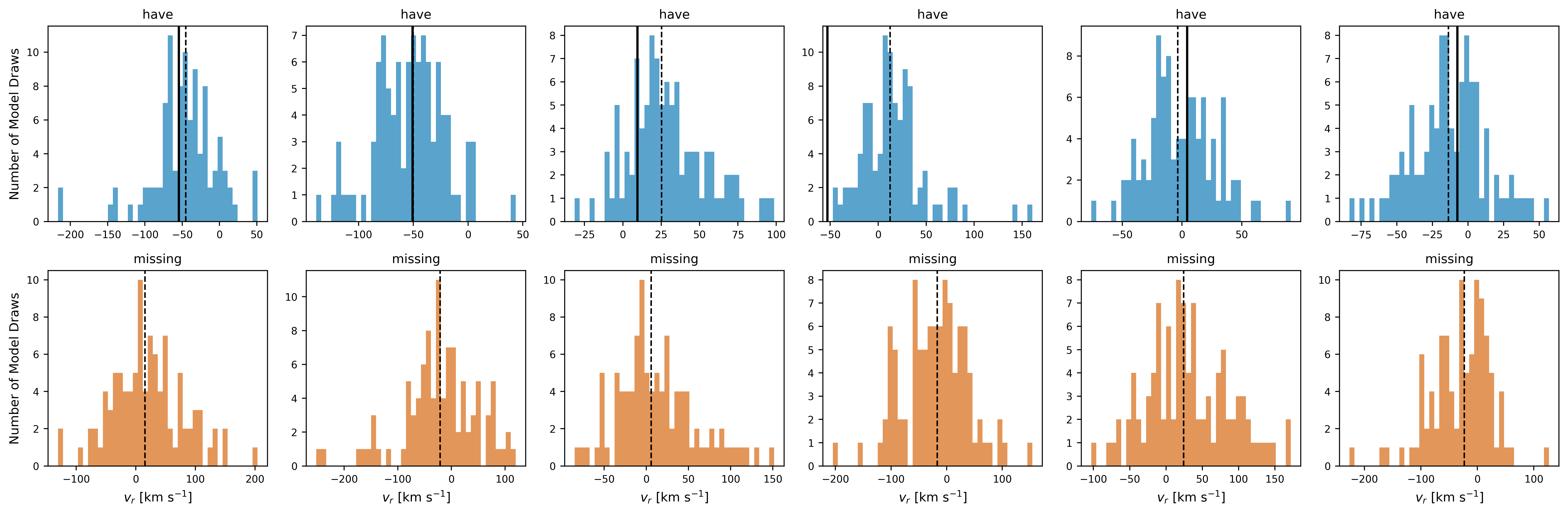}
\caption{Each plot shows 100 model generated radial velocities, $v_r$, conditioned on information ($\ell$, $b$, $\varpi$, $\mu_{\ell*}$, $\mu_{b}$, $G$, $G_{BP}$, $G_{RP}$) from a single source from the Gaia Source Catalog. The twelve sources were selected at random. In the top row, the distribution of model generated $v_r$ is shown as a blue histogram, with its median displayed as a black dashed line. The Gaia measured $v_r$ is shown as a black line. The bottom row displays the histograms of model generated $v_r$ for sources (in orange) without a Gaia measured $v_r$.}
\label{fig:single_imputation}
\end{figure*}

Similar to Section \ref{mw_field}, I test the model's imputation power by sampling the test set and using information from real sources as conditioning information for the model. Unlike Section \ref{mw_field}, I generate 100 samples per source to capture the simulated posterior distribution of the model (see Figure \ref{fig:single_imputation}). Also unlike Section \ref{mw_field}, I exclusively use sources that contain the complete pattern, e.g., in the case of drop $\varpi$ I only use sources with existing values for ($\ell$, $b$, $\mu_{\ell*}$, $\mu_{b}$, $G$, $G_{BP}$, $G_{RP}$, $v_r$). I use 1000 sources for each pattern I evaluate. I perform this test for two sets of sources, those that have the parameter I am asking the model to impute (``have"), and those that do not (``missing"). For example, given ($\ell$, $b$, $\varpi$, $\mu_{\ell*}$, $\mu_{b}$, $G$, $G_{BP}$, $G_{RP}$) (the drop $v_r$ pattern) I can use the model to simulate $v_r$ for sources with real $v_r$ values in the Gaia Source Catalog, and for sources with no $v_r$ values in the Gaia Source Catalog (see Figure \ref{fig:single_imputation}). These ``have" and ``missing" populations can have quite distinct parameter distributions for the other parameters, especially in photometry (see Figure \ref{fig:v_r_populations}). I use output masks ($m_{out}$) identically to the data for the ``have" case, but force the missing parameter in the ``missing" case to be unmasked. Thus imputation for the ``missing" case is an extrapolation using information from the ``have" case.

\begin{figure*}[ht!]
\plotone{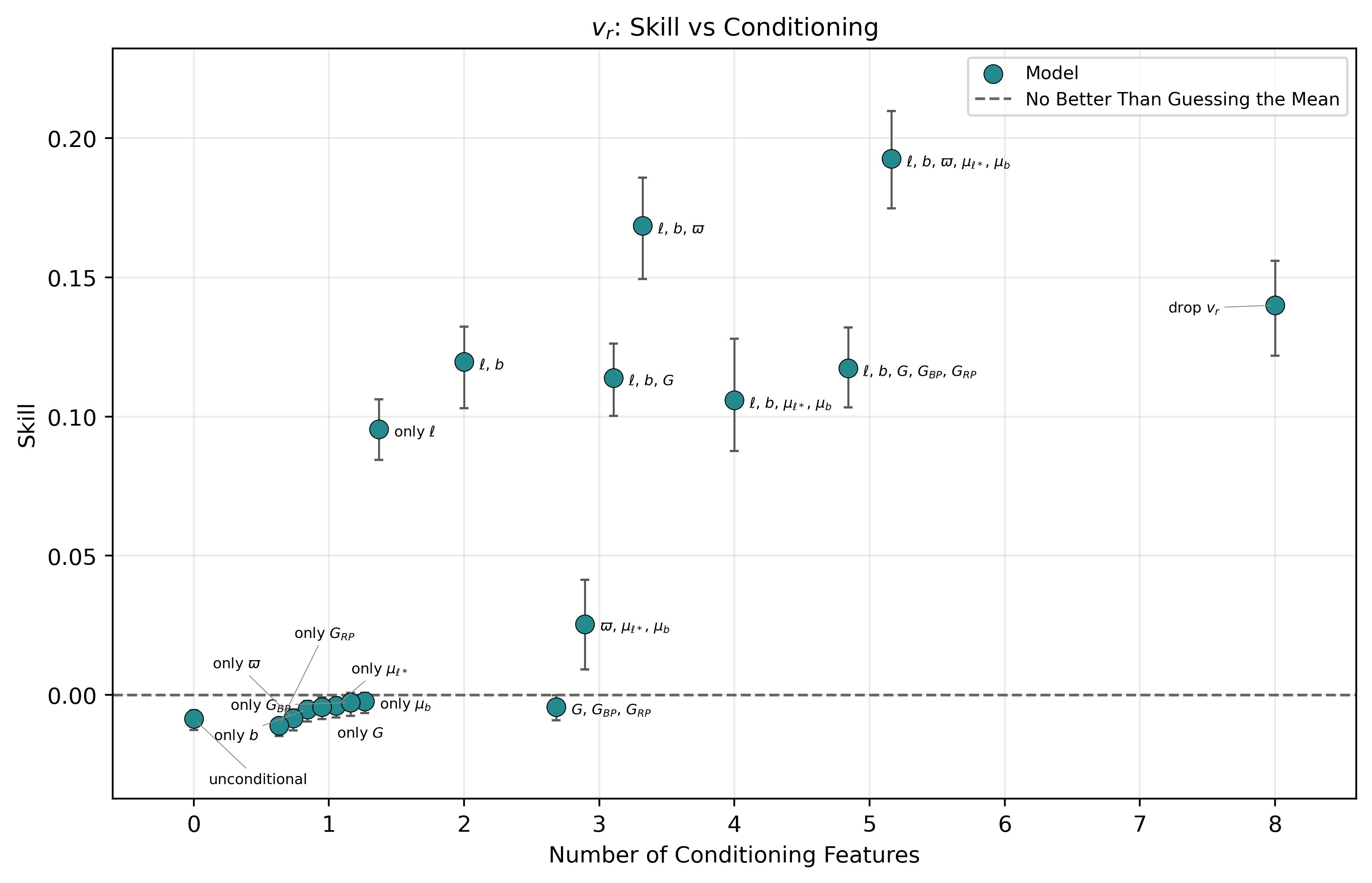}
\caption{The $v_r$ skill (defined in Equation \ref{eqn:skill}) is plotted against the number of conditioning features for different conditioning patterns. Error bars show the central $68$\% confidence interval of 1000 bootstraps, where each bootstrap is 1000 sources drawn from the original 1000 source set, but with replacement. A small offset is applied so that points do not all fall on integer x-axis values. Skill values greater than 0 imply the model imputes better than guessing the mean of the dataset. Note that the dataset in this case is 1000 sources used as conditioning information for the model. Since all sources must strictly satisfy the conditioning set, i.e., have all possible parameters in the conditioning set, this means that the sample changes for each conditioning pattern set. Nevertheless, it is apparent that conditioning patterns with the Galactic longitude, $\ell$, tend to provide additional constraining power, whereas those without do not. While calculated on different 1000 source sample, the results found in Table \ref{tab:skill:test:d} qualitatively agree with this result. The importance of $\ell$ is likely due to the relationship between $v_r$, $\ell$, the Solar system's motion within the Galaxy, and the Milky Way's rotation.}
\label{fig:impute_skill_and_confidence}
\end{figure*}

\begin{figure*}[ht!]
\includegraphics[width=\textwidth]{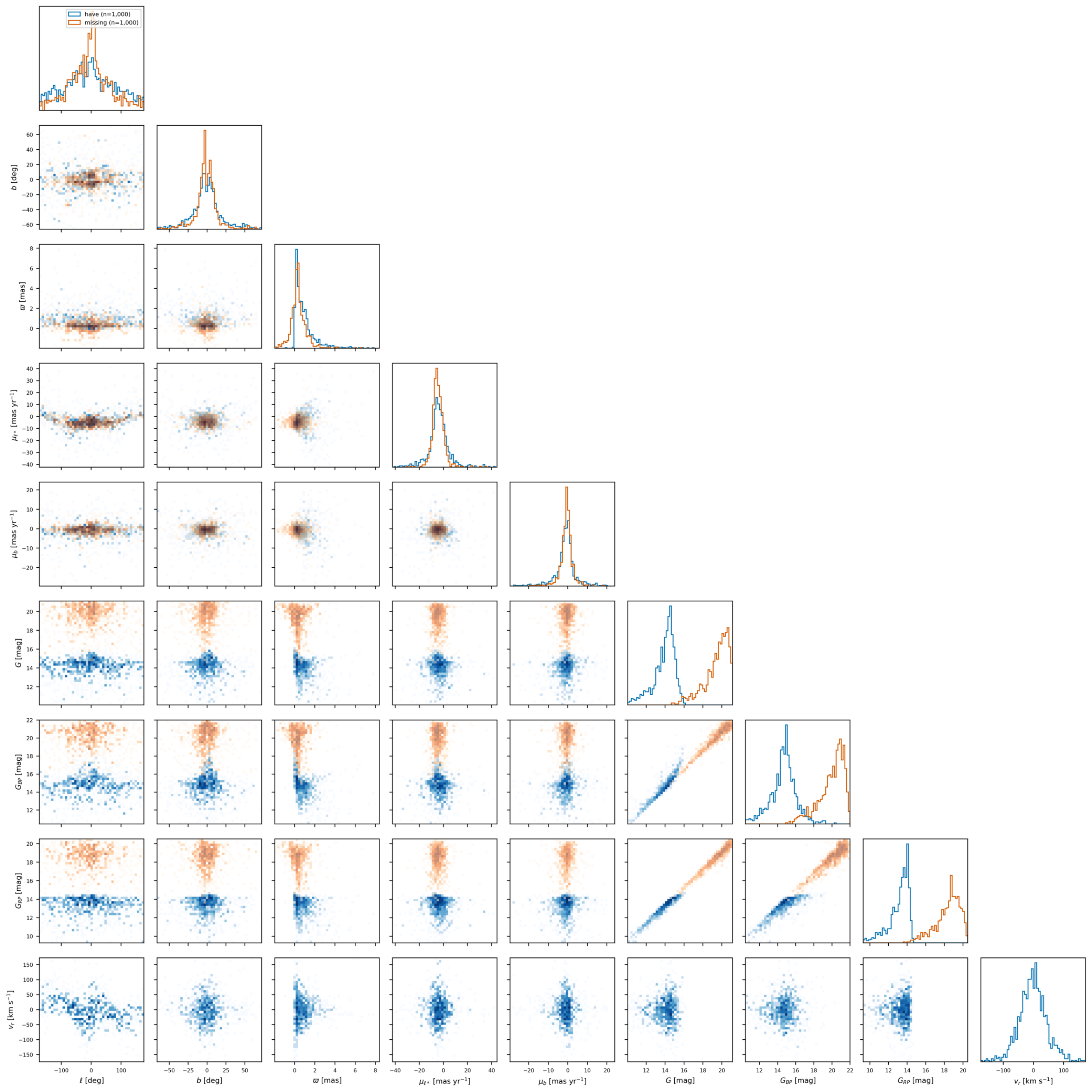}
\caption{Corner plot illustrating the differences with (``have") and without (``missing") radial velocity, $v_r$, as measured by Gaia. Note that while the RV missing population is similar (but narrower in spread) in position ($\ell$, $b$), parallax ($\varpi$), and proper motion ($\mu_{\ell*}$, $\mu_{b}$), they occupy distinct regions in photometry ($G$, $G_{BP}$, $G_{RP}$). Differences in parameter space limit the theoretical power of imputation.}
\label{fig:v_r_populations}
\end{figure*}

Figure \ref{fig:impute_skill_and_confidence} shows the ``skill" of the model for different conditioning patterns when imputing the radial velocity. The skill is defined by Equation \ref{eqn:skill}.
\begin{equation}\label{eqn:skill}
    Skill = 1-\frac{\mathrm{RMSE}}{\mathrm{RMSE}_{baseline}}
\end{equation}
The $\mathrm{RMSE}$ is defined as the RMSE of the mean of the generated $v_r$ for a source compared to the Gaia measured $v_r$. The $\mathrm{RMSE}_{baseline}$ is defined as the RMSE of the mean of all the sources used for conditioning compared to the Gaia measured $v_r$ for each source. In other words, the baseline is using the mean of the dataset one is presented with to impute all missing observations. A skill greater than 0 suggests that the model is providing additional information over guessing a single value for all missing sources, while a skill of $\leq0$ suggests the model is at best as good as guessing the mean for all missing values. Figure \ref{fig:impute_skill_and_confidence} reveals that conditional information can provide potentially useful information for imputation of $v_r$, which is important given the small fraction of sources with $v_r$ measurements in the Gaia Source Catalog. Moreover, the figure reveals the importance of Galactic longitude as a particularly decisive conditioning feature, which likely reflects the relationship between radial velocity, Galactic longitude, Solar reflex motion, and the Milky Way's rotation.

Further analysis of the model's imputation capabilities can be found in Appendix \ref{appendix_stats}. The model tends to impute best when given all but the imputed parameter (the drop $x$ pattern). Relationships between the parameters also strengthen imputation, especially the near deterministic relationship between photometric parameters. The model is typically much more effective at imputing Gaia photometry than astrometry (see Tables \ref{tab:skill:test:a}, \ref{tab:skill:test:b}, \ref{tab:skill:test:c}, and \ref{tab:skill:test:d}). The model is typically a more effective estimator of parallax when given proper motions, and more generally for sources with measured $v_r$, even when this is not included in the conditioning pattern (see the ($\ell$, $b$, $G$, $G_{BP}$, $G_{RP}$) and ($\ell$, $b$, $G_{BP}$, $G_{RP}$) patterns in Table \ref{tab:skill:test:d}).

\section{Conclusion}\label{conclusion}
The aim of this project was to train a generative machine learning model that could:
\begin{enumerate}
    \item Simulate the DR3 Gaia Source Catalog with and without conditioning information.
    \item Simulate source subpopulations like $\omega$ Cen.
    \item Impute missing parameters.
\end{enumerate}
The model succeeds in aim one, partially succeeds in aim three, but clearly fails in aim two. The model is able to reproduce a reasonable approximation of the full catalog with and without conditioning information, with variations in performance depending on the specific conditioning pattern used (see Appendix \ref{appendix_stats}). This means the model is capable of producing data-driven simulations of the Gaia survey, as a complement to existing forward model simulations \citep[e.g.,][]{Rybizki_2020}. However, analysis of model performance on $\omega$ Cen members strongly suggests that the model is unable to reproduce small distinct subpopulations. On aim three, the skill plots in Section \ref{im_RV} and tables in Appendix \ref{appendix_stats} show that the model can provide better point estimates than guessing, although the strength of that improvement depends strongly on the conditioning pattern and the imputed variable. Moreover, as expected from the results of Section \ref{om_cen}, the model is not able to impute most parameters that come from $\omega$ Cen,\footnote{Outside of the two types of patterns noted in Appendix \ref{appendix_stats}.} suggesting that imputation for this model is only reliable when the sources are drawn from the large scale global distribution of sources in the catalog.

The reason for the model's failure to simulate $\omega$ Cen is not clear. The model is clearly able to learn distinct structures, like the Magellanic Clouds and Milky Way disk, down to a certain scale. It is possible the model is simply insensitive to distinct structures that make up too small a share of the training data or are too compact in parameter space. Additional training of the model might also improve performance on subpopulations. Different model architectures, like using transformer based models \citep[e.g.,][]{GaiaFM_1}, and different training formats, like modifying the masking scheme, or training on local patches \citep[as in][]{Hallin_2025}, might also improve the model's capacity to reproduce subpopulations. Learning rate is important for learning finer features in the dataset,\footnote{A poor choice of learning rate can prevent the model from correctly distinguishing the Magellanic Clouds. Indeed, much of my time was focused on improving the unconditionally generated samples so that the Clouds were clearly distinguished in plots like Figure \ref{fig:MW_field_uncond_pmb}.} so a more optimal learning rate might improve subpopulation simulation. It is also possible that distinct subpopulations like $\omega$ Cen might be too small and distinct relative to the full catalog to be theoretically retrievable by a model.\footnote{Suggested by Marco Romanelli, private communication.} Further work is necessary to investigate this. Nevertheless, the model clearly has learned some local and distinct features, like the position and proper motions of the Magellanic Clouds (See Figure \ref{fig:MW_field_uncond_pmb}). This suggests that a larger model that is trained for longer on a greater proportion of the full catalog might learn and be able to reproduce smaller and more distinct subpopulations found within the catalog.

This work began with the hope of incorporating all of Gaia's available data products into the model, combining the source catalog with spectra, astrophysical parameters \citep{gaia_astrophys_1, gaia_astrophys_2, gaia_astrophys_3}, and more. Computational limitations resulted in an immediate narrowing of this work's scope. Despite these and other constraints, this work demonstrates that teaching a generative model the Gaia survey is possible, and plausibly useful. The upcoming Gaia Data Release 4 will improve existing observations, increase the number of observed sources with five and six parameter astrometry, and provide new data products like epoch astrometry.\footnote{\url{https://www.cosmos.esa.int/web/gaia/dr4}} This will provide an excellent dataset for building astronomy and Gaia-specific foundation models. This exploratory work, despite its limitations, enables further attempts at synthesizing Gaia data products into generative models.

\section*{Data/Code Availability}
The data is publicly available through the Gaia archive \url{https://gea.esac.esa.int/archive/}. \newline

The code for model training, sampling, testing, and analysis is available on \url{https://github.com/johnsoltis/Learning\_Gaia}.\newline

Model weights and optimizer information are available at 10.5281/zenodo.22946129.\newline

Model samples for Section \ref{mw_field} are available at 10.5281/zenodo.22946252.\newline

Test set subsample used in Section \ref{mw_field} and $\omega$ Cen member data used in Section \ref{om_cen} are available at 10.5281/zenodo.22946217.\newline

Model samples for Section \ref{om_cen} are available at 10.5281/zenodo.22946469.\newline

Imputation model samples for Section \ref{im_RV} are available at 10.5281/zenodo.22946654.\newline

The TARP and RMSE results used in the Appendix \ref{appendix_stats} are available at 10.5281/zenodo.23000102.

\begin{acknowledgments}
Conversations with Christina Lindberg, John Wu, Mikaeel Yunus, Ana Maria Delgado, Michelle Ntampaka, Roeland van der Marel, Doyeon (Avery) Kim, Carrie Filion, Matthew Ho, and Marco Romanelli provided important insights and motivation for this work. The full Gaia DR3 source catalog was easily accessible thanks to Adrian Price-Whelan.\newline

This work also served as an experiment in integrating commercial large language models into my research work flow. ChatGPT and Claude were both used to improve the training code, especially the data retrieval pipeline, and to write much of the analysis code and plotting code. Claude provided useful insights on relevant analysis metrics and suggested important changes to the model architecture, including adding a Gaussian Fourier time embedding. Code written by Claude was also used to automatically format the tables included in this work. All the writing is my own, however Claude was used to proofread this article for both clarity and accuracy.\newline

This work has made use of data from the European Space Agency (ESA) mission
{\it Gaia} (\url{https://www.cosmos.esa.int/gaia}), processed by the {\it Gaia}
Data Processing and Analysis Consortium (DPAC,
\url{https://www.cosmos.esa.int/web/gaia/dpac/consortium}). Funding for the DPAC
has been provided by national institutions, in particular the institutions
participating in the {\it Gaia} Multilateral Agreement.\newline

Initial work was carried out at the Advanced Research Computing at Hopkins (ARCH) core facility  (rockfish.jhu.edu), which is supported by the National Science Foundation (NSF) grant number OAC1920103.\newline

This work used SDSC Expanse at the San Diego Supercomputer Center through allocation PHY250370 from the Advanced Cyberinfrastructure Coordination Ecosystem: Services \& Support (ACCESS) program, which is supported by U.S. National Science Foundation grants \#2138259, \#2138286, \#2138307, \#2137603, and \#2138296. \citep{NSF_ACCESS}.\newline

This work made use of Astropy:\footnote{https://www.astropy.org} a community-developed core Python package and an ecosystem of tools and resources for astronomy \citep{astropy:2013, astropy:2018, astropy:2022}.
\end{acknowledgments}

\bibliography{sample701}{}

@ARTICLE{Tancik_2020,
       author = {{Tancik}, Matthew and {Srinivasan}, Pratul P. and {Mildenhall}, Ben and {Fridovich-Keil}, Sara and {Raghavan}, Nithin and {Singhal}, Utkarsh and {Ramamoorthi}, Ravi and {Barron}, Jonathan T. and {Ng}, Ren},
        title = "{Fourier Features Let Networks Learn High Frequency Functions in Low Dimensional Domains}",
      journal = {arXiv e-prints},
         year = 2020,
        month = jun,
          eid = {arXiv:2006.10739},
        pages = {arXiv:2006.10739},
          doi = {10.48550/arXiv.2006.10739},
archivePrefix = {arXiv},
       eprint = {2006.10739},
 primaryClass = {cs.CV},
       adsurl = {https://ui.adsabs.harvard.edu/abs/2020arXiv200610739T}
}

@ARTICLE{AstroCLIP,
       author = {{Parker}, Liam and {Lanusse}, Francois and {Golkar}, Siavash and {Sarra}, Leopoldo and {Cranmer}, Miles and {Bietti}, Alberto and {Eickenberg}, Michael and {Krawezik}, Geraud and {McCabe}, Michael and {Morel}, Rudy and {Ohana}, Ruben and {Pettee}, Mariel and {R{\'e}galdo-Saint Blancard}, Bruno and {Cho}, Kyunghyun and {Ho}, Shirley and {Polymathic AI Collaboration}},
        title = "{AstroCLIP: a cross-modal foundation model for galaxies}",
      journal = {\mnras},
         year = 2024,
        month = jul,
       volume = {531},
       number = {4},
        pages = {4990-5011},
          doi = {10.1093/mnras/stae1450},
archivePrefix = {arXiv},
       eprint = {2310.03024},
 primaryClass = {astro-ph.IM},
       adsurl = {https://ui.adsabs.harvard.edu/abs/2024MNRAS.531.4990P}
}

@ARTICLE{Historical_Review_AI_Astro,
       author = {{Smith}, Michael J. and {Geach}, James E.},
        title = "{Astronomia ex machina: a history, primer and outlook on neural networks in astronomy}",
      journal = {Royal Society Open Science},
         year = 2023,
        month = may,
       volume = {10},
       number = {5},
          eid = {221454},
        pages = {221454},
          doi = {10.1098/rsos.221454},
archivePrefix = {arXiv},
       eprint = {2211.03796},
 primaryClass = {astro-ph.IM},
       adsurl = {https://ui.adsabs.harvard.edu/abs/2023RSOS...1021454S}
}

@ARTICLE{RadioZooFM,
       author = {{Slijepcevic}, Inigo V. and {Scaife}, Anna M.~M. and {Walmsley}, Mike and {Bowles}, Micah and {Wong}, O. Ivy and {Shabala}, Stanislav S. and {White}, Sarah V.},
        title = "{Radio galaxy zoo: towards building the first multipurpose foundation model for radio astronomy with self-supervised learning}",
      journal = {RAS Techniques and Instruments},
         year = 2024,
        month = jan,
       volume = {3},
       number = {1},
        pages = {19-32},
          doi = {10.1093/rasti/rzad055},
archivePrefix = {arXiv},
       eprint = {2305.16127},
 primaryClass = {astro-ph.IM},
       adsurl = {https://ui.adsabs.harvard.edu/abs/2024RASTI...3...19S}
}

@ARTICLE{GaiaFM_1,
       author = {{Leung}, Henry W. and {Bovy}, Jo},
        title = "{Towards an astronomical foundation model for stars with a transformer-based model}",
      journal = {\mnras},
         year = 2024,
        month = jan,
       volume = {527},
       number = {1},
        pages = {1494-1520},
          doi = {10.1093/mnras/stad3015},
archivePrefix = {arXiv},
       eprint = {2308.10944},
 primaryClass = {astro-ph.IM},
       adsurl = {https://ui.adsabs.harvard.edu/abs/2024MNRAS.527.1494L}
}

@ARTICLE{AstroPT,
       author = {{Smith}, Michael J. and {Roberts}, Ryan J. and {Angeloudi}, Eirini and {Huertas-Company}, Marc},
        title = "{AstroPT: Scaling Large Observation Models for Astronomy}",
      journal = {arXiv e-prints},
         year = 2024,
        month = may,
          eid = {arXiv:2405.14930},
        pages = {arXiv:2405.14930},
          doi = {10.48550/arXiv.2405.14930},
archivePrefix = {arXiv},
       eprint = {2405.14930},
 primaryClass = {astro-ph.IM},
       adsurl = {https://ui.adsabs.harvard.edu/abs/2024arXiv240514930S}
}

@ARTICLE{AION,
       author = {{Parker}, Liam and {Lanusse}, Francois and {Shen}, Jeff and {Liu}, Ollie and {Hehir}, Tom and {Sarra}, Leopoldo and {Meyer}, Lucas and {Bowles}, Micah and {Wagner-Carena}, Sebastian and {Qu}, Helen and {Golkar}, Siavash and {Bietti}, Alberto and {Bourfoune}, Hatim and {Casserau}, Nathan and {Cornette}, Pierre and {Hirashima}, Keiya and {Krawezik}, Geraud and {Ohana}, Ruben and {Lourie}, Nicholas and {McCabe}, Michael and {Morel}, Rudy and {Mukhopadhyay}, Payel and {Pettee}, Mariel and {Regaldo-Saint Blancard}, Bruno and {Cho}, Kyunghyun and {Cranmer}, Miles and {Ho}, Shirley},
        title = "{AION-1: Omnimodal Foundation Model for Astronomical Sciences}",
      journal = {arXiv e-prints},
         year = 2025,
        month = oct,
          eid = {arXiv:2510.17960},
        pages = {arXiv:2510.17960},
          doi = {10.48550/arXiv.2510.17960},
archivePrefix = {arXiv},
       eprint = {2510.17960},
 primaryClass = {astro-ph.IM},
       adsurl = {https://ui.adsabs.harvard.edu/abs/2025arXiv251017960P}
}

@ARTICLE{SpectraFM,
       author = {{Koblischke}, Nolan and {Bovy}, Jo},
        title = "{SpectraFM: Tuning into Stellar Foundation Models}",
      journal = {arXiv e-prints},
         year = 2024,
        month = nov,
          eid = {arXiv:2411.04750},
        pages = {arXiv:2411.04750},
          doi = {10.48550/arXiv.2411.04750},
archivePrefix = {arXiv},
       eprint = {2411.04750},
 primaryClass = {astro-ph.IM},
       adsurl = {https://ui.adsabs.harvard.edu/abs/2024arXiv241104750K}
}

@ARTICLE{DR3,
       author = {{Gaia Collaboration} and {Vallenari}, A. and {Brown}, A.~G.~A. and {Prusti}, T. and {de Bruijne}, J.~H.~J. and {Arenou}, F. and {Babusiaux}, C. and {Biermann}, M. and {Creevey}, O.~L. and {Ducourant}, C. and {Evans}, D.~W. and {Eyer}, L. and {Guerra}, R. and {Hutton}, A. and {Jordi}, C. and {Klioner}, S.~A. and {Lammers}, U.~L. and {Lindegren}, L. and {Luri}, X. and {Mignard}, F. and {Panem}, C. and {Pourbaix}, D. and {Randich}, S. and {Sartoretti}, P. and {Soubiran}, C. and {Tanga}, P. and {Walton}, N.~A. and {Bailer-Jones}, C.~A.~L. and {Bastian}, U. and {Drimmel}, R. and {Jansen}, F. and {Katz}, D. and {Lattanzi}, M.~G. and {van Leeuwen}, F. and {Bakker}, J. and {Cacciari}, C. and {Casta{\~n}eda}, J. and {De Angeli}, F. and {Fabricius}, C. and {Fouesneau}, M. and {Fr{\'e}mat}, Y. and {Galluccio}, L. and {Guerrier}, A. and {Heiter}, U. and {Masana}, E. and {Messineo}, R. and {Mowlavi}, N. and {Nicolas}, C. and {Nienartowicz}, K. and {Pailler}, F. and {Panuzzo}, P. and {Riclet}, F. and {Roux}, W. and {Seabroke}, G.~M. and {Sordo}, R. and {Th{\'e}venin}, F. and {Gracia-Abril}, G. and {Portell}, J. and {Teyssier}, D. and {Altmann}, M. and {Andrae}, R. and {Audard}, M. and {Bellas-Velidis}, I. and {Benson}, K. and {Berthier}, J. and {Blomme}, R. and {Burgess}, P.~W. and {Busonero}, D. and {Busso}, G. and {C{\'a}novas}, H. and {Carry}, B. and {Cellino}, A. and {Cheek}, N. and {Clementini}, G. and {Damerdji}, Y. and {Davidson}, M. and {de Teodoro}, P. and {Nu{\~n}ez Campos}, M. and {Delchambre}, L. and {Dell'Oro}, A. and {Esquej}, P. and {Fern{\'a}ndez-Hern{\'a}ndez}, J. and {Fraile}, E. and {Garabato}, D. and {Garc{\'\i}a-Lario}, P. and {Gosset}, E. and {Haigron}, R. and {Halbwachs}, J.-L. and {Hambly}, N.~C. and {Harrison}, D.~L. and {Hern{\'a}ndez}, J. and {Hestroffer}, D. and {Hodgkin}, S.~T. and {Holl}, B. and {Jan{\ss}en}, K. and {Jevardat de Fombelle}, G. and {Jordan}, S. and {Krone-Martins}, A. and {Lanzafame}, A.~C. and {L{\"o}ffler}, W. and {Marchal}, O. and {Marrese}, P.~M. and {Moitinho}, A. and {Muinonen}, K. and {Osborne}, P. and {Pancino}, E. and {Pauwels}, T. and {Recio-Blanco}, A. and {Reyl{\'e}}, C. and {Riello}, M. and {Rimoldini}, L. and {Roegiers}, T. and {Rybizki}, J. and {Sarro}, L.~M. and {Siopis}, C. and {Smith}, M. and {Sozzetti}, A. and {Utrilla}, E. and {van Leeuwen}, M. and {Abbas}, U. and {{\'A}brah{\'a}m}, P. and {Abreu Aramburu}, A. and {Aerts}, C. and {Aguado}, J.~J. and {Ajaj}, M. and {Aldea-Montero}, F. and {Altavilla}, G. and {{\'A}lvarez}, M.~A. and {Alves}, J. and {Anders}, F. and {Anderson}, R.~I. and {Anglada Varela}, E. and {Antoja}, T. and {Baines}, D. and {Baker}, S.~G. and {Balaguer-N{\'u}{\~n}ez}, L. and {Balbinot}, E. and {Balog}, Z. and {Barache}, C. and {Barbato}, D. and {Barros}, M. and {Barstow}, M.~A. and {Bartolom{\'e}}, S. and {Bassilana}, J.-L. and {Bauchet}, N. and {Becciani}, U. and {Bellazzini}, M. and {Berihuete}, A. and {Bernet}, M. and {Bertone}, S. and {Bianchi}, L. and {Binnenfeld}, A. and {Blanco-Cuaresma}, S. and {Blazere}, A. and {Boch}, T. and {Bombrun}, A. and {Bossini}, D. and {Bouquillon}, S. and {Bragaglia}, A. and {Bramante}, L. and {Breedt}, E. and {Bressan}, A. and {Brouillet}, N. and {Brugaletta}, E. and {Bucciarelli}, B. and {Burlacu}, A. and {Butkevich}, A.~G. and {Buzzi}, R. and {Caffau}, E. and {Cancelliere}, R. and {Cantat-Gaudin}, T. and {Carballo}, R. and {Carlucci}, T. and {Carnerero}, M.~I. and {Carrasco}, J.~M. and {Casamiquela}, L. and {Castellani}, M. and {Castro-Ginard}, A. and {Chaoul}, L. and {Charlot}, P. and {Chemin}, L. and {Chiaramida}, V. and {Chiavassa}, A. and {Chornay}, N. and {Comoretto}, G. and {Contursi}, G. and {Cooper}, W.~J. and {Cornez}, T. and {Cowell}, S. and {Crifo}, F. and {Cropper}, M. and {Crosta}, M. and {Crowley}, C. and {Dafonte}, C. and {Dapergolas}, A. and {David}, M. and {David}, P. and {de Laverny}, P. and {De Luise}, F. and {De March}, R.},
        title = "{Gaia Data Release 3. Summary of the content and survey properties}",
      journal = {\aap},
         year = 2023,
        month = jun,
       volume = {674},
          eid = {A1},
        pages = {A1},
          doi = {10.1051/0004-6361/202243940},
archivePrefix = {arXiv},
       eprint = {2208.00211},
 primaryClass = {astro-ph.GA},
       adsurl = {https://ui.adsabs.harvard.edu/abs/2023A&A...674A...1G}
}

@ARTICLE{eDR3,
       author = {{Gaia Collaboration} and {Brown}, A.~G.~A. and {Vallenari}, A. and {Prusti}, T. and {de Bruijne}, J.~H.~J. and {Babusiaux}, C. and {Biermann}, M. and {Creevey}, O.~L. and {Evans}, D.~W. and {Eyer}, L. and {Hutton}, A. and {Jansen}, F. and {Jordi}, C. and {Klioner}, S.~A. and {Lammers}, U. and {Lindegren}, L. and {Luri}, X. and {Mignard}, F. and {Panem}, C. and {Pourbaix}, D. and {Randich}, S. and {Sartoretti}, P. and {Soubiran}, C. and {Walton}, N.~A. and {Arenou}, F. and {Bailer-Jones}, C.~A.~L. and {Bastian}, U. and {Cropper}, M. and {Drimmel}, R. and {Katz}, D. and {Lattanzi}, M.~G. and {van Leeuwen}, F. and {Bakker}, J. and {Cacciari}, C. and {Casta{\~n}eda}, J. and {De Angeli}, F. and {Ducourant}, C. and {Fabricius}, C. and {Fouesneau}, M. and {Fr{\'e}mat}, Y. and {Guerra}, R. and {Guerrier}, A. and {Guiraud}, J. and {Jean-Antoine Piccolo}, A. and {Masana}, E. and {Messineo}, R. and {Mowlavi}, N. and {Nicolas}, C. and {Nienartowicz}, K. and {Pailler}, F. and {Panuzzo}, P. and {Riclet}, F. and {Roux}, W. and {Seabroke}, G.~M. and {Sordo}, R. and {Tanga}, P. and {Th{\'e}venin}, F. and {Gracia-Abril}, G. and {Portell}, J. and {Teyssier}, D. and {Altmann}, M. and {Andrae}, R. and {Bellas-Velidis}, I. and {Benson}, K. and {Berthier}, J. and {Blomme}, R. and {Brugaletta}, E. and {Burgess}, P.~W. and {Busso}, G. and {Carry}, B. and {Cellino}, A. and {Cheek}, N. and {Clementini}, G. and {Damerdji}, Y. and {Davidson}, M. and {Delchambre}, L. and {Dell'Oro}, A. and {Fern{\'a}ndez-Hern{\'a}ndez}, J. and {Galluccio}, L. and {Garc{\'\i}a-Lario}, P. and {Garcia-Reinaldos}, M. and {Gonz{\'a}lez-N{\'u}{\~n}ez}, J. and {Gosset}, E. and {Haigron}, R. and {Halbwachs}, J.-L. and {Hambly}, N.~C. and {Harrison}, D.~L. and {Hatzidimitriou}, D. and {Heiter}, U. and {Hern{\'a}ndez}, J. and {Hestroffer}, D. and {Hodgkin}, S.~T. and {Holl}, B. and {Jan{\ss}en}, K. and {Jevardat de Fombelle}, G. and {Jordan}, S. and {Krone-Martins}, A. and {Lanzafame}, A.~C. and {L{\"o}ffler}, W. and {Lorca}, A. and {Manteiga}, M. and {Marchal}, O. and {Marrese}, P.~M. and {Moitinho}, A. and {Mora}, A. and {Muinonen}, K. and {Osborne}, P. and {Pancino}, E. and {Pauwels}, T. and {Petit}, J.-M. and {Recio-Blanco}, A. and {Richards}, P.~J. and {Riello}, M. and {Rimoldini}, L. and {Robin}, A.~C. and {Roegiers}, T. and {Rybizki}, J. and {Sarro}, L.~M. and {Siopis}, C. and {Smith}, M. and {Sozzetti}, A. and {Ulla}, A. and {Utrilla}, E. and {van Leeuwen}, M. and {van Reeven}, W. and {Abbas}, U. and {Abreu Aramburu}, A. and {Accart}, S. and {Aerts}, C. and {Aguado}, J.~J. and {Ajaj}, M. and {Altavilla}, G. and {{\'A}lvarez}, M.~A. and {{\'A}lvarez Cid-Fuentes}, J. and {Alves}, J. and {Anderson}, R.~I. and {Anglada Varela}, E. and {Antoja}, T. and {Audard}, M. and {Baines}, D. and {Baker}, S.~G. and {Balaguer-N{\'u}{\~n}ez}, L. and {Balbinot}, E. and {Balog}, Z. and {Barache}, C. and {Barbato}, D. and {Barros}, M. and {Barstow}, M.~A. and {Bartolom{\'e}}, S. and {Bassilana}, J.-L. and {Bauchet}, N. and {Baudesson-Stella}, A. and {Becciani}, U. and {Bellazzini}, M. and {Bernet}, M. and {Bertone}, S. and {Bianchi}, L. and {Blanco-Cuaresma}, S. and {Boch}, T. and {Bombrun}, A. and {Bossini}, D. and {Bouquillon}, S. and {Bragaglia}, A. and {Bramante}, L. and {Breedt}, E. and {Bressan}, A. and {Brouillet}, N. and {Bucciarelli}, B. and {Burlacu}, A. and {Busonero}, D. and {Butkevich}, A.~G. and {Buzzi}, R. and {Caffau}, E. and {Cancelliere}, R. and {C{\'a}novas}, H. and {Cantat-Gaudin}, T. and {Carballo}, R. and {Carlucci}, T. and {Carnerero}, M.~I. and {Carrasco}, J.~M. and {Casamiquela}, L. and {Castellani}, M. and {Castro-Ginard}, A. and {Castro Sampol}, P. and {Chaoul}, L. and {Charlot}, P. and {Chemin}, L. and {Chiavassa}, A. and {Cioni}, M.-R.~L. and {Comoretto}, G. and {Cooper}, W.~J. and {Cornez}, T. and {Cowell}, S. and {Crifo}, F. and {Crosta}, M. and {Crowley}, C. and {Dafonte}, C. and {Dapergolas}, A. and {David}, M. and {David}, P.},
        title = "{Gaia Early Data Release 3. Summary of the contents and survey properties}",
      journal = {\aap},
         year = 2021,
        month = may,
       volume = {649},
          eid = {A1},
        pages = {A1},
          doi = {10.1051/0004-6361/202039657},
archivePrefix = {arXiv},
       eprint = {2012.01533},
 primaryClass = {astro-ph.GA},
       adsurl = {https://ui.adsabs.harvard.edu/abs/2021A&A...649A...1G}
}

@ARTICLE{CFMI,
       author = {{Simkus}, Vaidotas and {Gutmann}, Michael U.},
        title = "{CFMI: Flow Matching for Missing Data Imputation}",
      journal = {arXiv e-prints},
         year = 2025,
        month = jun,
          eid = {arXiv:2506.09258},
        pages = {arXiv:2506.09258},
          doi = {10.48550/arXiv.2506.09258},
archivePrefix = {arXiv},
       eprint = {2506.09258},
 primaryClass = {stat.ML},
       adsurl = {https://ui.adsabs.harvard.edu/abs/2025arXiv250609258S}
}

@ARTICLE{DR3_RV,
       author = {{Katz}, D. and {Sartoretti}, P. and {Guerrier}, A. and {Panuzzo}, P. and {Seabroke}, G.~M. and {Th{\'e}venin}, F. and {Cropper}, M. and {Benson}, K. and {Blomme}, R. and {Haigron}, R. and {Marchal}, O. and {Smith}, M. and {Baker}, S. and {Chemin}, L. and {Damerdji}, Y. and {David}, M. and {Dolding}, C. and {Fr{\'e}mat}, Y. and {Gosset}, E. and {Jan{\ss}en}, K. and {Jasniewicz}, G. and {Lobel}, A. and {Plum}, G. and {Samaras}, N. and {Snaith}, O. and {Soubiran}, C. and {Vanel}, O. and {Zwitter}, T. and {Antoja}, T. and {Arenou}, F. and {Babusiaux}, C. and {Brouillet}, N. and {Caffau}, E. and {Di Matteo}, P. and {Fabre}, C. and {Fabricius}, C. and {Fragkoudi}, F. and {Haywood}, M. and {Huckle}, H.~E. and {Hottier}, C. and {Lasne}, Y. and {Leclerc}, N. and {Mastrobuono-Battisti}, A. and {Royer}, F. and {Teyssier}, D. and {Zorec}, J. and {Crifo}, F. and {Jean-Antoine Piccolo}, A. and {Turon}, C. and {Viala}, Y.},
        title = "{Gaia Data Release 3. Properties and validation of the radial velocities}",
      journal = {\aap},
         year = 2023,
        month = jun,
       volume = {674},
          eid = {A5},
        pages = {A5},
          doi = {10.1051/0004-6361/202244220},
archivePrefix = {arXiv},
       eprint = {2206.05902},
 primaryClass = {astro-ph.GA},
       adsurl = {https://ui.adsabs.harvard.edu/abs/2023A&A...674A...5K}
}

@ARTICLE{Gaia_XP_1,
       author = {{Carrasco}, J.~M. and {Weiler}, M. and {Jordi}, C. and {Fabricius}, C. and {De Angeli}, F. and {Evans}, D.~W. and {van Leeuwen}, F. and {Riello}, M. and {Montegriffo}, P.},
        title = "{Internal calibration of Gaia BP/RP low-resolution spectra}",
      journal = {\aap},
         year = 2021,
        month = aug,
       volume = {652},
          eid = {A86},
        pages = {A86},
          doi = {10.1051/0004-6361/202141249},
archivePrefix = {arXiv},
       eprint = {2106.01752},
 primaryClass = {astro-ph.IM},
       adsurl = {https://ui.adsabs.harvard.edu/abs/2021A&A...652A..86C}
}

@ARTICLE{Gaia_XP_2,
       author = {{De Angeli}, F. and {Weiler}, M. and {Montegriffo}, P. and {Evans}, D.~W. and {Riello}, M. and {Andrae}, R. and {Carrasco}, J.~M. and {Busso}, G. and {Burgess}, P.~W. and {Cacciari}, C. and {Davidson}, M. and {Harrison}, D.~L. and {Hodgkin}, S.~T. and {Jordi}, C. and {Osborne}, P.~J. and {Pancino}, E. and {Altavilla}, G. and {Barstow}, M.~A. and {Bailer-Jones}, C.~A.~L. and {Bellazzini}, M. and {Brown}, A.~G.~A. and {Castellani}, M. and {Cowell}, S. and {Delchambre}, L. and {De Luise}, F. and {Diener}, C. and {Fabricius}, C. and {Fouesneau}, M. and {Fr{\'e}mat}, Y. and {Gilmore}, G. and {Giuffrida}, G. and {Hambly}, N.~C. and {Hidalgo}, S. and {Holland}, G. and {Kostrzewa-Rutkowska}, Z. and {van Leeuwen}, F. and {Lobel}, A. and {Marinoni}, S. and {Miller}, N. and {Pagani}, C. and {Palaversa}, L. and {Piersimoni}, A.~M. and {Pulone}, L. and {Ragaini}, S. and {Rainer}, M. and {Richards}, P.~J. and {Rixon}, G.~T. and {Ruz-Mieres}, D. and {Sanna}, N. and {Sarro}, L.~M. and {Rowell}, N. and {Sordo}, R. and {Walton}, N.~A. and {Yoldas}, A.},
        title = "{Gaia Data Release 3. Processing and validation of BP/RP low-resolution spectral data}",
      journal = {\aap},
         year = 2023,
        month = jun,
       volume = {674},
          eid = {A2},
        pages = {A2},
          doi = {10.1051/0004-6361/202243680},
archivePrefix = {arXiv},
       eprint = {2206.06143},
 primaryClass = {astro-ph.IM},
       adsurl = {https://ui.adsabs.harvard.edu/abs/2023A&A...674A...2D}
}

@ARTICLE{Gaia_XP_3,
       author = {{Montegriffo}, P. and {De Angeli}, F. and {Andrae}, R. and {Riello}, M. and {Pancino}, E. and {Sanna}, N. and {Bellazzini}, M. and {Evans}, D.~W. and {Carrasco}, J.~M. and {Sordo}, R. and {Busso}, G. and {Cacciari}, C. and {Jordi}, C. and {van Leeuwen}, F. and {Vallenari}, A. and {Altavilla}, G. and {Barstow}, M.~A. and {Brown}, A.~G.~A. and {Burgess}, P.~W. and {Castellani}, M. and {Cowell}, S. and {Davidson}, M. and {De Luise}, F. and {Delchambre}, L. and {Diener}, C. and {Fabricius}, C. and {Fr{\'e}mat}, Y. and {Fouesneau}, M. and {Gilmore}, G. and {Giuffrida}, G. and {Hambly}, N.~C. and {Harrison}, D.~L. and {Hidalgo}, S. and {Hodgkin}, S.~T. and {Holland}, G. and {Marinoni}, S. and {Osborne}, P.~J. and {Pagani}, C. and {Palaversa}, L. and {Piersimoni}, A.~M. and {Pulone}, L. and {Ragaini}, S. and {Rainer}, M. and {Richards}, P.~J. and {Rowell}, N. and {Ruz-Mieres}, D. and {Sarro}, L.~M. and {Walton}, N.~A. and {Yoldas}, A.},
        title = "{Gaia Data Release 3. External calibration of BP/RP low-resolution spectroscopic data}",
      journal = {\aap},
         year = 2023,
        month = jun,
       volume = {674},
          eid = {A3},
        pages = {A3},
          doi = {10.1051/0004-6361/202243880},
archivePrefix = {arXiv},
       eprint = {2206.06205},
 primaryClass = {astro-ph.IM},
       adsurl = {https://ui.adsabs.harvard.edu/abs/2023A&A...674A...3M}
}

@ARTICLE{Gaia_ZP,
       author = {{Lindegren}, L. and {Bastian}, U. and {Biermann}, M. and {Bombrun}, A. and {de Torres}, A. and {Gerlach}, E. and {Geyer}, R. and {Hern{\'a}ndez}, J. and {Hilger}, T. and {Hobbs}, D. and {Klioner}, S.~A. and {Lammers}, U. and {McMillan}, P.~J. and {Ramos-Lerate}, M. and {Steidelm{\"u}ller}, H. and {Stephenson}, C.~A. and {van Leeuwen}, F.},
        title = "{Gaia Early Data Release 3. Parallax bias versus magnitude, colour, and position}",
      journal = {\aap},
         year = 2021,
        month = may,
       volume = {649},
          eid = {A4},
        pages = {A4},
          doi = {10.1051/0004-6361/202039653},
archivePrefix = {arXiv},
       eprint = {2012.01742},
 primaryClass = {astro-ph.IM},
       adsurl = {https://ui.adsabs.harvard.edu/abs/2021A&A...649A...4L}
}

@ARTICLE{CFM_OT,
       author = {{Tong}, Alexander and {Fatras}, Kilian and {Malkin}, Nikolay and {Huguet}, Guillaume and {Zhang}, Yanlei and {Rector-Brooks}, Jarrid and {Wolf}, Guy and {Bengio}, Yoshua},
        title = "{Improving and generalizing flow-based generative models with minibatch optimal transport}",
      journal = {arXiv e-prints},
         year = 2023,
        month = feb,
          eid = {arXiv:2302.00482},
        pages = {arXiv:2302.00482},
          doi = {10.48550/arXiv.2302.00482},
archivePrefix = {arXiv},
       eprint = {2302.00482},
 primaryClass = {cs.LG},
       adsurl = {https://ui.adsabs.harvard.edu/abs/2023arXiv230200482T}
}

@ARTICLE{CFM,
       author = {{Lipman}, Yaron and {Chen}, Ricky T.~Q. and {Ben-Hamu}, Heli and {Nickel}, Maximilian and {Le}, Matt},
        title = "{Flow Matching for Generative Modeling}",
      journal = {arXiv e-prints},
         year = 2022,
        month = oct,
          eid = {arXiv:2210.02747},
        pages = {arXiv:2210.02747},
          doi = {10.48550/arXiv.2210.02747},
archivePrefix = {arXiv},
       eprint = {2210.02747},
 primaryClass = {cs.LG},
       adsurl = {https://ui.adsabs.harvard.edu/abs/2022arXiv221002747L}
}

@ARTICLE{AdamW,
       author = {{Loshchilov}, Ilya and {Hutter}, Frank},
        title = "{Decoupled Weight Decay Regularization}",
      journal = {arXiv e-prints},
         year = 2017,
        month = nov,
          eid = {arXiv:1711.05101},
        pages = {arXiv:1711.05101},
          doi = {10.48550/arXiv.1711.05101},
archivePrefix = {arXiv},
       eprint = {1711.05101},
 primaryClass = {cs.LG},
       adsurl = {https://ui.adsabs.harvard.edu/abs/2017arXiv171105101L}
}

@ARTICLE{TARP,
       author = {{Lemos}, Pablo and {Coogan}, Adam and {Hezaveh}, Yashar and {Perreault-Levasseur}, Laurence},
        title = "{Sampling-Based Accuracy Testing of Posterior Estimators for General Inference}",
      journal = {40th International Conference on Machine Learning},
         year = 2023,
        month = jan,
       volume = {202},
        pages = {19256-19273},
          doi = {10.48550/arXiv.2302.03026},
archivePrefix = {arXiv},
       eprint = {2302.03026},
 primaryClass = {stat.ML},
       adsurl = {https://ui.adsabs.harvard.edu/abs/2023PMLR..20219256L}
}

@ARTICLE{LayerNorm,
       author = {{Ba}, Jimmy Lei and {Kiros}, Jamie Ryan and {Hinton}, Geoffrey E.},
        title = "{Layer Normalization}",
      journal = {arXiv e-prints},
         year = 2016,
        month = jul,
          eid = {arXiv:1607.06450},
        pages = {arXiv:1607.06450},
          doi = {10.48550/arXiv.1607.06450},
archivePrefix = {arXiv},
       eprint = {1607.06450},
 primaryClass = {stat.ML},
       adsurl = {https://ui.adsabs.harvard.edu/abs/2016arXiv160706450L}
}

@ARTICLE{GELU,
       author = {{Hendrycks}, Dan and {Gimpel}, Kevin},
        title = "{Gaussian Error Linear Units (GELUs)}",
      journal = {arXiv e-prints},
         year = 2016,
        month = jun,
          eid = {arXiv:1606.08415},
        pages = {arXiv:1606.08415},
          doi = {10.48550/arXiv.1606.08415},
archivePrefix = {arXiv},
       eprint = {1606.08415},
 primaryClass = {cs.LG},
       adsurl = {https://ui.adsabs.harvard.edu/abs/2016arXiv160608415H}
}

@ARTICLE{Soltis_21,
       author = {{Soltis}, John and {Casertano}, Stefano and {Riess}, Adam G.},
        title = "{The Parallax of {\ensuremath{\omega}} Centauri Measured from Gaia EDR3 and a Direct, Geometric Calibration of the Tip of the Red Giant Branch and the Hubble Constant}",
      journal = {\apjl},
         year = 2021,
        month = feb,
       volume = {908},
       number = {1},
          eid = {L5},
        pages = {L5},
          doi = {10.3847/2041-8213/abdbad},
archivePrefix = {arXiv},
       eprint = {2012.09196},
 primaryClass = {astro-ph.GA},
       adsurl = {https://ui.adsabs.harvard.edu/abs/2021ApJ...908L...5S}
}

@ARTICLE{Rybizki_2020,
       author = {{Rybizki}, Jan and {Demleitner}, Markus and {Bailer-Jones}, Coryn and {Tio}, Piero Dal and {Cantat-Gaudin}, Tristan and {Fouesneau}, Morgan and {Chen}, Yang and {Andrae}, Ren{\'e} and {Girardi}, L{\'e}o and {Sharma}, Sanjib},
        title = "{A Gaia Early DR3 Mock Stellar Catalog: Galactic Prior and Selection Function}",
      journal = {\pasp},
         year = 2020,
        month = jul,
       volume = {132},
       number = {1013},
          eid = {074501},
        pages = {074501},
          doi = {10.1088/1538-3873/ab8cb0},
archivePrefix = {arXiv},
       eprint = {2004.09991},
 primaryClass = {astro-ph.IM},
       adsurl = {https://ui.adsabs.harvard.edu/abs/2020PASP..132g4501R}
}

@ARTICLE{Song_2020,
       author = {{Song}, Yang and {Sohl-Dickstein}, Jascha and {Kingma}, Diederik P. and {Kumar}, Abhishek and {Ermon}, Stefano and {Poole}, Ben},
        title = "{Score-Based Generative Modeling through Stochastic Differential Equations}",
      journal = {arXiv e-prints},
         year = 2020,
        month = nov,
          eid = {arXiv:2011.13456},
        pages = {arXiv:2011.13456},
          doi = {10.48550/arXiv.2011.13456},
archivePrefix = {arXiv},
       eprint = {2011.13456},
 primaryClass = {cs.LG},
       adsurl = {https://ui.adsabs.harvard.edu/abs/2020arXiv201113456S}
}

@inproceedings{NSF_ACCESS,
author = {Boerner, Timothy J. and Deems, Stephen and Furlani, Thomas R. and Knuth, Shelley L. and Towns, John},
title = {ACCESS: Advancing Innovation: NSF’s Advanced Cyberinfrastructure Coordination Ecosystem: Services \& Support},
year = {2023},
isbn = {9781450399852},
publisher = {Association for Computing Machinery},
address = {New York, NY, USA},
url = {https://doi.org/10.1145/3569951.3597559},
doi = {10.1145/3569951.3597559},
booktitle = {Practice and Experience in Advanced Research Computing 2023: Computing for the Common Good},
pages = {173–176},
numpages = {4},
location = {Portland, OR, USA},
series = {PEARC '23}
}

@ARTICLE{Gaia_Mission,
       author = {{Gaia Collaboration} and {Prusti}, T. and {de Bruijne}, J.~H.~J. and {Brown}, A.~G.~A. and {Vallenari}, A. and {Babusiaux}, C. and {Bailer-Jones}, C.~A.~L. and {Bastian}, U. and {Biermann}, M. and {Evans}, D.~W. and {Eyer}, L. and {Jansen}, F. and {Jordi}, C. and {Klioner}, S.~A. and {Lammers}, U. and {Lindegren}, L. and {Luri}, X. and {Mignard}, F. and {Milligan}, D.~J. and {Panem}, C. and {Poinsignon}, V. and {Pourbaix}, D. and {Randich}, S. and {Sarri}, G. and {Sartoretti}, P. and {Siddiqui}, H.~I. and {Soubiran}, C. and {Valette}, V. and {van Leeuwen}, F. and {Walton}, N.~A. and {Aerts}, C. and {Arenou}, F. and {Cropper}, M. and {Drimmel}, R. and {H{\o}g}, E. and {Katz}, D. and {Lattanzi}, M.~G. and {O'Mullane}, W. and {Grebel}, E.~K. and {Holland}, A.~D. and {Huc}, C. and {Passot}, X. and {Bramante}, L. and {Cacciari}, C. and {Casta{\~n}eda}, J. and {Chaoul}, L. and {Cheek}, N. and {De Angeli}, F. and {Fabricius}, C. and {Guerra}, R. and {Hern{\'a}ndez}, J. and {Jean-Antoine-Piccolo}, A. and {Masana}, E. and {Messineo}, R. and {Mowlavi}, N. and {Nienartowicz}, K. and {Ord{\'o}{\~n}ez-Blanco}, D. and {Panuzzo}, P. and {Portell}, J. and {Richards}, P.~J. and {Riello}, M. and {Seabroke}, G.~M. and {Tanga}, P. and {Th{\'e}venin}, F. and {Torra}, J. and {Els}, S.~G. and {Gracia-Abril}, G. and {Comoretto}, G. and {Garcia-Reinaldos}, M. and {Lock}, T. and {Mercier}, E. and {Altmann}, M. and {Andrae}, R. and {Astraatmadja}, T.~L. and {Bellas-Velidis}, I. and {Benson}, K. and {Berthier}, J. and {Blomme}, R. and {Busso}, G. and {Carry}, B. and {Cellino}, A. and {Clementini}, G. and {Cowell}, S. and {Creevey}, O. and {Cuypers}, J. and {Davidson}, M. and {De Ridder}, J. and {de Torres}, A. and {Delchambre}, L. and {Dell'Oro}, A. and {Ducourant}, C. and {Fr{\'e}mat}, Y. and {Garc{\'\i}a-Torres}, M. and {Gosset}, E. and {Halbwachs}, J.-L. and {Hambly}, N.~C. and {Harrison}, D.~L. and {Hauser}, M. and {Hestroffer}, D. and {Hodgkin}, S.~T. and {Huckle}, H.~E. and {Hutton}, A. and {Jasniewicz}, G. and {Jordan}, S. and {Kontizas}, M. and {Korn}, A.~J. and {Lanzafame}, A.~C. and {Manteiga}, M. and {Moitinho}, A. and {Muinonen}, K. and {Osinde}, J. and {Pancino}, E. and {Pauwels}, T. and {Petit}, J.-M. and {Recio-Blanco}, A. and {Robin}, A.~C. and {Sarro}, L.~M. and {Siopis}, C. and {Smith}, M. and {Smith}, K.~W. and {Sozzetti}, A. and {Thuillot}, W. and {van Reeven}, W. and {Viala}, Y. and {Abbas}, U. and {Abreu Aramburu}, A. and {Accart}, S. and {Aguado}, J.~J. and {Allan}, P.~M. and {Allasia}, W. and {Altavilla}, G. and {{\'A}lvarez}, M.~A. and {Alves}, J. and {Anderson}, R.~I. and {Andrei}, A.~H. and {Anglada Varela}, E. and {Antiche}, E. and {Antoja}, T. and {Ant{\'o}n}, S. and {Arcay}, B. and {Atzei}, A. and {Ayache}, L. and {Bach}, N. and {Baker}, S.~G. and {Balaguer-N{\'u}{\~n}ez}, L. and {Barache}, C. and {Barata}, C. and {Barbier}, A. and {Barblan}, F. and {Baroni}, M. and {Barrado y Navascu{\'e}s}, D. and {Barros}, M. and {Barstow}, M.~A. and {Becciani}, U. and {Bellazzini}, M. and {Bellei}, G. and {Bello Garc{\'\i}a}, A. and {Belokurov}, V. and {Bendjoya}, P. and {Berihuete}, A. and {Bianchi}, L. and {Bienaym{\'e}}, O. and {Billebaud}, F. and {Blagorodnova}, N. and {Blanco-Cuaresma}, S. and {Boch}, T. and {Bombrun}, A. and {Borrachero}, R. and {Bouquillon}, S. and {Bourda}, G. and {Bouy}, H. and {Bragaglia}, A. and {Breddels}, M.~A. and {Brouillet}, N. and {Br{\"u}semeister}, T. and {Bucciarelli}, B. and {Budnik}, F. and {Burgess}, P. and {Burgon}, R. and {Burlacu}, A. and {Busonero}, D. and {Buzzi}, R. and {Caffau}, E. and {Cambras}, J. and {Campbell}, H. and {Cancelliere}, R. and {Cantat-Gaudin}, T. and {Carlucci}, T. and {Carrasco}, J.~M. and {Castellani}, M. and {Charlot}, P. and {Charnas}, J. and {Charvet}, P. and {Chassat}, F. and {Chiavassa}, A. and {Clotet}, M. and {Cocozza}, G. and {Collins}, R.~S. and {Collins}, P. and {Costigan}, G.},
        title = "{The Gaia mission}",
      journal = {\aap},
         year = 2016,
        month = nov,
       volume = {595},
          eid = {A1},
        pages = {A1},
          doi = {10.1051/0004-6361/201629272},
archivePrefix = {arXiv},
       eprint = {1609.04153},
 primaryClass = {astro-ph.IM},
       adsurl = {https://ui.adsabs.harvard.edu/abs/2016A&A...595A...1G}
}

@ARTICLE{Pytorch,
       author = {{Paszke}, Adam and {Gross}, Sam and {Massa}, Francisco and {Lerer}, Adam and {Bradbury}, James and {Chanan}, Gregory and {Killeen}, Trevor and {Lin}, Zeming and {Gimelshein}, Natalia and {Antiga}, Luca and {Desmaison}, Alban and {K{\"o}pf}, Andreas and {Yang}, Edward and {DeVito}, Zach and {Raison}, Martin and {Tejani}, Alykhan and {Chilamkurthy}, Sasank and {Steiner}, Benoit and {Fang}, Lu and {Bai}, Junjie and {Chintala}, Soumith},
        title = "{PyTorch: An Imperative Style, High-Performance Deep Learning Library}",
      journal = {arXiv e-prints},
         year = 2019,
        month = dec,
          eid = {arXiv:1912.01703},
        pages = {arXiv:1912.01703},
          doi = {10.48550/arXiv.1912.01703},
archivePrefix = {arXiv},
       eprint = {1912.01703},
 primaryClass = {cs.LG},
       adsurl = {https://ui.adsabs.harvard.edu/abs/2019arXiv191201703P}
}

@ARTICLE{astropy:2013,
       author = {{Astropy Collaboration} and {Robitaille}, Thomas P. and {Tollerud}, Erik J. and {Greenfield}, Perry and {Droettboom}, Michael and {Bray}, Erik and {Aldcroft}, Tom and {Davis}, Matt and {Ginsburg}, Adam and {Price-Whelan}, Adrian M. and {Kerzendorf}, Wolfgang E. and {Conley}, Alexander and {Crighton}, Neil and {Barbary}, Kyle and {Muna}, Demitri and {Ferguson}, Henry and {Grollier}, Fr{\'e}d{\'e}ric and {Parikh}, Madhura M. and {Nair}, Prasanth H. and {Unther}, Hans M. and {Deil}, Christoph and {Woillez}, Julien and {Conseil}, Simon and {Kramer}, Roban and {Turner}, James E.~H. and {Singer}, Leo and {Fox}, Ryan and {Weaver}, Benjamin A. and {Zabalza}, Victor and {Edwards}, Zachary I. and {Azalee Bostroem}, K. and {Burke}, D.~J. and {Casey}, Andrew R. and {Crawford}, Steven M. and {Dencheva}, Nadia and {Ely}, Justin and {Jenness}, Tim and {Labrie}, Kathleen and {Lim}, Pey Lian and {Pierfederici}, Francesco and {Pontzen}, Andrew and {Ptak}, Andy and {Refsdal}, Brian and {Servillat}, Mathieu and {Streicher}, Ole},
        title = "{Astropy: A community Python package for astronomy}",
      journal = {\aap},
         year = 2013,
        month = oct,
       volume = {558},
          eid = {A33},
        pages = {A33},
          doi = {10.1051/0004-6361/201322068},
archivePrefix = {arXiv},
       eprint = {1307.6212},
 primaryClass = {astro-ph.IM},
       adsurl = {https://ui.adsabs.harvard.edu/abs/2013A&A...558A..33A}
}

@ARTICLE{astropy:2018,
       author = {{Astropy Collaboration} and {Price-Whelan}, A.~M. and {Sip{\H{o}}cz}, B.~M. and {G{\"u}nther}, H.~M. and {Lim}, P.~L. and {Crawford}, S.~M. and {Conseil}, S. and {Shupe}, D.~L. and {Craig}, M.~W. and {Dencheva}, N. and {Ginsburg}, A. and {VanderPlas}, J.~T. and {Bradley}, L.~D. and {P{\'e}rez-Su{\'a}rez}, D. and {de Val-Borro}, M. and {Aldcroft}, T.~L. and {Cruz}, K.~L. and {Robitaille}, T.~P. and {Tollerud}, E.~J. and {Ardelean}, C. and {Babej}, T. and {Bach}, Y.~P. and {Bachetti}, M. and {Bakanov}, A.~V. and {Bamford}, S.~P. and {Barentsen}, G. and {Barmby}, P. and {Baumbach}, A. and {Berry}, K.~L. and {Biscani}, F. and {Boquien}, M. and {Bostroem}, K.~A. and {Bouma}, L.~G. and {Brammer}, G.~B. and {Bray}, E.~M. and {Breytenbach}, H. and {Buddelmeijer}, H. and {Burke}, D.~J. and {Calderone}, G. and {Cano Rodr{\'\i}guez}, J.~L. and {Cara}, M. and {Cardoso}, J.~V.~M. and {Cheedella}, S. and {Copin}, Y. and {Corrales}, L. and {Crichton}, D. and {D'Avella}, D. and {Deil}, C. and {Depagne}, {\'E}. and {Dietrich}, J.~P. and {Donath}, A. and {Droettboom}, M. and {Earl}, N. and {Erben}, T. and {Fabbro}, S. and {Ferreira}, L.~A. and {Finethy}, T. and {Fox}, R.~T. and {Garrison}, L.~H. and {Gibbons}, S.~L.~J. and {Goldstein}, D.~A. and {Gommers}, R. and {Greco}, J.~P. and {Greenfield}, P. and {Groener}, A.~M. and {Grollier}, F. and {Hagen}, A. and {Hirst}, P. and {Homeier}, D. and {Horton}, A.~J. and {Hosseinzadeh}, G. and {Hu}, L. and {Hunkeler}, J.~S. and {Ivezi{\'c}}, {\v{Z}}. and {Jain}, A. and {Jenness}, T. and {Kanarek}, G. and {Kendrew}, S. and {Kern}, N.~S. and {Kerzendorf}, W.~E. and {Khvalko}, A. and {King}, J. and {Kirkby}, D. and {Kulkarni}, A.~M. and {Kumar}, A. and {Lee}, A. and {Lenz}, D. and {Littlefair}, S.~P. and {Ma}, Z. and {Macleod}, D.~M. and {Mastropietro}, M. and {McCully}, C. and {Montagnac}, S. and {Morris}, B.~M. and {Mueller}, M. and {Mumford}, S.~J. and {Muna}, D. and {Murphy}, N.~A. and {Nelson}, S. and {Nguyen}, G.~H. and {Ninan}, J.~P. and {N{\"o}the}, M. and {Ogaz}, S. and {Oh}, S. and {Parejko}, J.~K. and {Parley}, N. and {Pascual}, S. and {Patil}, R. and {Patil}, A.~A. and {Plunkett}, A.~L. and {Prochaska}, J.~X. and {Rastogi}, T. and {Reddy Janga}, V. and {Sabater}, J. and {Sakurikar}, P. and {Seifert}, M. and {Sherbert}, L.~E. and {Sherwood-Taylor}, H. and {Shih}, A.~Y. and {Sick}, J. and {Silbiger}, M.~T. and {Singanamalla}, S. and {Singer}, L.~P. and {Sladen}, P.~H. and {Sooley}, K.~A. and {Sornarajah}, S. and {Streicher}, O. and {Teuben}, P. and {Thomas}, S.~W. and {Tremblay}, G.~R. and {Turner}, J.~E.~H. and {Terr{\'o}n}, V. and {van Kerkwijk}, M.~H. and {de la Vega}, A. and {Watkins}, L.~L. and {Weaver}, B.~A. and {Whitmore}, J.~B. and {Woillez}, J. and {Zabalza}, V. and {Astropy Contributors}},
        title = "{The Astropy Project: Building an Open-science Project and Status of the v2.0 Core Package}",
      journal = {\aj},
         year = 2018,
        month = sep,
       volume = {156},
       number = {3},
          eid = {123},
        pages = {123},
          doi = {10.3847/1538-3881/aabc4f},
archivePrefix = {arXiv},
       eprint = {1801.02634},
 primaryClass = {astro-ph.IM},
       adsurl = {https://ui.adsabs.harvard.edu/abs/2018AJ....156..123A}
}

@ARTICLE{astropy:2022,
       author = {{Astropy Collaboration} and {Price-Whelan}, Adrian M. and {Lim}, Pey Lian and {Earl}, Nicholas and {Starkman}, Nathaniel and {Bradley}, Larry and {Shupe}, David L. and {Patil}, Aarya A. and {Corrales}, Lia and {Brasseur}, C.~E. and {N{\"o}the}, Maximilian and {Donath}, Axel and {Tollerud}, Erik and {Morris}, Brett M. and {Ginsburg}, Adam and {Vaher}, Eero and {Weaver}, Benjamin A. and {Tocknell}, James and {Jamieson}, William and {van Kerkwijk}, Marten H. and {Robitaille}, Thomas P. and {Merry}, Bruce and {Bachetti}, Matteo and {G{\"u}nther}, H. Moritz and {Aldcroft}, Thomas L. and {Alvarado-Montes}, Jaime A. and {Archibald}, Anne M. and {B{\'o}di}, Attila and {Bapat}, Shreyas and {Barentsen}, Geert and {Baz{\'a}n}, Juanjo and {Biswas}, Manish and {Boquien}, M{\'e}d{\'e}ric and {Burke}, D.~J. and {Cara}, Daria and {Cara}, Mihai and {Conroy}, Kyle E. and {Conseil}, Simon and {Craig}, Matthew W. and {Cross}, Robert M. and {Cruz}, Kelle L. and {D'Eugenio}, Francesco and {Dencheva}, Nadia and {Devillepoix}, Hadrien A.~R. and {Dietrich}, J{\"o}rg P. and {Eigenbrot}, Arthur Davis and {Erben}, Thomas and {Ferreira}, Leonardo and {Foreman-Mackey}, Daniel and {Fox}, Ryan and {Freij}, Nabil and {Garg}, Suyog and {Geda}, Robel and {Glattly}, Lauren and {Gondhalekar}, Yash and {Gordon}, Karl D. and {Grant}, David and {Greenfield}, Perry and {Groener}, Austen M. and {Guest}, Steve and {Gurovich}, Sebastian and {Handberg}, Rasmus and {Hart}, Akeem and {Hatfield-Dodds}, Zac and {Homeier}, Derek and {Hosseinzadeh}, Griffin and {Jenness}, Tim and {Jones}, Craig K. and {Joseph}, Prajwel and {Kalmbach}, J. Bryce and {Karamehmetoglu}, Emir and {Ka{\l}uszy{\'n}ski}, Miko{\l}aj and {Kelley}, Michael S.~P. and {Kern}, Nicholas and {Kerzendorf}, Wolfgang E. and {Koch}, Eric W. and {Kulumani}, Shankar and {Lee}, Antony and {Ly}, Chun and {Ma}, Zhiyuan and {MacBride}, Conor and {Maljaars}, Jakob M. and {Muna}, Demitri and {Murphy}, N.~A. and {Norman}, Henrik and {O'Steen}, Richard and {Oman}, Kyle A. and {Pacifici}, Camilla and {Pascual}, Sergio and {Pascual-Granado}, J. and {Patil}, Rohit R. and {Perren}, Gabriel I. and {Pickering}, Timothy E. and {Rastogi}, Tanuj and {Roulston}, Benjamin R. and {Ryan}, Daniel F. and {Rykoff}, Eli S. and {Sabater}, Jose and {Sakurikar}, Parikshit and {Salgado}, Jes{\'u}s and {Sanghi}, Aniket and {Saunders}, Nicholas and {Savchenko}, Volodymyr and {Schwardt}, Ludwig and {Seifert-Eckert}, Michael and {Shih}, Albert Y. and {Jain}, Anany Shrey and {Shukla}, Gyanendra and {Sick}, Jonathan and {Simpson}, Chris and {Singanamalla}, Sudheesh and {Singer}, Leo P. and {Singhal}, Jaladh and {Sinha}, Manodeep and {Sip{\H{o}}cz}, Brigitta M. and {Spitler}, Lee R. and {Stansby}, David and {Streicher}, Ole and {{\v{S}}umak}, Jani and {Swinbank}, John D. and {Taranu}, Dan S. and {Tewary}, Nikita and {Tremblay}, Grant R. and {de Val-Borro}, Miguel and {Van Kooten}, Samuel J. and {Vasovi{\'c}}, Zlatan and {Verma}, Shresth and {de Miranda Cardoso}, Jos{\'e} Vin{\'\i}cius and {Williams}, Peter K.~G. and {Wilson}, Tom J. and {Winkel}, Benjamin and {Wood-Vasey}, W.~M. and {Xue}, Rui and {Yoachim}, Peter and {Zhang}, Chen and {Zonca}, Andrea and {Astropy Project Contributors}},
        title = "{The Astropy Project: Sustaining and Growing a Community-oriented Open-source Project and the Latest Major Release (v5.0) of the Core Package}",
      journal = {\apj},
         year = 2022,
        month = aug,
       volume = {935},
       number = {2},
          eid = {167},
        pages = {167},
          doi = {10.3847/1538-4357/ac7c74},
archivePrefix = {arXiv},
       eprint = {2206.14220},
 primaryClass = {astro-ph.IM},
       adsurl = {https://ui.adsabs.harvard.edu/abs/2022ApJ...935..167A}
}

@article{kingma2017adammethodstochasticoptimization,
  title={Adam: A method for stochastic optimization},
  author={Kingma, Diederik P and Ba, Jimmy},
  journal={arXiv preprint arXiv:1412.6980},
  year={2014}
}

@ARTICLE{eDR3_phot,
       author = {{Riello}, M. and {De Angeli}, F. and {Evans}, D.~W. and {Montegriffo}, P. and {Carrasco}, J.~M. and {Busso}, G. and {Palaversa}, L. and {Burgess}, P.~W. and {Diener}, C. and {Davidson}, M. and {Rowell}, N. and {Fabricius}, C. and {Jordi}, C. and {Bellazzini}, M. and {Pancino}, E. and {Harrison}, D.~L. and {Cacciari}, C. and {van Leeuwen}, F. and {Hambly}, N.~C. and {Hodgkin}, S.~T. and {Osborne}, P.~J. and {Altavilla}, G. and {Barstow}, M.~A. and {Brown}, A.~G.~A. and {Castellani}, M. and {Cowell}, S. and {De Luise}, F. and {Gilmore}, G. and {Giuffrida}, G. and {Hidalgo}, S. and {Holland}, G. and {Marinoni}, S. and {Pagani}, C. and {Piersimoni}, A.~M. and {Pulone}, L. and {Ragaini}, S. and {Rainer}, M. and {Richards}, P.~J. and {Sanna}, N. and {Walton}, N.~A. and {Weiler}, M. and {Yoldas}, A.},
        title = "{Gaia Early Data Release 3. Photometric content and validation}",
      journal = {\aap},
         year = 2021,
        month = may,
       volume = {649},
          eid = {A3},
        pages = {A3},
          doi = {10.1051/0004-6361/202039587},
archivePrefix = {arXiv},
       eprint = {2012.01916},
 primaryClass = {astro-ph.IM},
       adsurl = {https://ui.adsabs.harvard.edu/abs/2021A&A...649A...3R}
}

@article{Rubin_76,
    author = {Rubin, Donald B.},
    title = {Inference and missing data},
    journal = {Biometrika},
    volume = {63},
    number = {3},
    pages = {581-592},
    year = {1976},
    month = {12},
    issn = {0006-3444},
    doi = {10.1093/biomet/63.3.581},
    url = {https://doi.org/10.1093/biomet/63.3.581},
    eprint = {https://academic.oup.com/biomet/article-pdf/63/3/581/756166/63-3-581.pdf},
}

@ARTICLE{Castro_Ginard_2023,
       author = {{Castro-Ginard}, Alfred and {Brown}, Anthony G.~A. and {Kostrzewa-Rutkowska}, Zuzanna and {Cantat-Gaudin}, Tristan and {Drimmel}, Ronald and {Oh}, Semyeong and {Belokurov}, Vasily and {Casey}, Andrew R. and {Fouesneau}, Morgan and {Khanna}, Shourya and {Price-Whelan}, Adrian M. and {Rix}, Hans-Walter},
        title = "{Estimating the selection function of Gaia DR3 subsamples}",
      journal = {\aap},
         year = 2023,
        month = sep,
       volume = {677},
          eid = {A37},
        pages = {A37},
          doi = {10.1051/0004-6361/202346547},
archivePrefix = {arXiv},
       eprint = {2303.17738},
 primaryClass = {astro-ph.GA},
       adsurl = {https://ui.adsabs.harvard.edu/abs/2023A&A...677A..37C}
}

@ARTICLE{Everall_2022,
       author = {{Everall}, Andrew and {Boubert}, Douglas},
        title = "{Completeness of the Gaia verse - V. Astrometry and radial velocity sample selection functions in Gaia EDR3}",
      journal = {\mnras},
         year = 2022,
        month = feb,
       volume = {509},
       number = {4},
        pages = {6205-6224},
          doi = {10.1093/mnras/stab3262},
archivePrefix = {arXiv},
       eprint = {2111.04127},
 primaryClass = {astro-ph.GA},
       adsurl = {https://ui.adsabs.harvard.edu/abs/2022MNRAS.509.6205E}
}

@ARTICLE{Zhang_2026,
       author = {{Zhang}, Mengmeng and {Bu}, Yude and {Wang}, Siqi and {Li}, Shanshan and {Zhang}, Jiangchuan and {Sun}, Jingzhen and {Zhang}, Yuhang and {Wang}, Ke and {Liu}, Jian and {Yan}, Hongliang and {Yi}, Zhenping and {Liu}, Meng and {Kong}, Xiaoming},
        title = "{A Generalist Model Including Evolved Star Mass and Age}",
      journal = {arXiv e-prints},
         year = 2026,
        month = mar,
          eid = {arXiv:2603.03732},
        pages = {arXiv:2603.03732},
          doi = {10.48550/arXiv.2603.03732},
archivePrefix = {arXiv},
       eprint = {2603.03732},
 primaryClass = {astro-ph.SR},
       adsurl = {https://ui.adsabs.harvard.edu/abs/2026arXiv260303732Z}
}

@ARTICLE{Zhao_2026,
       author = {{Zhao}, Xiaosheng and {Huang}, Yang and {Xue}, Guirong and {Kong}, Xiao and {Liu}, Jifeng and {Tang}, Xiaoyu and {Beers}, Timothy C. and {Ting}, Yuan-Sen and {Luo}, A.-Li},
        title = "{SpecCLIP: Aligning and Translating Spectroscopic Measurements for Stars}",
      journal = {\apj},
         year = 2026,
        month = feb,
       volume = {998},
       number = {2},
          eid = {189},
        pages = {189},
          doi = {10.3847/1538-4357/ae2c7e},
archivePrefix = {arXiv},
       eprint = {2507.01939},
 primaryClass = {astro-ph.IM},
       adsurl = {https://ui.adsabs.harvard.edu/abs/2026ApJ...998..189Z}
}

@ARTICLE{gaia_astrophys_1,
       author = {{Creevey}, O.~L. and {Sordo}, R. and {Pailler}, F. and {Fr{\'e}mat}, Y. and {Heiter}, U. and {Th{\'e}venin}, F. and {Andrae}, R. and {Fouesneau}, M. and {Lobel}, A. and {Bailer-Jones}, C.~A.~L. and {Garabato}, D. and {Bellas-Velidis}, I. and {Brugaletta}, E. and {Lorca}, A. and {Ordenovic}, C. and {Palicio}, P.~A. and {Sarro}, L.~M. and {Delchambre}, L. and {Drimmel}, R. and {Rybizki}, J. and {Torralba Elipe}, G. and {Korn}, A.~J. and {Recio-Blanco}, A. and {Schultheis}, M.~S. and {De Angeli}, F. and {Montegriffo}, P. and {Abreu Aramburu}, A. and {Accart}, S. and {{\'A}lvarez}, M.~A. and {Bakker}, J. and {Brouillet}, N. and {Burlacu}, A. and {Carballo}, R. and {Casamiquela}, L. and {Chiavassa}, A. and {Contursi}, G. and {Cooper}, W.~J. and {Dafonte}, C. and {Dapergolas}, A. and {de Laverny}, P. and {Dharmawardena}, T.~E. and {Edvardsson}, B. and {Le Fustec}, Y. and {Garc{\'\i}a-Lario}, P. and {Garc{\'\i}a-Torres}, M. and {Gomez}, A. and {Gonz{\'a}lez-Santamar{\'\i}a}, I. and {Hatzidimitriou}, D. and {Jean-Antoine Piccolo}, A. and {Kontiza}, M. and {Kordopatis}, G. and {Lanzafame}, A.~C. and {Lebreton}, Y. and {Licata}, E.~L. and {Lindstr{\o}m}, H.~E.~P. and {Livanou}, E. and {Magdaleno Romeo}, A. and {Manteiga}, M. and {Marocco}, F. and {Marshall}, D.~J. and {Mary}, N. and {Nicolas}, C. and {Pallas-Quintela}, L. and {Panem}, C. and {Pichon}, B. and {Poggio}, E. and {Riclet}, F. and {Robin}, C. and {Santove{\~n}a}, R. and {Silvelo}, A. and {Slezak}, I. and {Smart}, R.~L. and {Soubiran}, C. and {S{\"u}veges}, M. and {Ulla}, A. and {Utrilla}, E. and {Vallenari}, A. and {Zhao}, H. and {Zorec}, J. and {Barrado}, D. and {Bijaoui}, A. and {Bouret}, J.-C. and {Blomme}, R. and {Brott}, I. and {Cassisi}, S. and {Kochukhov}, O. and {Martayan}, C. and {Shulyak}, D. and {Silvester}, J.},
        title = "{Gaia Data Release 3. Astrophysical parameters inference system (Apsis). I. Methods and content overview}",
      journal = {\aap},
         year = 2023,
        month = jun,
       volume = {674},
          eid = {A26},
        pages = {A26},
          doi = {10.1051/0004-6361/202243688},
archivePrefix = {arXiv},
       eprint = {2206.05864},
 primaryClass = {astro-ph.GA},
       adsurl = {https://ui.adsabs.harvard.edu/abs/2023A&A...674A..26C}
}

@ARTICLE{gaia_astrophys_2,
       author = {{Fouesneau}, M. and {Fr{\'e}mat}, Y. and {Andrae}, R. and {Korn}, A.~J. and {Soubiran}, C. and {Kordopatis}, G. and {Vallenari}, A. and {Heiter}, U. and {Creevey}, O.~L. and {Sarro}, L.~M. and {de Laverny}, P. and {Lanzafame}, A.~C. and {Lobel}, A. and {Sordo}, R. and {Rybizki}, J. and {Slezak}, I. and {{\'A}lvarez}, M.~A. and {Drimmel}, R. and {Garabato}, D. and {Delchambre}, L. and {Bailer-Jones}, C.~A.~L. and {Hatzidimitriou}, D. and {Lorca}, A. and {Le Fustec}, Y. and {Pailler}, F. and {Mary}, N. and {Robin}, C. and {Utrilla}, E. and {Abreu Aramburu}, A. and {Bakker}, J. and {Bellas-Velidis}, I. and {Bijaoui}, A. and {Blomme}, R. and {Bouret}, J.-C. and {Brouillet}, N. and {Brugaletta}, E. and {Burlacu}, A. and {Carballo}, R. and {Casamiquela}, L. and {Chaoul}, L. and {Chiavassa}, A. and {Contursi}, G. and {Cooper}, W.~J. and {Dafonte}, C. and {Demouchy}, C. and {Dharmawardena}, T.~E. and {Garc{\'\i}a-Lario}, P. and {Garc{\'\i}a-Torres}, M. and {Gomez}, A. and {Gonz{\'a}lez-Santamar{\'\i}a}, I. and {Jean-Antoine Piccolo}, A. and {Kontizas}, M. and {Lebreton}, Y. and {Licata}, E.~L. and {Lindstr{\o}m}, H.~E.~P. and {Livanou}, E. and {Magdaleno Romeo}, A. and {Manteiga}, M. and {Marocco}, F. and {Martayan}, C. and {Marshall}, D.~J. and {Nicolas}, C. and {Ordenovic}, C. and {Palicio}, P.~A. and {Pallas-Quintela}, L. and {Pichon}, B. and {Poggio}, E. and {Recio-Blanco}, A. and {Riclet}, F. and {Santove{\~n}a}, R. and {Schultheis}, M.~S. and {Segol}, M. and {Silvelo}, A. and {Smart}, R.~L. and {S{\"u}veges}, M. and {Th{\'e}venin}, F. and {Torralba Elipe}, G. and {Ulla}, A. and {van Dillen}, E. and {Zhao}, H. and {Zorec}, J.},
        title = "{Gaia Data Release 3. Apsis. II. Stellar parameters}",
      journal = {\aap},
         year = 2023,
        month = jun,
       volume = {674},
          eid = {A28},
        pages = {A28},
          doi = {10.1051/0004-6361/202243919},
archivePrefix = {arXiv},
       eprint = {2206.05992},
 primaryClass = {astro-ph.SR},
       adsurl = {https://ui.adsabs.harvard.edu/abs/2023A&A...674A..28F}
}

@ARTICLE{gaia_astrophys_3,
       author = {{Delchambre}, L. and {Bailer-Jones}, C.~A.~L. and {Bellas-Velidis}, I. and {Drimmel}, R. and {Garabato}, D. and {Carballo}, R. and {Hatzidimitriou}, D. and {Marshall}, D.~J. and {Andrae}, R. and {Dafonte}, C. and {Livanou}, E. and {Fouesneau}, M. and {Licata}, E.~L. and {Lindstr{\o}m}, H.~E.~P. and {Manteiga}, M. and {Robin}, C. and {Silvelo}, A. and {Abreu Aramburu}, A. and {{\'A}lvarez}, M.~A. and {Bakker}, J. and {Bijaoui}, A. and {Brouillet}, N. and {Brugaletta}, E. and {Burlacu}, A. and {Casamiquela}, L. and {Chaoul}, L. and {Chiavassa}, A. and {Contursi}, G. and {Cooper}, W.~J. and {Creevey}, O.~L. and {Dapergolas}, A. and {de Laverny}, P. and {Demouchy}, C. and {Dharmawardena}, T.~E. and {Edvardsson}, B. and {Fr{\'e}mat}, Y. and {Garc{\'\i}a-Lario}, P. and {Garc{\'\i}a-Torres}, M. and {Gavel}, A. and {Gomez}, A. and {Gonz{\'a}lez-Santamar{\'\i}a}, I. and {Heiter}, U. and {Jean-Antoine Piccolo}, A. and {Kontizas}, M. and {Kordopatis}, G. and {Korn}, A.~J. and {Lanzafame}, A.~C. and {Lebreton}, Y. and {Lobel}, A. and {Lorca}, A. and {Magdaleno Romeo}, A. and {Marocco}, F. and {Mary}, N. and {Nicolas}, C. and {Ordenovic}, C. and {Pailler}, F. and {Palicio}, P.~A. and {Pallas-Quintela}, L. and {Panem}, C. and {Pichon}, B. and {Poggio}, E. and {Recio-Blanco}, A. and {Riclet}, F. and {Rybizki}, J. and {Santove{\~n}a}, R. and {Sarro}, L.~M. and {Schultheis}, M.~S. and {Segol}, M. and {Slezak}, I. and {Smart}, R.~L. and {Sordo}, R. and {Soubiran}, C. and {S{\"u}veges}, M. and {Th{\'e}venin}, F. and {Torralba Elipe}, G. and {Ulla}, A. and {Utrilla}, E. and {Vallenari}, A. and {van Dillen}, E. and {Zhao}, H. and {Zorec}, J.},
        title = "{Gaia Data Release 3. Apsis. III. Non-stellar content and source classification}",
      journal = {\aap},
         year = 2023,
        month = jun,
       volume = {674},
          eid = {A31},
        pages = {A31},
          doi = {10.1051/0004-6361/202243423},
archivePrefix = {arXiv},
       eprint = {2206.06710},
 primaryClass = {astro-ph.GA},
       adsurl = {https://ui.adsabs.harvard.edu/abs/2023A&A...674A..31D}
}

@ARTICLE{Poggio_2025,
       author = {{Poggio}, E. and {Khanna}, S. and {Drimmel}, R. and {Zari}, E. and {D'Onghia}, E. and {Lattanzi}, M.~G. and {Palicio}, P.~A. and {Recio-Blanco}, A. and {Thulasidharan}, L.},
        title = "{The great wave: Evidence of a large-scale vertical corrugation propagating outwards in the Galactic disc}",
      journal = {\aap},
         year = 2025,
        month = jul,
       volume = {699},
          eid = {A199},
        pages = {A199},
          doi = {10.1051/0004-6361/202451668},
archivePrefix = {arXiv},
       eprint = {2407.18659},
 primaryClass = {astro-ph.GA},
       adsurl = {https://ui.adsabs.harvard.edu/abs/2025A&A...699A.199P}
}

@ARTICLE{Reyes_2023,
       author = {{Cruz Reyes}, Mauricio and {Anderson}, Richard I.},
        title = "{A 0.9\% calibration of the Galactic Cepheid luminosity scale based on Gaia DR3 data of open clusters and Cepheids}",
      journal = {\aap},
         year = 2023,
        month = apr,
       volume = {672},
          eid = {A85},
        pages = {A85},
          doi = {10.1051/0004-6361/202244775},
archivePrefix = {arXiv},
       eprint = {2208.09403},
 primaryClass = {astro-ph.GA},
       adsurl = {https://ui.adsabs.harvard.edu/abs/2023A&A...672A..85C}
}

@ARTICLE{Dubus_2024,
       author = {{Dubus}, Guillaume and {Babusiaux}, Carine},
        title = "{Cataclysmic variables and the disc instability model in the Gaia DR3 colour-magnitude diagram}",
      journal = {\aap},
         year = 2024,
        month = mar,
       volume = {683},
          eid = {A247},
        pages = {A247},
          doi = {10.1051/0004-6361/202348510},
archivePrefix = {arXiv},
       eprint = {2401.12206},
 primaryClass = {astro-ph.SR},
       adsurl = {https://ui.adsabs.harvard.edu/abs/2024A&A...683A.247D}
}

@ARTICLE{Dantas_2026,
       author = {{Dantas}, M.~L.~L. and {Garc{\'\i}a-Delgado}, J.~J. and {Rebollido}, I. and {Smiljanic}, R.},
        title = "{Galactic archaeology meets exoplanets: linking stellar birth radii to exoplanet demographics}",
      journal = {arXiv e-prints},
         year = 2026,
        month = sep,
          eid = {arXiv:2609.19274},
        pages = {arXiv:2609.19274},
          doi = {10.48550/arXiv.2609.19274},
archivePrefix = {arXiv},
       eprint = {2609.19274},
 primaryClass = {astro-ph.GA},
       adsurl = {https://ui.adsabs.harvard.edu/abs/2026arXiv260919274D}
}

@ARTICLE{Dropulic_2023,
       author = {{Dropulic}, Adriana and {Liu}, Hongwan and {Ostdiek}, Bryan and {Lisanti}, Mariangela},
        title = "{Revealing the Milky Way's most recent major merger with a Gaia EDR3 catalogue of machine-learned line-of-sight velocities}",
      journal = {\mnras},
         year = 2023,
        month = may,
       volume = {521},
       number = {2},
        pages = {1633-1645},
          doi = {10.1093/mnras/stad209},
archivePrefix = {arXiv},
       eprint = {2205.12278},
 primaryClass = {astro-ph.GA},
       adsurl = {https://ui.adsabs.harvard.edu/abs/2023MNRAS.521.1633D}
}

@ARTICLE{Naik_2024,
       author = {{Naik}, Aneesh P. and {Widmark}, Axel},
        title = "{The missing radial velocities of Gaia: a catalogue of Bayesian estimates for DR3}",
      journal = {\mnras},
         year = 2024,
        month = feb,
       volume = {527},
       number = {4},
        pages = {11559-11574},
          doi = {10.1093/mnras/stad3822},
archivePrefix = {arXiv},
       eprint = {2307.13398},
 primaryClass = {astro-ph.GA},
       adsurl = {https://ui.adsabs.harvard.edu/abs/2024MNRAS.52711559N}
}

@ARTICLE{Brown_2021,
       author = {{Brown}, Anthony G.~A.},
        title = "{Microarcsecond Astrometry: Science Highlights from Gaia}",
      journal = {\araa},
         year = 2021,
        month = sep,
       volume = {59},
        pages = {59-115},
          doi = {10.1146/annurev-astro-112320-035628},
archivePrefix = {arXiv},
       eprint = {2102.11712},
 primaryClass = {astro-ph.IM},
       adsurl = {https://ui.adsabs.harvard.edu/abs/2021ARA&A..59...59B}
}

@ARTICLE{Hallin_2025,
       author = {{Hallin}, Anna and {Shih}, David and {Krause}, Claudius and {Buckley}, Matthew R.},
        title = "{Via Machinae 3.0: A search for stellar streams in Gaia with the CATHODE algorithm}",
      journal = {arXiv e-prints},
         year = 2025,
        month = sep,
          eid = {arXiv:2509.08064},
        pages = {arXiv:2509.08064},
          doi = {10.48550/arXiv.2509.08064},
archivePrefix = {arXiv},
       eprint = {2509.08064},
 primaryClass = {astro-ph.GA},
       adsurl = {https://ui.adsabs.harvard.edu/abs/2025arXiv250908064H}
}
\bibliographystyle{aasjournalv7}

\appendix

\section{TARP and RMSE Across Conditioning Patterns}\label{appendix_stats}
\begin{figure*}[ht!]
\includegraphics[width=\textwidth]{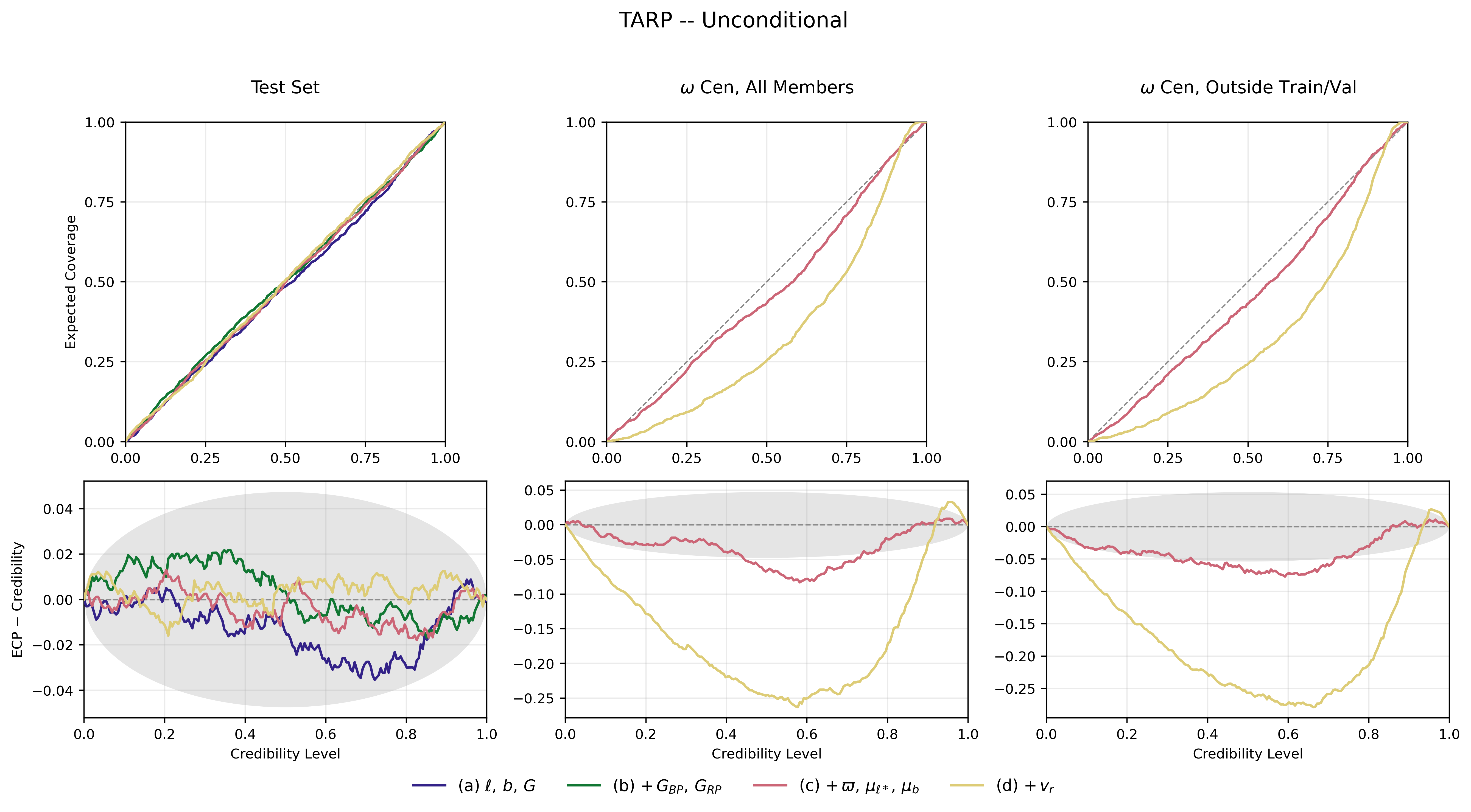}
\caption{TARP results for the unconditional case. The 4 curves correspond to sources with different numbers of measured parameters in the source catalog (see text of Appendix \ref{appendix_stats} for more details). The left column corresponds to results for the test set. The center column corresponds to results for the full $\omega$ Cen member set, as in Section \ref{om_cen}. The rightmost column displays results for members of $\omega$ Cen that were not included in the training or validation set. The top row shows the expected coverage of the posterior estimator as calculated following the procedure outlined in \cite{TARP} compared to the credibility level. The bottom row shows the difference between the expected coverage probability ($\rm{ECP}$) and the credibility level. The gray shaded region corresponds to three times the TARP error, defined in Equation \ref{eqn:TARP_error}. For the test set and $\omega$ Cen ``All Members" set each shown class has 1000 sources, however for the $\omega$ Cen ``Outside Train/Val" set class (d) only has 781 sources. TARP values are calculated using 200 model draws per source. Note that for the test set data, which corresponds to the full catalog distribution, the results suggest the model could be an accurate posterior estimator. For both sets of $\omega$ Cen data, the model is demonstrably biased.}
\label{fig:TARP_uncond}
\end{figure*}

Tests of Accuracy with Random Points (TARP) \citep{TARP}, provides a metric for evaluating the accuracy of generative posterior estimators. A TARP curve that hews closely to the diagonal line (see Figure \ref{fig:TARP_uncond}, for example) suggests that the posterior estimator (the model) produces a distribution of samples that matches the true posterior. I perform this analysis for 47 conditioning patterns, for both the test set and $\omega$ Cen members. I include two $\omega$ Cen sets, one with all members found by \citet{Soltis_21},\footnote{As in Section \ref{om_cen}, I do not include the CMD cut.} and one with only those members not found in the model training or validation set. For each conditioning pattern, four TARP curves are plotted. Each curve corresponds to the results for sources that have valid measurements for the listed parameters, and only those parameters: (a) $\ell$, $b$, $G$; (b) $\ell$, $b$, $G$, $G_{BP}$, $G_{RP}$; (c) $\ell$, $b$, $G$, $G_{BP}$, $G_{RP}$, $\varpi$, $\mu_{\ell*}$, $\mu_{b}$; (d) $\ell$, $b$, $G$, $G_{BP}$, $G_{RP}$, $\varpi$, $\mu_{\ell*}$, $\mu_{b}$, $v_r$. These are referred to as ``classes". Where the conditioning pattern contains more than the listed parameters, that class is excluded. Only classes (c) and (d) are available for the $\omega$ Cen data due to the selection criteria used to determine cluster membership. For each class, if possible, 1000 sources are chosen that satisfy both the curve requirements and the conditioning pattern requirements (strictly, as in Section \ref{im_RV}). For the test set and $\omega$ Cen ``All Members" set each shown class has 1000 sources, however for the $\omega$ Cen ``Outside Train/Val" set class (d) only has 781 sources. For each source 200 model samples are generated. These samples are then used to calculate the TARP curve. Reference points are drawn from a uniform distribution, $\mathcal{U}\left(-1,1\right)$, in normalized units. TARP is calculated using only the valid parameters of the corresponding class that are not included in the conditioning pattern. An error on the TARP measurement, corresponding to the standard deviation of a binomial distribution with $n$ trials and probability $c$ of success, is $\sigma_{B} = \sqrt{nc(1-c)}$ divided by the number of independent trials: 

\begin{equation}\label{eqn:TARP_error}
    \sigma_{\rm{TARP}} = \sqrt{\frac{c(1-c)}{n}}
\end{equation}

\begin{figure*}[ht!]
\includegraphics[width=\textwidth]{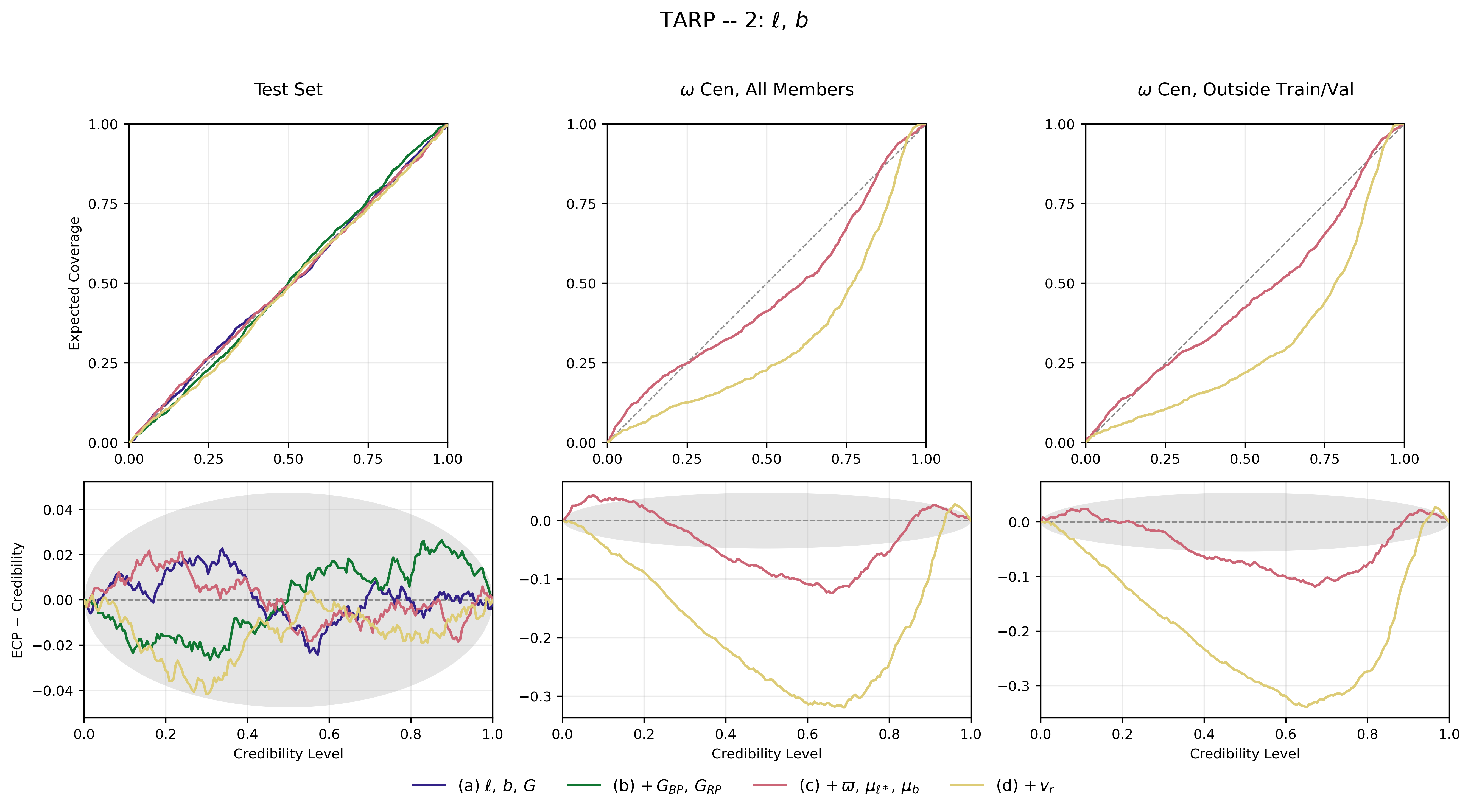}
\caption{TARP results for the model conditioned on source position ($\ell$, $b$). See Figure \ref{fig:TARP_uncond} and text for a description of the methodology. The gray shaded region corresponds to three times the TARP error, defined in Equation \ref{eqn:TARP_error}. Here, as in Figure \ref{fig:TARP_uncond}, the test set results suggest that the model could be an accurate posterior estimator for the test set. The $\omega$ Cen results strongly suggest the model is an inaccurate posterior estimator for cluster members. These results corroborate the results found in Sections \ref{mw_field} and \ref{om_cen}.}
\label{fig:TARP_l_b}
\end{figure*}

\begin{figure*}[ht!]
\includegraphics[width=\textwidth]{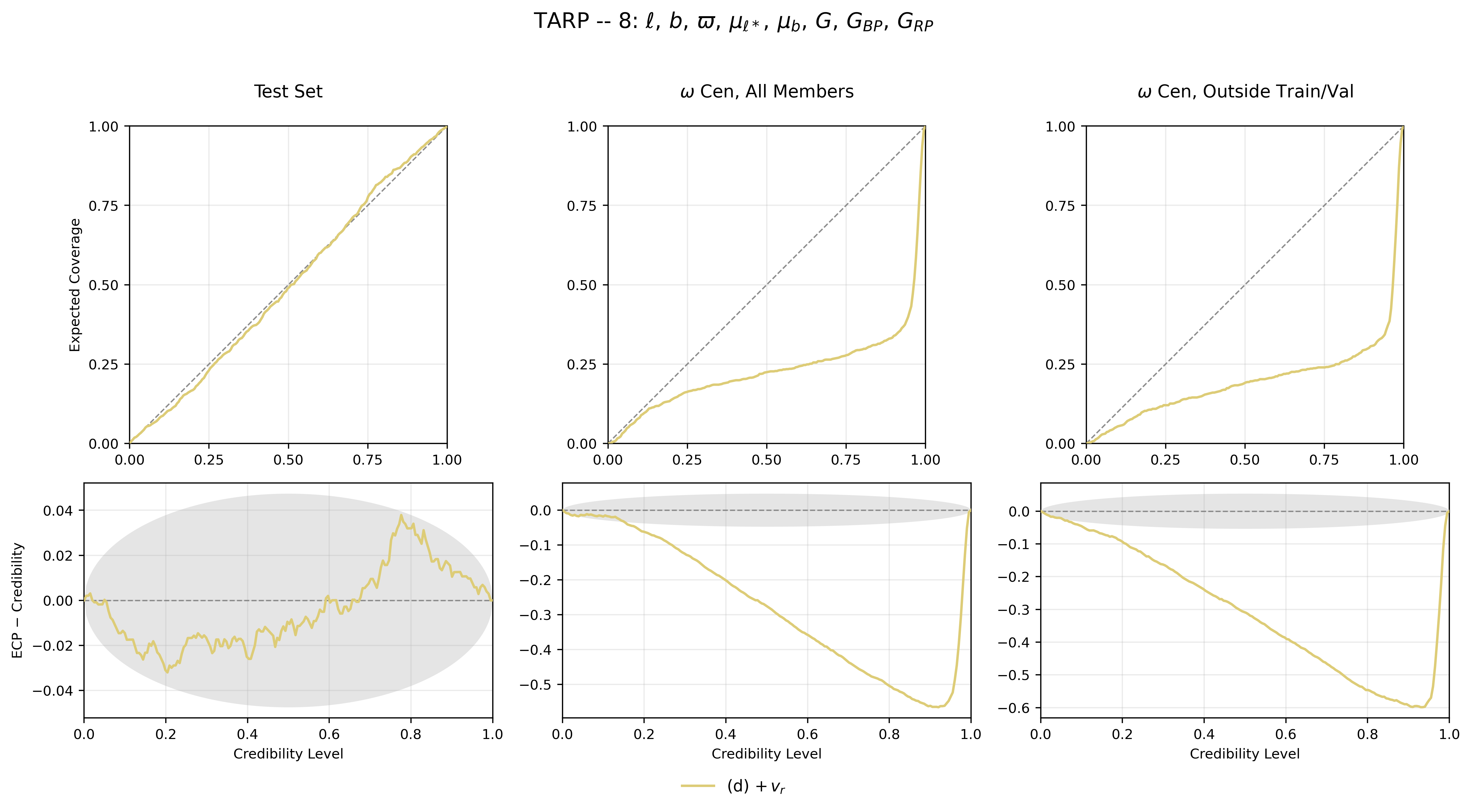}
\caption{TARP results for the model conditioned on all parameters but the radial velocity ($v_r$). See Figure \ref{fig:TARP_uncond} and text for a description of the methodology. The gray shaded region corresponds to three times the TARP error, defined in Equation \ref{eqn:TARP_error}. The results provide additional evidence that the model is potentially useful for imputing missing radial velocities, as suggested by Section \ref{im_RV}. While within the bounds of the error, the pattern of the curve in the leftmost plot (below the diagonal below credibility = 0.5, above the diagonal above credibility = 0.5) suggests the model could be underconfident. On the other hand, the results for $\omega$ Cen demonstrate the model is not an accurate posterior estimator for radial velocities of cluster members.}
\label{fig:TARP_drop_RV}
\end{figure*}

\begin{figure*}[ht!]
\includegraphics[width=\textwidth]{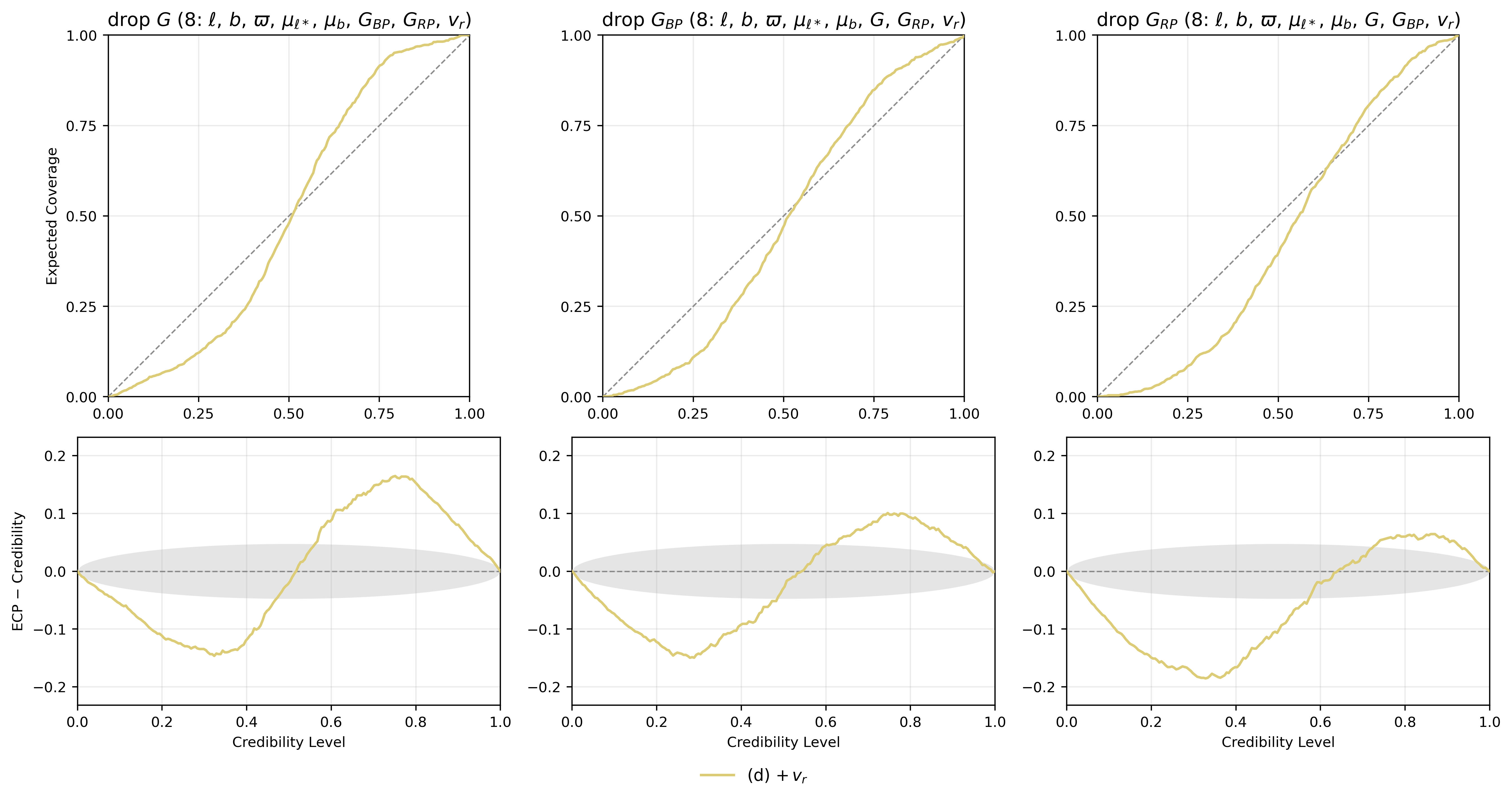}
\caption{TARP test set results for the model conditioned on all parameters but $G$ (left), $G_{BP}$ (center), or $G_{RP}$ (right). Unlike most of the other test set conditioning patterns investigated, the results for the three patterns shown clearly show the model to be a poor posterior estimator. The TARP curves suggest the model is strongly underconfident. The flux in the $G$ band is closely related to the sum of the flux in the $BP$ band and the flux in the $RP$ band. The model fails to learn the near deterministic relationship between $G$, $G_{BP}$, and $G_{RP}$.}
\label{fig:TARP_drop_phot_comp}
\end{figure*}

Results for the TARP analysis for both the test set and $\omega$ Cen are shown in Figures \ref{fig:TARP_uncond}, \ref{fig:TARP_l_b}, and \ref{fig:TARP_drop_RV}. For each figure, the gray region corresponds to three times the error defined in Equation \ref{eqn:TARP_error}. Because the reference points were chosen independently of the conditioning parameters, this version of TARP is blind to the case where the posterior estimator is equal to the prior \citep[][see their Figure 12]{TARP}. Thus having an expected coverage probability ($\rm{ECP}$) as measured following the procedure in \citet{TARP} equivalent to the theoretical credibility ($c$, identical to the probability of success in Equation \ref{eqn:TARP_error}) is a necessary but not sufficient condition for the model to be an accurate posterior estimator. I attempt to break this degeneracy later by testing the RMSE of the model estimates and comparing them to the RMSE of guessing the mean. For Figures \ref{fig:TARP_uncond}, \ref{fig:TARP_l_b}, and \ref{fig:TARP_drop_RV}, the results suggest that (a) the model could be an effective posterior estimator for the test set data, as suggested in Section \ref{mw_field} and (b) the model is a biased posterior estimator for $\omega$ Cen members as suggested in Section \ref{om_cen}. In Figure \ref{fig:TARP_drop_RV}, I investigate the conditioning pattern corresponding to imputing a missing radial velocity. The results agree with Figures \ref{fig:TARP_uncond} and \ref{fig:TARP_l_b}, as well as Section \ref{im_RV}. In Tables \ref{tab:tarp:test:a}, \ref{tab:tarp:test:b}, \ref{tab:tarp:test:c}, and \ref{tab:tarp:test:d}, I share the TARP absolute maximum deviations, defined as
\begin{equation}\label{eqn:max_dev_tarp}
    |\Delta| \equiv |\rm{ECP} - c|
\end{equation}
I share this for 47 conditioning patterns, all on the test set data, each using 1000 sources. These results reveal that most patterns stay within $\lesssim3\sigma_{\rm{TARP}}$,\footnote{I use $3\sigma_{\rm{TARP}}$ only as a rough heuristic.} albeit with important exceptions. One such exception is for patterns that contain some of the photometric parameters ($G$, $G_{BP}$, $G_{RP}$) and are asked to predict only the other photometric parameters (See the last row of Table \ref{tab:tarp:test:b} and the ($\ell$, $b$, $\mu_{b}$, $\varpi$, $\mu_{\ell*}$, $G$, $v_r$) and drop $G$/$G_{BP}$/$G_{RP}$ rows in Table \ref{tab:tarp:test:d}). In these cases the TARP curves reveal the model is strongly underconfident, as seen in Figure \ref{fig:TARP_drop_phot_comp}. Tables \ref{tab:tarp:omega_cen_not_A:c} and \ref{tab:tarp:omega_cen_not_A:d} show the class c and d max deviation results for $\omega$ Cen members, excluding those in the training and validation sets. Unlike for the test set results, the max deviation typically exceeds $3 \sigma_{\rm TARP}$, suggesting that the model is not a reliable posterior density estimator for $\omega$ Cen members.

Complementary to the TARP analysis, and using the same sources and model simulations per source, I also calculate the RMSE of the mean model prediction of a parameter compared to the true source value. As in Section \ref{im_RV}, I compare this RMSE to a baseline of guessing the mean parameter value of the dataset (in this case the class-specific sources) for every source. I then calculate a ``skill", as defined in Equation \ref{eqn:skill}. This provides an easily interpretable estimate of how effective the model is as a point estimate predictor of parameters.

Test set results for all 47 patterns, and all four classes (defined in TARP description above), are shown in Tables \ref{tab:skill:test:a}, \ref{tab:skill:test:b}, \ref{tab:skill:test:c}, and \ref{tab:skill:test:d}. Results for class c and class d for $\omega$ Cen (excluding sources in the training and validation datasets) are shown in Tables \ref{tab:skill:omega_cen_not_A:c} and \ref{tab:skill:omega_cen_not_A:d}. A couple trends merit mentioning. First, the results suggest the model is a powerful point estimator of $G$, $G_{BP}$, and $G_{RP}$ when the conditioning pattern contains at least one of those three. While the model underestimates the strength of their relationship (see Figure \ref{fig:TARP_drop_phot_comp}), it is clearly still learning it to some degree. This represents a real imputation use case, when considering sources without $G_{BP}$ or $G_{RP}$ (e.g., one could impute $G_{BP}$ and $G_{RP}$ for the sources found in class (a) in Table \ref{tab:skill:test:a}). Second, the results (especially Table \ref{tab:skill:test:d}) demonstrate that the model tends to be a poor point estimator for astrometric parameters and radial velocity ($\varpi$, $\mu_b$, $\mu_{\ell*}$, $v_r$) when other astrometric parameters or radial velocity are not provided in the conditioning information. These results show that the model's ability to use non-astrometric information to predict astrometric parameters in the Gaia Source Catalog is limited. Finally, Tables \ref{tab:skill:omega_cen_not_A:c} and \ref{tab:skill:omega_cen_not_A:d} show that the model is not an effective point estimator for $\omega$ Cen sources, except when imputing photometry with partial photometry information, or $\varpi$ with proper motion and radial velocity information. The latter suggests that the model has learned important information about the 6-d kinematic distribution of the Gaia survey, information about correlations in astrometric parameter estimation, the importance of the radial velocity selection function, or some combination of the three. 

\begingroup\setlength{\tabcolsep}{4pt}
\begin{longtable}{l rrrr}
\caption{TARP deviation for 1000 sources from the test set. Calculated on class (a), i.e., sources with measured $\ell,\,b,\,G$ only. Cond.\ Pattern denotes the specific set of conditioning information provided to the model when sampling. $\max|\Delta|$ is the maximum deviation, where the deviation is defined in Equation~\ref{eqn:max_dev_tarp}; its sign is that of ECP~$-$~$c$ at the maximum. $c$ is the credibility corresponding to the maximum deviation. $\sigma_{\rm TARP}$ is the error on the TARP measurement at $c$, defined in Equation~\ref{eqn:TARP_error}. $|\Delta|/\sigma_{\rm TARP}$ is the maximum deviation in units of that error.}\label{tab:tarp:test:a}\\
\toprule
Cond.\ Pattern & $\max |\Delta|$ & $c$ & $\sigma_{\rm TARP}$ & $\frac{|\Delta|}{\sigma_{\rm TARP}}$ \\
\midrule
\endfirsthead
\caption[]{TARP deviation for the test set, class (a), 1000 sources with measured $\ell,\,b,\,G$ only (continued)}\\
\toprule
Cond.\ Pattern & $\max |\Delta|$ & $c$ & $\sigma_{\rm TARP}$ & $\frac{|\Delta|}{\sigma_{\rm TARP}}$ \\
\midrule
\endhead
\midrule
\multicolumn{5}{r}{continued}\\
\endfoot
\bottomrule
\endlastfoot
uncond & $-$0.035 & 0.72 & 0.014 & 2.5 \\
$\ell,\,b$ & $-$0.024 & 0.57 & 0.016 & 1.5 \\
only $\ell$ & $+$0.027 & 0.82 & 0.012 & 2.2 \\
only $b$ & $+$0.029 & 0.81 & 0.012 & 2.3 \\
only $G$ & $+$0.020 & 0.34 & 0.015 & 1.3 \\
\end{longtable}
\endgroup

\begingroup\setlength{\tabcolsep}{4pt}
\begin{longtable}{l rrrr}
\caption{TARP deviation for 1000 sources from the test set. Calculated on class (b), i.e., sources with measured $\ell,\,b,\,G,\,G_{BP},\,G_{RP}$ only. See the caption of Table~\ref{tab:tarp:test:a} for a full description of the columns.}\label{tab:tarp:test:b}\\
\toprule
Cond.\ Pattern & $\max |\Delta|$ & $c$ & $\sigma_{\rm TARP}$ & $\frac{|\Delta|}{\sigma_{\rm TARP}}$ \\
\midrule
\endfirsthead
\caption[]{TARP deviation for the test set, class (b), 1000 sources with measured $\ell,\,b,\,G,\,G_{BP},\,G_{RP}$ only (continued)}\\
\toprule
Cond.\ Pattern & $\max |\Delta|$ & $c$ & $\sigma_{\rm TARP}$ & $\frac{|\Delta|}{\sigma_{\rm TARP}}$ \\
\midrule
\endhead
\midrule
\multicolumn{5}{r}{continued}\\
\endfoot
\bottomrule
\endlastfoot
uncond & $+$0.022 & 0.36 & 0.015 & 1.4 \\
$\ell,\,b$ & $-$0.026 & 0.31 & 0.015 & 1.8 \\
$\ell,\,b,\,G$ & $-$0.026 & 0.36 & 0.015 & 1.7 \\
$G,\,G_{BP},\,G_{RP}$ & $-$0.025 & 0.36 & 0.015 & 1.7 \\
only $\ell$ & $-$0.018 & 0.74 & 0.014 & 1.3 \\
only $b$ & $+$0.017 & 0.30 & 0.015 & 1.1 \\
only $G$ & $+$0.034 & 0.78 & 0.013 & 2.6 \\
only $G_{BP}$ & $-$0.020 & 0.72 & 0.014 & 1.4 \\
only $G_{RP}$ & $+$0.022 & 0.22 & 0.013 & 1.7 \\
$G_{BP},\,G_{RP}$ & $-$0.029 & 0.51 & 0.016 & 1.9 \\
$\ell,\,b,\,G_{BP},\,G_{RP}$ & $+$0.056 & 0.76 & 0.014 & 4.1 \\
\end{longtable}
\endgroup

\begingroup\setlength{\tabcolsep}{4pt}
\begin{longtable}{l rrrr}
\caption{TARP deviation for 1000 sources from the test set. Calculated on class (c), i.e., sources with measured $\ell,\,b,\,\varpi,\,\mu_{\ell*},\,\mu_b,\,G,\,G_{BP},\,G_{RP}$ only. See the caption of Table~\ref{tab:tarp:test:a} for a full description of the columns.}\label{tab:tarp:test:c}\\
\toprule
Cond.\ Pattern & $\max |\Delta|$ & $c$ & $\sigma_{\rm TARP}$ & $\frac{|\Delta|}{\sigma_{\rm TARP}}$ \\
\midrule
\endfirsthead
\caption[]{TARP deviation for the test set, class (c), 1000 sources with measured $\ell,\,b,\,\varpi,\,\mu_{\ell*},\,\mu_b,\,G,\,G_{BP},\,G_{RP}$ only (continued)}\\
\toprule
Cond.\ Pattern & $\max |\Delta|$ & $c$ & $\sigma_{\rm TARP}$ & $\frac{|\Delta|}{\sigma_{\rm TARP}}$ \\
\midrule
\endhead
\midrule
\multicolumn{5}{r}{continued}\\
\endfoot
\bottomrule
\endlastfoot
uncond & $-$0.018 & 0.83 & 0.012 & 1.5 \\
$\ell,\,b$ & $+$0.022 & 0.16 & 0.012 & 1.9 \\
only $\varpi$ & $-$0.017 & 0.07 & 0.008 & 2.0 \\
$\ell,\,b,\,\varpi$ & $+$0.045 & 0.51 & 0.016 & 2.8 \\
$\ell,\,b,\,G$ & $+$0.019 & 0.10 & 0.010 & 1.9 \\
$\ell,\,b,\,\varpi,\,\mu_{\ell*},\,\mu_b$ & $+$0.048 & 0.61 & 0.015 & 3.1 \\
$G,\,G_{BP},\,G_{RP}$ & $-$0.032 & 0.69 & 0.015 & 2.2 \\
$\ell,\,b,\,G,\,G_{BP},\,G_{RP}$ & $+$0.017 & 0.35 & 0.015 & 1.1 \\
only $\ell$ & $-$0.031 & 0.54 & 0.016 & 2.0 \\
only $b$ & $+$0.021 & 0.18 & 0.012 & 1.7 \\
only $\mu_{\ell*}$ & $-$0.015 & 0.05 & 0.007 & 2.1 \\
only $\mu_b$ & $-$0.028 & 0.84 & 0.012 & 2.4 \\
only $G$ & $-$0.014 & 0.58 & 0.016 & 0.9 \\
only $G_{BP}$ & $-$0.022 & 0.81 & 0.013 & 1.8 \\
only $G_{RP}$ & $+$0.028 & 0.49 & 0.016 & 1.8 \\
$\mu_{\ell*},\,\mu_b$ & $-$0.018 & 0.27 & 0.014 & 1.3 \\
$G_{BP},\,G_{RP}$ & $-$0.018 & 0.10 & 0.010 & 1.9 \\
$\ell,\,b,\,G_{BP},\,G_{RP}$ & $+$0.039 & 0.42 & 0.016 & 2.5 \\
$\varpi,\,G$ & $+$0.030 & 0.18 & 0.012 & 2.4 \\
$\varpi,\,G_{BP},\,G_{RP}$ & $+$0.018 & 0.41 & 0.016 & 1.2 \\
$\varpi,\,G,\,G_{BP},\,G_{RP}$ & $+$0.029 & 0.80 & 0.013 & 2.3 \\
$\ell,\,b,\,\varpi,\,G$ & $-$0.032 & 0.37 & 0.015 & 2.1 \\
$\ell,\,b,\,\varpi,\,G_{BP},\,G_{RP}$ & $-$0.026 & 0.76 & 0.014 & 1.9 \\
$\ell,\,b,\,\varpi,\,G,\,G_{BP},\,G_{RP}$ & $-$0.039 & 0.36 & 0.015 & 2.6 \\
$\ell,\,b,\,\mu_{\ell*},\,\mu_b$ & $-$0.020 & 0.15 & 0.011 & 1.8 \\
$\mu_{\ell*},\,\mu_b,\,G,\,G_{BP},\,G_{RP}$ & $-$0.034 & 0.81 & 0.013 & 2.7 \\
$\varpi,\,\mu_{\ell*},\,\mu_b,\,G,\,G_{BP},\,G_{RP}$ & $-$0.033 & 0.39 & 0.015 & 2.1 \\
$\varpi,\,\mu_{\ell*},\,\mu_b$ & $+$0.021 & 0.14 & 0.011 & 1.9 \\
$\ell,\,b,\,\mu_{\ell*},\,\mu_b,\,G,\,G_{BP},\,G_{RP}$ & $+$0.027 & 0.61 & 0.015 & 1.8 \\
\end{longtable}
\endgroup

\begingroup\setlength{\tabcolsep}{4pt}
\begin{longtable}{l rrrr}
\caption{TARP deviation for 1000 sources from the test set. Calculated on class (d), i.e., sources with measured $\ell,\,b,\,\varpi,\,\mu_{\ell*},\,\mu_b,\,G,\,G_{BP},\,G_{RP},\,v_r$ only. See the caption of Table~\ref{tab:tarp:test:a} for a full description of the columns.}\label{tab:tarp:test:d}\\
\toprule
Cond.\ Pattern & $\max |\Delta|$ & $c$ & $\sigma_{\rm TARP}$ & $\frac{|\Delta|}{\sigma_{\rm TARP}}$ \\
\midrule
\endfirsthead
\caption[]{TARP deviation for the test set, class (d), 1000 sources with measured $\ell,\,b,\,\varpi,\,\mu_{\ell*},\,\mu_b,\,G,\,G_{BP},\,G_{RP},\,v_r$ only (continued)}\\
\toprule
Cond.\ Pattern & $\max |\Delta|$ & $c$ & $\sigma_{\rm TARP}$ & $\frac{|\Delta|}{\sigma_{\rm TARP}}$ \\
\midrule
\endhead
\midrule
\multicolumn{5}{r}{continued}\\
\endfoot
\bottomrule
\endlastfoot
uncond & $-$0.016 & 0.21 & 0.013 & 1.2 \\
$\ell,\,b$ & $-$0.042 & 0.30 & 0.014 & 2.9 \\
only $\varpi$ & $+$0.016 & 0.91 & 0.009 & 1.7 \\
$\ell,\,b,\,\varpi$ & $-$0.034 & 0.50 & 0.016 & 2.1 \\
$\ell,\,b,\,G$ & $+$0.022 & 0.92 & 0.009 & 2.5 \\
$\ell,\,b,\,\varpi,\,\mu_{\ell*},\,\mu_b$ & $-$0.043 & 0.21 & 0.013 & 3.3 \\
$G,\,G_{BP},\,G_{RP}$ & $+$0.029 & 0.29 & 0.014 & 2.1 \\
$\ell,\,b,\,G,\,G_{BP},\,G_{RP}$ & $+$0.047 & 0.49 & 0.016 & 3.0 \\
$\varpi,\,\mu_{\ell*},\,\mu_b,\,G,\,G_{BP},\,G_{RP},\,v_r$ & $-$0.029 & 0.34 & 0.015 & 2.0 \\
$\mu_{\ell*},\,\mu_b,\,G,\,G_{BP},\,G_{RP},\,v_r$ & $+$0.030 & 0.75 & 0.014 & 2.2 \\
$\ell,\,b,\,\varpi,\,G,\,G_{BP},\,G_{RP},\,v_r$ & $-$0.024 & 0.74 & 0.014 & 1.8 \\
$\ell,\,b,\,\varpi,\,\mu_{\ell*},\,\mu_b,\,G,\,v_r$ & $+$0.081 & 0.78 & 0.013 & 6.1 \\
$\ell,\,b,\,\varpi,\,\mu_{\ell*},\,\mu_b,\,v_r$ & $-$0.013 & 0.21 & 0.013 & 1.0 \\
drop $\ell$ & $-$0.035 & 0.16 & 0.012 & 3.0 \\
drop $b$ & $-$0.029 & 0.46 & 0.016 & 1.8 \\
drop $\varpi$ & $+$0.043 & 0.87 & 0.011 & 4.0 \\
drop $\mu_{\ell*}$ & $-$0.017 & 0.43 & 0.016 & 1.1 \\
drop $\mu_b$ & $-$0.041 & 0.34 & 0.015 & 2.8 \\
drop $G$ & $+$0.165 & 0.75 & 0.014 & 12.1 \\
drop $G_{BP}$ & $-$0.150 & 0.28 & 0.014 & 10.6 \\
drop $G_{RP}$ & $-$0.185 & 0.33 & 0.015 & 12.5 \\
drop $v_r$ & $+$0.038 & 0.78 & 0.013 & 2.9 \\
only $\ell$ & $+$0.044 & 0.44 & 0.016 & 2.8 \\
only $b$ & $+$0.022 & 0.59 & 0.016 & 1.4 \\
only $\mu_{\ell*}$ & $-$0.020 & 0.47 & 0.016 & 1.2 \\
only $\mu_b$ & $+$0.026 & 0.51 & 0.016 & 1.6 \\
only $G$ & $+$0.033 & 0.78 & 0.013 & 2.5 \\
only $G_{BP}$ & $-$0.022 & 0.63 & 0.015 & 1.4 \\
only $G_{RP}$ & $-$0.044 & 0.53 & 0.016 & 2.8 \\
only $v_r$ & $+$0.035 & 0.38 & 0.015 & 2.3 \\
$\mu_{\ell*},\,\mu_b$ & $+$0.016 & 0.65 & 0.015 & 1.1 \\
$G_{BP},\,G_{RP}$ & $-$0.023 & 0.78 & 0.013 & 1.8 \\
$\ell,\,b,\,G_{BP},\,G_{RP}$ & $-$0.026 & 0.45 & 0.016 & 1.6 \\
$\varpi,\,G$ & $-$0.018 & 0.76 & 0.013 & 1.3 \\
$\varpi,\,G_{BP},\,G_{RP}$ & $+$0.026 & 0.58 & 0.016 & 1.7 \\
$\varpi,\,G,\,G_{BP},\,G_{RP}$ & $-$0.013 & 0.53 & 0.016 & 0.8 \\
$\ell,\,b,\,\varpi,\,G$ & $+$0.014 & 0.76 & 0.014 & 1.0 \\
$\ell,\,b,\,\varpi,\,G_{BP},\,G_{RP}$ & $-$0.027 & 0.36 & 0.015 & 1.8 \\
$\ell,\,b,\,\varpi,\,G,\,G_{BP},\,G_{RP}$ & $-$0.035 & 0.50 & 0.016 & 2.2 \\
$\mu_{\ell*},\,\mu_b,\,v_r$ & $-$0.023 & 0.41 & 0.016 & 1.5 \\
$\ell,\,b,\,\mu_{\ell*},\,\mu_b,\,v_r$ & $+$0.028 & 0.29 & 0.014 & 2.0 \\
$\ell,\,b,\,\mu_{\ell*},\,\mu_b$ & $-$0.018 & 0.35 & 0.015 & 1.2 \\
$\mu_{\ell*},\,\mu_b,\,G,\,G_{BP},\,G_{RP}$ & $+$0.021 & 0.74 & 0.014 & 1.5 \\
$\varpi,\,\mu_{\ell*},\,\mu_b,\,G,\,G_{BP},\,G_{RP}$ & $-$0.032 & 0.53 & 0.016 & 2.0 \\
$\varpi,\,\mu_{\ell*},\,\mu_b$ & $-$0.022 & 0.75 & 0.014 & 1.6 \\
$\ell,\,b,\,\mu_{\ell*},\,\mu_b,\,G,\,G_{BP},\,G_{RP}$ & $+$0.037 & 0.73 & 0.014 & 2.6 \\
$\ell,\,b,\,v_r$ & $-$0.019 & 0.88 & 0.010 & 1.8 \\
\end{longtable}
\endgroup

\begingroup\setlength{\tabcolsep}{4pt}
\begin{longtable}{l rrrr}
\caption{TARP deviation for 1000 sources from $\omega$~Cen members outside the training and validation sets. Calculated on class (c), i.e., sources with measured $\ell,\,b,\,\varpi,\,\mu_{\ell*},\,\mu_b,\,G,\,G_{BP},\,G_{RP}$ only. See the caption of Table~\ref{tab:tarp:test:a} for a full description.}\label{tab:tarp:omega_cen_not_A:c}\\
\toprule
Cond.\ Pattern & $\max |\Delta|$ & $c$ & $\sigma_{\rm TARP}$ & $\frac{|\Delta|}{\sigma_{\rm TARP}}$ \\
\midrule
\endfirsthead
\caption[]{TARP deviation for $\omega$~Cen members outside the training and validation sets, class (c), 1000 sources with measured $\ell,\,b,\,\varpi,\,\mu_{\ell*},\,\mu_b,\,G,\,G_{BP},\,G_{RP}$ only (continued)}\\
\toprule
Cond.\ Pattern & $\max |\Delta|$ & $c$ & $\sigma_{\rm TARP}$ & $\frac{|\Delta|}{\sigma_{\rm TARP}}$ \\
\midrule
\endhead
\midrule
\multicolumn{5}{r}{continued}\\
\endfoot
\bottomrule
\endlastfoot
uncond & $-$0.077 & 0.59 & 0.016 & 5.0 \\
$\ell,\,b$ & $-$0.119 & 0.67 & 0.015 & 8.0 \\
only $\varpi$ & $-$0.091 & 0.71 & 0.014 & 6.4 \\
$\ell,\,b,\,\varpi$ & $-$0.180 & 0.71 & 0.014 & 12.6 \\
$\ell,\,b,\,G$ & $-$0.216 & 0.82 & 0.012 & 17.8 \\
$\ell,\,b,\,\varpi,\,\mu_{\ell*},\,\mu_b$ & $-$0.088 & 0.90 & 0.010 & 9.0 \\
$G,\,G_{BP},\,G_{RP}$ & $-$0.036 & 0.16 & 0.012 & 3.1 \\
$\ell,\,b,\,G,\,G_{BP},\,G_{RP}$ & $-$0.261 & 0.85 & 0.011 & 23.1 \\
only $\ell$ & $-$0.177 & 0.69 & 0.015 & 12.1 \\
only $b$ & $-$0.062 & 0.60 & 0.016 & 4.0 \\
only $\mu_{\ell*}$ & $-$0.122 & 0.70 & 0.014 & 8.5 \\
only $\mu_b$ & $+$0.095 & 0.80 & 0.013 & 7.5 \\
only $G$ & $-$0.059 & 0.66 & 0.015 & 3.9 \\
only $G_{BP}$ & $-$0.028 & 0.20 & 0.013 & 2.2 \\
only $G_{RP}$ & $-$0.058 & 0.67 & 0.015 & 3.9 \\
$\mu_{\ell*},\,\mu_b$ & $+$0.068 & 0.25 & 0.014 & 5.0 \\
$G_{BP},\,G_{RP}$ & $-$0.054 & 0.29 & 0.014 & 3.7 \\
$\ell,\,b,\,G_{BP},\,G_{RP}$ & $-$0.300 & 0.88 & 0.010 & 29.2 \\
$\varpi,\,G$ & $-$0.103 & 0.69 & 0.015 & 7.0 \\
$\varpi,\,G_{BP},\,G_{RP}$ & $-$0.051 & 0.24 & 0.013 & 3.8 \\
$\varpi,\,G,\,G_{BP},\,G_{RP}$ & $-$0.053 & 0.25 & 0.014 & 3.8 \\
$\ell,\,b,\,\varpi,\,G$ & $-$0.273 & 0.85 & 0.011 & 24.2 \\
$\ell,\,b,\,\varpi,\,G_{BP},\,G_{RP}$ & $-$0.290 & 0.86 & 0.011 & 26.1 \\
$\ell,\,b,\,\varpi,\,G,\,G_{BP},\,G_{RP}$ & $-$0.288 & 0.87 & 0.011 & 26.7 \\
$\ell,\,b,\,\mu_{\ell*},\,\mu_b$ & $+$0.093 & 0.76 & 0.014 & 6.8 \\
$\mu_{\ell*},\,\mu_b,\,G,\,G_{BP},\,G_{RP}$ & $+$0.170 & 0.22 & 0.013 & 12.9 \\
$\varpi,\,\mu_{\ell*},\,\mu_b,\,G,\,G_{BP},\,G_{RP}$ & $+$0.192 & 0.23 & 0.013 & 14.5 \\
$\varpi,\,\mu_{\ell*},\,\mu_b$ & $+$0.111 & 0.23 & 0.013 & 8.3 \\
$\ell,\,b,\,\mu_{\ell*},\,\mu_b,\,G,\,G_{BP},\,G_{RP}$ & $+$0.057 & 0.17 & 0.012 & 4.7 \\
\end{longtable}
\endgroup

\begingroup\setlength{\tabcolsep}{4pt}
\begin{longtable}{l rrrr}
\caption{TARP deviation for 781 sources from $\omega$~Cen members outside the training and validation sets. Calculated on class (d), i.e., sources with measured $\ell,\,b,\,\varpi,\,\mu_{\ell*},\,\mu_b,\,G,\,G_{BP},\,G_{RP},\,v_r$ only. See the caption of Table~\ref{tab:tarp:test:a} for a full description of the columns.}\label{tab:tarp:omega_cen_not_A:d}\\
\toprule
Cond.\ Pattern & $\max |\Delta|$ & $c$ & $\sigma_{\rm TARP}$ & $\frac{|\Delta|}{\sigma_{\rm TARP}}$ \\
\midrule
\endfirsthead
\caption[]{TARP deviation for $\omega$~Cen members outside the training and validation sets, class (d), 781 sources with measured $\ell,\,b,\,\varpi,\,\mu_{\ell*},\,\mu_b,\,G,\,G_{BP},\,G_{RP},\,v_r$ only (continued)}\\
\toprule
Cond.\ Pattern & $\max |\Delta|$ & $c$ & $\sigma_{\rm TARP}$ & $\frac{|\Delta|}{\sigma_{\rm TARP}}$ \\
\midrule
\endhead
\midrule
\multicolumn{5}{r}{continued}\\
\endfoot
\bottomrule
\endlastfoot
uncond & $-$0.279 & 0.66 & 0.017 & 16.5 \\
$\ell,\,b$ & $-$0.339 & 0.65 & 0.017 & 19.9 \\
only $\varpi$ & $-$0.477 & 0.80 & 0.014 & 33.1 \\
$\ell,\,b,\,\varpi$ & $-$0.529 & 0.85 & 0.013 & 41.5 \\
$\ell,\,b,\,G$ & $-$0.399 & 0.80 & 0.014 & 27.9 \\
$\ell,\,b,\,\varpi,\,\mu_{\ell*},\,\mu_b$ & $-$0.509 & 0.85 & 0.013 & 39.9 \\
$G,\,G_{BP},\,G_{RP}$ & $-$0.198 & 0.77 & 0.015 & 13.1 \\
$\ell,\,b,\,G,\,G_{BP},\,G_{RP}$ & $-$0.437 & 0.80 & 0.014 & 30.6 \\
$\varpi,\,\mu_{\ell*},\,\mu_b,\,G,\,G_{BP},\,G_{RP},\,v_r$ & $+$0.184 & 0.59 & 0.018 & 10.4 \\
$\mu_{\ell*},\,\mu_b,\,G,\,G_{BP},\,G_{RP},\,v_r$ & $+$0.178 & 0.57 & 0.018 & 10.1 \\
$\ell,\,b,\,\varpi,\,G,\,G_{BP},\,G_{RP},\,v_r$ & $-$0.140 & 0.68 & 0.017 & 8.4 \\
$\ell,\,b,\,\varpi,\,\mu_{\ell*},\,\mu_b,\,G,\,v_r$ & $-$0.298 & 0.82 & 0.014 & 21.8 \\
$\ell,\,b,\,\varpi,\,\mu_{\ell*},\,\mu_b,\,v_r$ & $-$0.127 & 0.38 & 0.017 & 7.3 \\
drop $\ell$ & $+$0.176 & 0.63 & 0.017 & 10.2 \\
drop $b$ & $+$0.343 & 0.64 & 0.017 & 19.9 \\
drop $\varpi$ & $+$0.150 & 0.75 & 0.016 & 9.6 \\
drop $\mu_{\ell*}$ & $+$0.327 & 0.63 & 0.017 & 19.0 \\
drop $\mu_b$ & $-$0.353 & 0.80 & 0.014 & 24.5 \\
drop $G$ & $-$0.240 & 0.44 & 0.018 & 13.5 \\
drop $G_{BP}$ & $-$0.784 & 0.87 & 0.012 & 65.3 \\
drop $G_{RP}$ & $+$0.457 & 0.34 & 0.017 & 27.0 \\
drop $v_r$ & $-$0.599 & 0.93 & 0.009 & 65.7 \\
only $\ell$ & $-$0.394 & 0.70 & 0.016 & 24.1 \\
only $b$ & $-$0.188 & 0.57 & 0.018 & 10.6 \\
only $\mu_{\ell*}$ & $-$0.381 & 0.73 & 0.016 & 23.9 \\
only $\mu_b$ & $-$0.146 & 0.70 & 0.016 & 8.9 \\
only $G$ & $-$0.249 & 0.71 & 0.016 & 15.3 \\
only $G_{BP}$ & $-$0.220 & 0.66 & 0.017 & 13.0 \\
only $G_{RP}$ & $-$0.256 & 0.58 & 0.018 & 14.5 \\
only $v_r$ & $-$0.265 & 0.48 & 0.018 & 14.8 \\
$\mu_{\ell*},\,\mu_b$ & $-$0.293 & 0.71 & 0.016 & 18.1 \\
$G_{BP},\,G_{RP}$ & $-$0.226 & 0.78 & 0.015 & 15.1 \\
$\ell,\,b,\,G_{BP},\,G_{RP}$ & $-$0.449 & 0.83 & 0.013 & 33.5 \\
$\varpi,\,G$ & $-$0.380 & 0.90 & 0.011 & 34.7 \\
$\varpi,\,G_{BP},\,G_{RP}$ & $-$0.335 & 0.87 & 0.012 & 27.9 \\
$\varpi,\,G,\,G_{BP},\,G_{RP}$ & $-$0.311 & 0.88 & 0.012 & 26.3 \\
$\ell,\,b,\,\varpi,\,G$ & $-$0.573 & 0.94 & 0.009 & 65.1 \\
$\ell,\,b,\,\varpi,\,G_{BP},\,G_{RP}$ & $-$0.556 & 0.93 & 0.009 & 61.1 \\
$\ell,\,b,\,\varpi,\,G,\,G_{BP},\,G_{RP}$ & $-$0.585 & 0.93 & 0.009 & 62.2 \\
$\mu_{\ell*},\,\mu_b,\,v_r$ & $-$0.178 & 0.48 & 0.018 & 10.0 \\
$\ell,\,b,\,\mu_{\ell*},\,\mu_b,\,v_r$ & $-$0.173 & 0.42 & 0.018 & 9.8 \\
$\ell,\,b,\,\mu_{\ell*},\,\mu_b$ & $-$0.429 & 0.78 & 0.015 & 29.0 \\
$\mu_{\ell*},\,\mu_b,\,G,\,G_{BP},\,G_{RP}$ & $-$0.322 & 0.82 & 0.014 & 23.2 \\
$\varpi,\,\mu_{\ell*},\,\mu_b,\,G,\,G_{BP},\,G_{RP}$ & $-$0.338 & 0.87 & 0.012 & 28.1 \\
$\varpi,\,\mu_{\ell*},\,\mu_b$ & $-$0.445 & 0.76 & 0.015 & 29.0 \\
$\ell,\,b,\,\mu_{\ell*},\,\mu_b,\,G,\,G_{BP},\,G_{RP}$ & $-$0.634 & 0.94 & 0.009 & 72.1 \\
$\ell,\,b,\,v_r$ & $-$0.185 & 0.45 & 0.018 & 10.4 \\
\end{longtable}
\endgroup

\begingroup\setlength{\tabcolsep}{4pt}
\begin{longtable}{l rrr}
\caption{Skill ($1 -$ RMSE / RMSE$_{\rm baseline}$) for the test set, class (a), 1000 sources with measured $\ell,\,b,\,G$ only. Each column beyond Cond.\ Pattern corresponds to a single parameter, and each row to a conditioning pattern. The RMSE and the baseline are computed on the same sources and model samples as the TARP curve for this class. When a parameter is included in the conditioning pattern, no RMSE is calculated and its entry is marked with ``--''.}\label{tab:skill:test:a}\\
\toprule
Cond.\ Pattern & $\ell$ & $b$ & $G$ \\
\midrule
\endfirsthead
\caption[]{Skill ($1 -$ RMSE / RMSE$_{\rm baseline}$) for the test set, class (a), 1000 sources with measured $\ell,\,b,\,G$ only (continued)}\\
\toprule
Cond.\ Pattern & $\ell$ & $b$ & $G$ \\
\midrule
\endhead
\midrule
\multicolumn{4}{r}{continued}\\
\endfoot
\bottomrule
\endlastfoot
uncond & 0.00 & $-$0.01 & 0.00 \\
$\ell,\,b$ & -- & -- & $+$0.15 \\
only $\ell$ & -- & $+$0.10 & $+$0.05 \\
only $b$ & $+$0.04 & -- & $+$0.04 \\
only $G$ & $+$0.02 & $+$0.01 & -- \\
\end{longtable}
\endgroup

\begingroup\setlength{\tabcolsep}{4pt}
\begin{longtable}{l rrrrr}
\caption{Skill ($1 -$ RMSE / RMSE$_{\rm baseline}$) for the test set, class (b), 1000 sources with measured $\ell,\,b,\,G,\,G_{BP},\,G_{RP}$ only. See the caption of Table~\ref{tab:skill:test:a} for a full description.}\label{tab:skill:test:b}\\
\toprule
Cond.\ Pattern & $\ell$ & $b$ & $G$ & $G_{BP}$ & $G_{RP}$ \\
\midrule
\endfirsthead
\caption[]{Skill ($1 -$ RMSE / RMSE$_{\rm baseline}$) for the test set, class (b), 1000 sources with measured $\ell,\,b,\,G,\,G_{BP},\,G_{RP}$ only (continued)}\\
\toprule
Cond.\ Pattern & $\ell$ & $b$ & $G$ & $G_{BP}$ & $G_{RP}$ \\
\midrule
\endhead
\midrule
\multicolumn{6}{r}{continued}\\
\endfoot
\bottomrule
\endlastfoot
uncond & 0.00 & 0.00 & $-$0.01 & $-$0.01 & $-$0.01 \\
$\ell,\,b$ & -- & -- & $+$0.10 & $+$0.07 & $+$0.08 \\
$\ell,\,b,\,G$ & -- & -- & -- & $+$0.30 & $+$0.40 \\
$G,\,G_{BP},\,G_{RP}$ & 0.00 & 0.00 & -- & -- & -- \\
only $\ell$ & -- & $+$0.01 & $+$0.05 & $+$0.01 & $+$0.02 \\
only $b$ & $+$0.01 & -- & $+$0.01 & $+$0.02 & $+$0.01 \\
only $G$ & 0.00 & 0.00 & -- & $+$0.33 & $+$0.39 \\
only $G_{BP}$ & 0.00 & 0.00 & $+$0.33 & -- & $+$0.32 \\
only $G_{RP}$ & 0.00 & $-$0.01 & $+$0.26 & $+$0.30 & -- \\
$G_{BP},\,G_{RP}$ & 0.00 & 0.00 & $+$0.38 & -- & -- \\
$\ell,\,b,\,G_{BP},\,G_{RP}$ & -- & -- & $+$0.63 & -- & -- \\
\end{longtable}
\endgroup

\begingroup\setlength{\tabcolsep}{4pt}
\begin{longtable}{l rrrrrrrr}
\caption{Skill ($1 -$ RMSE / RMSE$_{\rm baseline}$) for the test set, class (c), 1000 sources with measured $\ell,\,b,\,\varpi,\,\mu_{\ell*},\,\mu_b,\,G,\,G_{BP},\,G_{RP}$ only. See the caption of Table~\ref{tab:skill:test:a} for a full description.}\label{tab:skill:test:c}\\
\toprule
Cond.\ Pattern & $\ell$ & $b$ & $\varpi$ & $\mu_{\ell*}$ & $\mu_b$ & $G$ & $G_{BP}$ & $G_{RP}$ \\
\midrule
\endfirsthead
\caption[]{Skill ($1 -$ RMSE / RMSE$_{\rm baseline}$) for the test set, class (c), 1000 sources with measured $\ell,\,b,\,\varpi,\,\mu_{\ell*},\,\mu_b,\,G,\,G_{BP},\,G_{RP}$ only (continued)}\\
\toprule
Cond.\ Pattern & $\ell$ & $b$ & $\varpi$ & $\mu_{\ell*}$ & $\mu_b$ & $G$ & $G_{BP}$ & $G_{RP}$ \\
\midrule
\endhead
\midrule
\multicolumn{9}{r}{continued}\\
\endfoot
\bottomrule
\endlastfoot
uncond & 0.00 & 0.00 & 0.00 & $-$0.01 & 0.00 & $-$0.01 & $-$0.01 & $-$0.01 \\
$\ell,\,b$ & -- & -- & $+$0.01 & $+$0.11 & $+$0.02 & 0.00 & $+$0.01 & $+$0.02 \\
only $\varpi$ & 0.00 & 0.00 & -- & $-$0.01 & 0.00 & $+$0.07 & $+$0.07 & $+$0.06 \\
$\ell,\,b,\,\varpi$ & -- & -- & -- & $+$0.13 & $-$0.03 & $+$0.11 & $+$0.11 & $+$0.10 \\
$\ell,\,b,\,G$ & -- & -- & 0.00 & $+$0.12 & $+$0.07 & -- & $+$0.74 & $+$0.80 \\
$\ell,\,b,\,\varpi,\,\mu_{\ell*},\,\mu_b$ & -- & -- & -- & -- & -- & $+$0.15 & $+$0.16 & $+$0.14 \\
$G,\,G_{BP},\,G_{RP}$ & 0.00 & 0.00 & $+$0.02 & 0.00 & $+$0.01 & -- & -- & -- \\
$\ell,\,b,\,G,\,G_{BP},\,G_{RP}$ & -- & -- & $+$0.04 & $+$0.16 & $+$0.06 & -- & -- & -- \\
only $\ell$ & -- & $+$0.01 & $+$0.01 & $+$0.11 & 0.00 & 0.00 & 0.00 & $+$0.01 \\
only $b$ & 0.00 & -- & 0.00 & 0.00 & 0.00 & 0.00 & $+$0.01 & 0.00 \\
only $\mu_{\ell*}$ & $+$0.12 & 0.00 & $+$0.04 & -- & $-$0.01 & 0.00 & 0.00 & 0.00 \\
only $\mu_b$ & 0.00 & $-$0.01 & 0.00 & 0.00 & -- & 0.00 & 0.00 & 0.00 \\
only $G$ & 0.00 & 0.00 & 0.00 & 0.00 & 0.00 & -- & $+$0.70 & $+$0.77 \\
only $G_{BP}$ & 0.00 & 0.00 & 0.00 & 0.00 & 0.00 & $+$0.72 & -- & $+$0.59 \\
only $G_{RP}$ & 0.00 & 0.00 & 0.00 & $-$0.01 & 0.00 & $+$0.79 & $+$0.57 & -- \\
$\mu_{\ell*},\,\mu_b$ & $+$0.13 & 0.00 & $+$0.02 & -- & -- & $+$0.01 & $+$0.01 & $+$0.01 \\
$G_{BP},\,G_{RP}$ & $+$0.01 & 0.00 & $+$0.02 & 0.00 & 0.00 & $+$0.86 & -- & -- \\
$\ell,\,b,\,G_{BP},\,G_{RP}$ & -- & -- & $+$0.10 & $+$0.13 & $+$0.07 & $+$0.88 & -- & -- \\
$\varpi,\,G$ & 0.00 & 0.00 & -- & $-$0.01 & $-$0.01 & -- & $+$0.67 & $+$0.74 \\
$\varpi,\,G_{BP},\,G_{RP}$ & $+$0.01 & $+$0.01 & -- & $+$0.01 & $-$0.02 & $+$0.87 & -- & -- \\
$\varpi,\,G,\,G_{BP},\,G_{RP}$ & 0.00 & 0.00 & -- & $+$0.01 & $+$0.01 & -- & -- & -- \\
$\ell,\,b,\,\varpi,\,G$ & -- & -- & -- & $+$0.13 & $+$0.07 & -- & $+$0.78 & $+$0.84 \\
$\ell,\,b,\,\varpi,\,G_{BP},\,G_{RP}$ & -- & -- & -- & $+$0.12 & $+$0.09 & $+$0.89 & -- & -- \\
$\ell,\,b,\,\varpi,\,G,\,G_{BP},\,G_{RP}$ & -- & -- & -- & $+$0.16 & $-$0.02 & -- & -- & -- \\
$\ell,\,b,\,\mu_{\ell*},\,\mu_b$ & -- & -- & $+$0.04 & -- & -- & $+$0.02 & $+$0.02 & $+$0.03 \\
$\mu_{\ell*},\,\mu_b,\,G,\,G_{BP},\,G_{RP}$ & $+$0.12 & $+$0.03 & $+$0.19 & -- & -- & -- & -- & -- \\
$\varpi,\,\mu_{\ell*},\,\mu_b,\,G,\,G_{BP},\,G_{RP}$ & $+$0.16 & $+$0.02 & -- & -- & -- & -- & -- & -- \\
$\varpi,\,\mu_{\ell*},\,\mu_b$ & $+$0.13 & $+$0.01 & -- & -- & -- & $+$0.10 & $+$0.09 & $+$0.09 \\
$\ell,\,b,\,\mu_{\ell*},\,\mu_b,\,G,\,G_{BP},\,G_{RP}$ & -- & -- & $+$0.06 & -- & -- & -- & -- & -- \\
\end{longtable}
\endgroup

\begingroup\setlength{\tabcolsep}{4pt}
\begin{longtable}{l rrrrrrrrr}
\caption{Skill ($1 -$ RMSE / RMSE$_{\rm baseline}$) for the test set, class (d), 1000 sources with measured $\ell,\,b,\,\varpi,\,\mu_{\ell*},\,\mu_b,\,G,\,G_{BP},\,G_{RP},\,v_r$ only. See the caption of Table~\ref{tab:skill:test:a} for a full description. Note that the $v_r$ skill values in Figure \ref{fig:impute_skill_and_confidence} and those found in this table are calculated using different sources.}\label{tab:skill:test:d}\\
\toprule
Cond.\ Pattern & $\ell$ & $b$ & $\varpi$ & $\mu_{\ell*}$ & $\mu_b$ & $G$ & $G_{BP}$ & $G_{RP}$ & $v_r$ \\
\midrule
\endfirsthead
\caption[]{Skill ($1 -$ RMSE / RMSE$_{\rm baseline}$) for the test set, class (d), 1000 sources with measured $\ell,\,b,\,\varpi,\,\mu_{\ell*},\,\mu_b,\,G,\,G_{BP},\,G_{RP},\,v_r$ only (continued)}\\
\toprule
Cond.\ Pattern & $\ell$ & $b$ & $\varpi$ & $\mu_{\ell*}$ & $\mu_b$ & $G$ & $G_{BP}$ & $G_{RP}$ & $v_r$ \\
\midrule
\endhead
\midrule
\multicolumn{10}{r}{continued}\\
\endfoot
\bottomrule
\endlastfoot
uncond & $-$0.01 & 0.00 & $-$0.01 & 0.00 & 0.00 & 0.00 & 0.00 & 0.00 & 0.00 \\
$\ell,\,b$ & -- & -- & $+$0.07 & $+$0.10 & $+$0.05 & $+$0.03 & $+$0.10 & 0.00 & $+$0.12 \\
only $\varpi$ & $-$0.01 & 0.00 & -- & 0.00 & $+$0.04 & $+$0.07 & $+$0.14 & $+$0.04 & 0.00 \\
$\ell,\,b,\,\varpi$ & -- & -- & -- & $+$0.11 & $+$0.04 & $+$0.12 & $+$0.26 & $+$0.07 & $+$0.17 \\
$\ell,\,b,\,G$ & -- & -- & $+$0.09 & $+$0.08 & $+$0.03 & -- & $+$0.72 & $+$0.83 & $+$0.14 \\
$\ell,\,b,\,\varpi,\,\mu_{\ell*},\,\mu_b$ & -- & -- & -- & -- & -- & $+$0.07 & $+$0.19 & $+$0.03 & $+$0.23 \\
$G,\,G_{BP},\,G_{RP}$ & $-$0.01 & 0.00 & $+$0.09 & 0.00 & $+$0.01 & -- & -- & -- & $+$0.01 \\
$\ell,\,b,\,G,\,G_{BP},\,G_{RP}$ & -- & -- & $+$0.26 & $+$0.07 & $+$0.07 & -- & -- & -- & $+$0.14 \\
$\varpi,\,\mu_{\ell*},\,\mu_b,\,G,\,G_{BP},\,G_{RP},\,v_r$ & $+$0.30 & $+$0.09 & -- & -- & -- & -- & -- & -- & -- \\
$\mu_{\ell*},\,\mu_b,\,G,\,G_{BP},\,G_{RP},\,v_r$ & $+$0.25 & $+$0.07 & $+$0.41 & -- & -- & -- & -- & -- & -- \\
$\ell,\,b,\,\varpi,\,G,\,G_{BP},\,G_{RP},\,v_r$ & -- & -- & -- & $+$0.18 & $+$0.14 & -- & -- & -- & -- \\
$\ell,\,b,\,\varpi,\,\mu_{\ell*},\,\mu_b,\,G,\,v_r$ & -- & -- & -- & -- & -- & -- & $+$0.79 & $+$0.89 & -- \\
$\ell,\,b,\,\varpi,\,\mu_{\ell*},\,\mu_b,\,v_r$ & -- & -- & -- & -- & -- & $+$0.10 & $+$0.20 & $+$0.05 & -- \\
drop $\ell$ & $+$0.31 & -- & -- & -- & -- & -- & -- & -- & -- \\
drop $b$ & -- & $+$0.09 & -- & -- & -- & -- & -- & -- & -- \\
drop $\varpi$ & -- & -- & $+$0.59 & -- & -- & -- & -- & -- & -- \\
drop $\mu_{\ell*}$ & -- & -- & -- & $+$0.13 & -- & -- & -- & -- & -- \\
drop $\mu_b$ & -- & -- & -- & -- & $+$0.17 & -- & -- & -- & -- \\
drop $G$ & -- & -- & -- & -- & -- & $+$0.96 & -- & -- & -- \\
drop $G_{BP}$ & -- & -- & -- & -- & -- & -- & $+$0.91 & -- & -- \\
drop $G_{RP}$ & -- & -- & -- & -- & -- & -- & -- & $+$0.95 & -- \\
drop $v_r$ & -- & -- & -- & -- & -- & -- & -- & -- & $+$0.23 \\
only $\ell$ & -- & 0.00 & $+$0.01 & $+$0.07 & $-$0.01 & $+$0.01 & $+$0.03 & 0.00 & $+$0.11 \\
only $b$ & $-$0.01 & -- & $+$0.05 & 0.00 & 0.00 & $+$0.02 & $+$0.07 & $+$0.01 & 0.00 \\
only $\mu_{\ell*}$ & $+$0.16 & 0.00 & $+$0.21 & -- & $-$0.04 & $+$0.01 & $+$0.03 & 0.00 & 0.00 \\
only $\mu_b$ & $-$0.01 & 0.00 & $+$0.29 & 0.00 & -- & $+$0.03 & $+$0.05 & $+$0.02 & 0.00 \\
only $G$ & 0.00 & 0.00 & $+$0.04 & 0.00 & $-$0.01 & -- & $+$0.68 & $+$0.81 & 0.00 \\
only $G_{BP}$ & 0.00 & $-$0.01 & $+$0.05 & $-$0.01 & 0.00 & $+$0.74 & -- & $+$0.60 & 0.00 \\
only $G_{RP}$ & 0.00 & 0.00 & $+$0.03 & 0.00 & 0.00 & $+$0.79 & $+$0.49 & -- & 0.00 \\
only $v_r$ & $+$0.10 & 0.00 & $+$0.02 & $-$0.01 & 0.00 & $+$0.01 & $+$0.02 & 0.00 & -- \\
$\mu_{\ell*},\,\mu_b$ & $+$0.13 & $-$0.01 & $+$0.39 & -- & -- & $+$0.02 & $+$0.05 & $+$0.01 & $+$0.01 \\
$G_{BP},\,G_{RP}$ & $+$0.01 & 0.00 & $+$0.09 & 0.00 & $-$0.01 & $+$0.96 & -- & -- & 0.00 \\
$\ell,\,b,\,G_{BP},\,G_{RP}$ & -- & -- & $+$0.27 & $+$0.09 & $+$0.02 & $+$0.96 & -- & -- & $+$0.15 \\
$\varpi,\,G$ & $-$0.01 & $-$0.01 & -- & $+$0.01 & $-$0.04 & -- & $+$0.73 & $+$0.87 & $-$0.01 \\
$\varpi,\,G_{BP},\,G_{RP}$ & $+$0.03 & 0.00 & -- & $-$0.01 & $+$0.03 & $+$0.97 & -- & -- & 0.00 \\
$\varpi,\,G,\,G_{BP},\,G_{RP}$ & $+$0.01 & $-$0.01 & -- & $+$0.07 & $-$0.11 & -- & -- & -- & 0.00 \\
$\ell,\,b,\,\varpi,\,G$ & -- & -- & -- & $+$0.14 & $+$0.06 & -- & $+$0.81 & $+$0.90 & $+$0.17 \\
$\ell,\,b,\,\varpi,\,G_{BP},\,G_{RP}$ & -- & -- & -- & $+$0.10 & $+$0.11 & $+$0.97 & -- & -- & $+$0.14 \\
$\ell,\,b,\,\varpi,\,G,\,G_{BP},\,G_{RP}$ & -- & -- & -- & $+$0.11 & $+$0.12 & -- & -- & -- & $+$0.17 \\
$\mu_{\ell*},\,\mu_b,\,v_r$ & $+$0.26 & $+$0.04 & $+$0.35 & -- & -- & $+$0.05 & $+$0.08 & $+$0.02 & -- \\
$\ell,\,b,\,\mu_{\ell*},\,\mu_b,\,v_r$ & -- & -- & $+$0.24 & -- & -- & $+$0.04 & $+$0.13 & $+$0.01 & -- \\
$\ell,\,b,\,\mu_{\ell*},\,\mu_b$ & -- & -- & $+$0.45 & -- & -- & $+$0.07 & $+$0.15 & $+$0.03 & $+$0.16 \\
$\mu_{\ell*},\,\mu_b,\,G,\,G_{BP},\,G_{RP}$ & $+$0.15 & $-$0.01 & $+$0.34 & -- & -- & -- & -- & -- & 0.00 \\
$\varpi,\,\mu_{\ell*},\,\mu_b,\,G,\,G_{BP},\,G_{RP}$ & $+$0.13 & 0.00 & -- & -- & -- & -- & -- & -- & 0.00 \\
$\varpi,\,\mu_{\ell*},\,\mu_b$ & $+$0.14 & 0.00 & -- & -- & -- & $+$0.07 & $+$0.14 & $+$0.04 & 0.00 \\
$\ell,\,b,\,\mu_{\ell*},\,\mu_b,\,G,\,G_{BP},\,G_{RP}$ & -- & -- & $+$0.55 & -- & -- & -- & -- & -- & $+$0.18 \\
$\ell,\,b,\,v_r$ & -- & -- & $+$0.08 & $+$0.08 & $+$0.03 & $+$0.06 & $+$0.14 & $+$0.03 & -- \\
\end{longtable}
\endgroup

\begingroup\setlength{\tabcolsep}{4pt}
\begin{longtable}{l rrrrrrrr}
\caption{Skill ($1 -$ RMSE / RMSE$_{\rm baseline}$) for $\omega$~Cen members outside the training and validation sets, class (c), 1000 sources with measured $\ell,\,b,\,\varpi,\,\mu_{\ell*},\,\mu_b,\,G,\,G_{BP},\,G_{RP}$ only. See the caption of Table~\ref{tab:skill:test:a} for a full description. Very negative values imply the model is a strongly biased point estimator, however, note that $\omega$ Cen members have a very narrow distribution in position. When the model is not given explicit position information, one should expect the baseline RMSE will be much lower than even an ideal model (making the skill extremely negative). This is also true to a lesser extent for parallax and proper motion. Even when given position information, for example, the model is still predicting proper motions from a distribution that includes non-members removed by \citet{Soltis_21}.}\label{tab:skill:omega_cen_not_A:c}\\
\toprule
Cond.\ Pattern & $\ell$ & $b$ & $\varpi$ & $\mu_{\ell*}$ & $\mu_b$ & $G$ & $G_{BP}$ & $G_{RP}$ \\
\midrule
\endfirsthead
\caption[]{Skill ($1 -$ RMSE / RMSE$_{\rm baseline}$) for $\omega$~Cen members outside the training and validation sets, class (c), 1000 sources with measured $\ell,\,b,\,\varpi,\,\mu_{\ell*},\,\mu_b,\,G,\,G_{BP},\,G_{RP}$ only (continued)}\\
\toprule
Cond.\ Pattern & $\ell$ & $b$ & $\varpi$ & $\mu_{\ell*}$ & $\mu_b$ & $G$ & $G_{BP}$ & $G_{RP}$ \\
\midrule
\endhead
\midrule
\multicolumn{9}{r}{continued}\\
\endfoot
\bottomrule
\endlastfoot
uncond & $-$227.08 & $-$84.50 & $-$0.14 & $-$0.41 & $-$9.98 & $-$0.25 & $-$0.52 & $-$0.11 \\
$\ell,\,b$ & -- & -- & $-$0.24 & $-$4.95 & $-$9.55 & $-$0.22 & $-$0.41 & $-$0.15 \\
only $\varpi$ & $-$212.36 & $-$88.17 & -- & $-$0.35 & $-$10.28 & $-$0.13 & $-$0.38 & $-$0.03 \\
$\ell,\,b,\,\varpi$ & -- & -- & -- & $-$4.07 & $-$9.94 & $-$0.17 & $-$0.29 & $-$0.12 \\
$\ell,\,b,\,G$ & -- & -- & $-$0.29 & $-$5.09 & $-$9.23 & -- & $+$0.65 & $+$0.78 \\
$\ell,\,b,\,\varpi,\,\mu_{\ell*},\,\mu_b$ & -- & -- & -- & -- & -- & $-$0.27 & $-$0.44 & $-$0.20 \\
$G,\,G_{BP},\,G_{RP}$ & $-$168.75 & $-$102.64 & $-$0.08 & $-$0.44 & $-$9.49 & -- & -- & -- \\
$\ell,\,b,\,G,\,G_{BP},\,G_{RP}$ & -- & -- & $-$0.09 & $-$3.49 & $-$10.00 & -- & -- & -- \\
only $\ell$ & -- & $-$83.84 & $-$0.12 & $-$4.51 & $-$10.07 & $-$0.31 & $-$0.59 & $-$0.14 \\
only $b$ & $-$249.26 & -- & $-$0.28 & $-$0.38 & $-$9.86 & $-$0.21 & $-$0.41 & $-$0.12 \\
only $\mu_{\ell*}$ & $-$242.88 & $-$85.98 & $-$0.04 & -- & $-$9.89 & $-$0.31 & $-$0.61 & $-$0.14 \\
only $\mu_b$ & $-$228.25 & $-$86.56 & $-$0.64 & $-$0.59 & -- & $-$0.16 & $-$0.41 & $-$0.04 \\
only $G$ & $-$232.77 & $-$89.16 & $-$0.23 & $-$0.47 & $-$10.33 & -- & $+$0.56 & $+$0.71 \\
only $G_{BP}$ & $-$210.86 & $-$85.85 & $-$0.19 & $-$0.45 & $-$9.83 & $+$0.64 & -- & $+$0.48 \\
only $G_{RP}$ & $-$226.80 & $-$87.68 & $-$0.24 & $-$0.44 & $-$8.81 & $+$0.63 & $+$0.25 & -- \\
$\mu_{\ell*},\,\mu_b$ & $-$242.89 & $-$86.78 & $-$0.29 & -- & -- & $-$0.19 & $-$0.43 & $-$0.06 \\
$G_{BP},\,G_{RP}$ & $-$165.58 & $-$94.94 & $-$0.10 & $-$0.31 & $-$10.43 & $+$0.82 & -- & -- \\
$\ell,\,b,\,G_{BP},\,G_{RP}$ & -- & -- & $-$0.08 & $-$3.33 & $-$9.56 & $+$0.83 & -- & -- \\
$\varpi,\,G$ & $-$225.44 & $-$88.90 & -- & $-$0.47 & $-$10.03 & -- & $+$0.54 & $+$0.70 \\
$\varpi,\,G_{BP},\,G_{RP}$ & $-$162.26 & $-$101.06 & -- & $-$0.46 & $-$10.19 & $+$0.84 & -- & -- \\
$\varpi,\,G,\,G_{BP},\,G_{RP}$ & $-$177.49 & $-$97.29 & -- & $-$0.46 & $-$9.99 & -- & -- & -- \\
$\ell,\,b,\,\varpi,\,G$ & -- & -- & -- & $-$3.87 & $-$8.88 & -- & $+$0.69 & $+$0.79 \\
$\ell,\,b,\,\varpi,\,G_{BP},\,G_{RP}$ & -- & -- & -- & $-$3.10 & $-$9.18 & $+$0.83 & -- & -- \\
$\ell,\,b,\,\varpi,\,G,\,G_{BP},\,G_{RP}$ & -- & -- & -- & $-$3.23 & $-$9.84 & -- & -- & -- \\
$\ell,\,b,\,\mu_{\ell*},\,\mu_b$ & -- & -- & $-$0.27 & -- & -- & $-$0.18 & $-$0.37 & $-$0.08 \\
$\mu_{\ell*},\,\mu_b,\,G,\,G_{BP},\,G_{RP}$ & $-$223.25 & $-$102.14 & $-$0.13 & -- & -- & -- & -- & -- \\
$\varpi,\,\mu_{\ell*},\,\mu_b,\,G,\,G_{BP},\,G_{RP}$ & $-$223.06 & $-$96.28 & -- & -- & -- & -- & -- & -- \\
$\varpi,\,\mu_{\ell*},\,\mu_b$ & $-$233.23 & $-$86.79 & -- & -- & -- & $-$0.18 & $-$0.37 & $-$0.09 \\
$\ell,\,b,\,\mu_{\ell*},\,\mu_b,\,G,\,G_{BP},\,G_{RP}$ & -- & -- & $-$0.07 & -- & -- & -- & -- & -- \\
\end{longtable}
\endgroup

\begingroup\setlength{\tabcolsep}{4pt}
\begin{longtable}{l rrrrrrrrr}
\caption{Skill ($1 -$ RMSE / RMSE$_{\rm baseline}$) for $\omega$~Cen members outside the training and validation sets, class (d), 781 sources with measured $\ell,\,b,\,\varpi,\,\mu_{\ell*},\,\mu_b,\,G,\,G_{BP},\,G_{RP},\,v_r$ only. See the caption of Table~\ref{tab:skill:test:a} for a full description. Note the importance of the membership and selection criteria for $\omega$ Cen, as noted in Table \ref{tab:skill:omega_cen_not_A:c}. Even with that caveat, the negative skill drop $\mu_x$ results suggest that model has failed to learn $\omega$ Cen information. Despite being presented with 8 out of 9 parameters in those cases, the model is a poor point estimator of the proper motion.} \label{tab:skill:omega_cen_not_A:d}\\
\toprule
Cond.\ Pattern & $\ell$ & $b$ & $\varpi$ & $\mu_{\ell*}$ & $\mu_b$ & $G$ & $G_{BP}$ & $G_{RP}$ & $v_r$ \\
\midrule
\endfirsthead
\caption[]{Skill ($1 -$ RMSE / RMSE$_{\rm baseline}$) for $\omega$~Cen members outside the training and validation sets, class (d), 781 sources with measured $\ell,\,b,\,\varpi,\,\mu_{\ell*},\,\mu_b,\,G,\,G_{BP},\,G_{RP},\,v_r$ only (continued)}\\
\toprule
Cond.\ Pattern & $\ell$ & $b$ & $\varpi$ & $\mu_{\ell*}$ & $\mu_b$ & $G$ & $G_{BP}$ & $G_{RP}$ & $v_r$ \\
\midrule
\endhead
\midrule
\multicolumn{10}{r}{continued}\\
\endfoot
\bottomrule
\endlastfoot
uncond & $-$269.44 & $-$90.75 & $-$6.07 & $-$1.58 & $-$8.88 & $-$0.03 & $-$0.13 & $-$0.02 & $-$5.03 \\
$\ell,\,b$ & -- & -- & $-$6.25 & $-$10.89 & $-$8.10 & $-$0.01 & $-$0.01 & $-$0.02 & $-$5.04 \\
only $\varpi$ & $-$273.38 & $-$90.32 & -- & $-$0.54 & $-$11.34 & $-$0.39 & $-$1.05 & $-$0.16 & $-$5.08 \\
$\ell,\,b,\,\varpi$ & -- & -- & -- & $-$3.76 & $-$10.85 & $-$0.12 & $-$0.25 & $-$0.07 & $-$5.05 \\
$\ell,\,b,\,G$ & -- & -- & $-$6.39 & $-$10.63 & $-$8.45 & -- & $+$0.90 & $+$0.91 & $-$5.03 \\
$\ell,\,b,\,\varpi,\,\mu_{\ell*},\,\mu_b$ & -- & -- & -- & -- & -- & $-$0.12 & $-$0.25 & $-$0.08 & $-$4.32 \\
$G,\,G_{BP},\,G_{RP}$ & $-$275.80 & $-$93.29 & $-$6.37 & $-$1.26 & $-$8.64 & -- & -- & -- & $-$5.00 \\
$\ell,\,b,\,G,\,G_{BP},\,G_{RP}$ & -- & -- & $-$4.33 & $-$9.59 & $-$9.24 & -- & -- & -- & $-$5.11 \\
$\varpi,\,\mu_{\ell*},\,\mu_b,\,G,\,G_{BP},\,G_{RP},\,v_r$ & $-$55.50 & $-$26.65 & -- & -- & -- & -- & -- & -- & -- \\
$\mu_{\ell*},\,\mu_b,\,G,\,G_{BP},\,G_{RP},\,v_r$ & $-$66.19 & $-$30.03 & $+$0.13 & -- & -- & -- & -- & -- & -- \\
$\ell,\,b,\,\varpi,\,G,\,G_{BP},\,G_{RP},\,v_r$ & -- & -- & -- & $-$3.48 & $-$8.30 & -- & -- & -- & -- \\
$\ell,\,b,\,\varpi,\,\mu_{\ell*},\,\mu_b,\,G,\,v_r$ & -- & -- & -- & -- & -- & -- & $+$0.50 & $+$0.75 & -- \\
$\ell,\,b,\,\varpi,\,\mu_{\ell*},\,\mu_b,\,v_r$ & -- & -- & -- & -- & -- & $-$0.01 & $-$0.18 & 0.00 & -- \\
drop $\ell$ & $-$68.07 & -- & -- & -- & -- & -- & -- & -- & -- \\
drop $b$ & -- & $-$46.19 & -- & -- & -- & -- & -- & -- & -- \\
drop $\varpi$ & -- & -- & $+$0.36 & -- & -- & -- & -- & -- & -- \\
drop $\mu_{\ell*}$ & -- & -- & -- & $-$2.14 & -- & -- & -- & -- & -- \\
drop $\mu_b$ & -- & -- & -- & -- & $-$8.14 & -- & -- & -- & -- \\
drop $G$ & -- & -- & -- & -- & -- & $+$0.97 & -- & -- & -- \\
drop $G_{BP}$ & -- & -- & -- & -- & -- & -- & $+$0.83 & -- & -- \\
drop $G_{RP}$ & -- & -- & -- & -- & -- & -- & -- & $+$0.91 & -- \\
drop $v_r$ & -- & -- & -- & -- & -- & -- & -- & -- & $-$4.20 \\
only $\ell$ & -- & $-$91.51 & $-$4.74 & $-$9.27 & $-$9.01 & $-$0.08 & $-$0.22 & $-$0.04 & $-$5.18 \\
only $b$ & $-$301.71 & -- & $-$6.85 & $-$1.46 & $-$8.48 & $-$0.01 & $-$0.01 & $-$0.01 & $-$5.08 \\
only $\mu_{\ell*}$ & $-$287.24 & $-$90.83 & $-$2.20 & -- & $-$10.07 & $-$0.13 & $-$0.41 & $-$0.06 & $-$5.08 \\
only $\mu_b$ & $-$275.57 & $-$93.06 & $-$8.37 & $-$1.34 & -- & $-$0.01 & $-$0.01 & 0.00 & $-$5.03 \\
only $G$ & $-$249.09 & $-$90.67 & $-$6.45 & $-$1.42 & $-$8.65 & -- & $+$0.81 & $+$0.89 & $-$5.02 \\
only $G_{BP}$ & $-$231.65 & $-$92.78 & $-$6.38 & $-$1.52 & $-$8.78 & $+$0.85 & -- & $+$0.75 & $-$4.99 \\
only $G_{RP}$ & $-$273.21 & $-$89.57 & $-$6.25 & $-$1.51 & $-$8.81 & $+$0.88 & $+$0.63 & -- & $-$5.03 \\
only $v_r$ & $-$96.11 & $-$119.54 & $-$0.05 & $-$1.72 & $-$12.26 & $-$0.38 & $-$1.02 & $-$0.17 & -- \\
$\mu_{\ell*},\,\mu_b$ & $-$278.17 & $-$93.61 & $-$5.59 & -- & -- & $-$0.01 & $-$0.01 & $-$0.01 & $-$5.07 \\
$G_{BP},\,G_{RP}$ & $-$279.23 & $-$94.93 & $-$6.35 & $-$1.54 & $-$8.59 & $+$0.97 & -- & -- & $-$5.00 \\
$\ell,\,b,\,G_{BP},\,G_{RP}$ & -- & -- & $-$4.31 & $-$9.39 & $-$9.25 & $+$0.97 & -- & -- & $-$5.06 \\
$\varpi,\,G$ & $-$260.19 & $-$91.70 & -- & $-$0.50 & $-$11.25 & -- & $+$0.28 & $+$0.70 & $-$5.09 \\
$\varpi,\,G_{BP},\,G_{RP}$ & $-$184.78 & $-$103.06 & -- & $-$0.26 & $-$10.92 & $+$0.97 & -- & -- & $-$4.85 \\
$\varpi,\,G,\,G_{BP},\,G_{RP}$ & $-$189.59 & $-$105.49 & -- & $-$0.27 & $-$10.88 & -- & -- & -- & $-$4.83 \\
$\ell,\,b,\,\varpi,\,G$ & -- & -- & -- & $-$3.92 & $-$10.83 & -- & $+$0.72 & $+$0.86 & $-$4.99 \\
$\ell,\,b,\,\varpi,\,G_{BP},\,G_{RP}$ & -- & -- & -- & $-$4.47 & $-$10.61 & $+$0.97 & -- & -- & $-$5.03 \\
$\ell,\,b,\,\varpi,\,G,\,G_{BP},\,G_{RP}$ & -- & -- & -- & $-$4.17 & $-$10.69 & -- & -- & -- & $-$5.00 \\
$\mu_{\ell*},\,\mu_b,\,v_r$ & $-$86.57 & $-$43.95 & $+$0.30 & -- & -- & $-$0.17 & $-$0.55 & $-$0.07 & -- \\
$\ell,\,b,\,\mu_{\ell*},\,\mu_b,\,v_r$ & -- & -- & $+$0.26 & -- & -- & $-$0.05 & $-$0.28 & $-$0.01 & -- \\
$\ell,\,b,\,\mu_{\ell*},\,\mu_b$ & -- & -- & $-$5.13 & -- & -- & $-$0.01 & 0.00 & $-$0.02 & $-$4.74 \\
$\mu_{\ell*},\,\mu_b,\,G,\,G_{BP},\,G_{RP}$ & $-$301.27 & $-$96.20 & $-$5.58 & -- & -- & -- & -- & -- & $-$5.14 \\
$\varpi,\,\mu_{\ell*},\,\mu_b,\,G,\,G_{BP},\,G_{RP}$ & $-$259.14 & $-$116.81 & -- & -- & -- & -- & -- & -- & $-$5.11 \\
$\varpi,\,\mu_{\ell*},\,\mu_b$ & $-$275.12 & $-$95.39 & -- & -- & -- & $-$0.29 & $-$0.71 & $-$0.14 & $-$5.21 \\
$\ell,\,b,\,\mu_{\ell*},\,\mu_b,\,G,\,G_{BP},\,G_{RP}$ & -- & -- & $-$3.77 & -- & -- & -- & -- & -- & $-$4.79 \\
$\ell,\,b,\,v_r$ & -- & -- & $-$0.13 & $-$5.59 & $-$6.11 & $-$0.10 & $-$0.32 & $-$0.04 & -- \\
\end{longtable}
\endgroup

\end{document}